\documentclass[a4paper,11pt]{article}
\usepackage{fullpage}
\usepackage{todonotes,color}
\definecolor{lacre}{rgb}{0.67,0,0}
\usepackage[centertags]{amsmath}
\usepackage{amscd,amsfonts,amssymb,latexsym}
\usepackage{graphicx}
\usepackage{theorem}
\usepackage{enumerate}
\usepackage[toc,page]{appendix}
\usepackage{subcaption}
\usepackage{overpic}

\newcommand{\tm}{\tilde{\mu}}
\newcommand{\tb}{\tilde{b}}
\newcommand{\re}{\mbox{Re}}
\newcommand{\im}{\mbox{Im}}
\renewcommand{\d}{\mathrm{d}}

\newcommand{\al}{\alpha}

\newcommand{\ra}{\rightarrow}
\newcommand{\ee}{\mathrm{e}}
\newcommand{\ii}{\mathrm{i}}
\newcommand{\e}{\epsilon}
\newcommand{\R}{r^*}
\newcommand{\eps}{\epsilon}
\newcommand{\dd}{\partial}
\newcommand{\areax}{|\Omega|}
\newcommand{\m}{\mu}
\newcommand{\p}{\psi}
\newcommand{\del}{\nabla}
\newcommand{\delX}{{\nabla}}
\newcommand{\ki}{\chi}
\newcommand{\lap}{\nabla^2}
\newcommand{\lapX}{{\nabla}^2}
\newcommand{\order}{O}
\newcommand{\Vx}{V_x}
\newcommand{\Vy}{V_y}
\newcommand{\bX}{\mathbf{X}}

\newcommand{\bxi}{\boldsymbol{\xi}}
\newcommand{\bY}{\mathbf{Y}}
\newcommand{\oGpn}{\overline{G}'_{\mathrm{n}}}
\newcommand{\oGpd}{\overline{G}'_{\mathrm{d}}}
\newcommand{\oGpnreg}{\overline{G}'_{\mathrm{n,reg}}}
\newcommand{\oGpdreg}{\overline{G}'_{\mathrm{d,reg}}}
\newcommand{\Gpn}{G'_{\mathrm{n}}}
\newcommand{\Gpd}{G'_{\mathrm{d}}}
\newcommand{\Gn}{G_{\mathrm{n}}}

\newcommand{\Gnreg}{G_{\mathrm{n,reg}}}

\newcommand{\Gpnreg}{G'_{\mathrm{n,reg}}}
\newcommand{\Gpdreg}{G'_{\mathrm{d,reg}}}
\newcommand{\oGn}{\overline{G}_{\mathrm{n}}}
\newcommand{\oGd}{\overline{G}_{\mathrm{d}}}
\newcommand{\oH}{\overline{H}}
\newcommand{\oGnreg}{\overline{G}_{\mathrm{n,reg}}}
\newcommand{\oGdreg}{\overline{G}_{\mathrm{d,reg}}}
\newcommand{\oHreg}{\overline{H}_{\mathrm{reg}}}
\newcommand{\GG}{{\cal G}}

\newcommand{\oGG}{\overline{\cal G}}
\newcommand{\bx}{\mathbf{x}}
\newcommand{\by}{\mathbf{y}}
\newcommand{\inx}{\bar{\mathbf{x}}}

\newcommand{\bv}{\mathbf{v}}

\newcommand{\beq}{\begin{equation}}
\newcommand{\eeq}{\end{equation}}
\newcommand{\beqas}{\begin{eqnarray*}}
\newcommand{\eeqas}{\end{eqnarray*}}
\newcommand{\beqa}{\begin{eqnarray}}
\newcommand{\eeqa}{\end{eqnarray}}
\newcommand{\lx}{L_x}
\newcommand{\ly}{L_y}
\newcommand{\ph}{\hat{\psi}}
\newcommand{\finit}{\hat{f}}
\newcommand{\chiinit}{\hat{\chi}}

\theoremstyle{plain}

\newtheorem{rem}{Remark}

\author{M. Aguareles \and
  S.J. Chapman \and T. Witelski}
\title{Dynamics of spiral waves in the complex Ginzburg-Landau equation in bounded domains}

\begin{document}
\maketitle

\begin{abstract}
Multiple-spiral-wave solutions of the general cubic complex
Ginzburg-Landau equation in bounded domains are considered. We
investigate the effect of the boundaries on spiral motion under
homogeneous Neumann boundary conditions, for small values of the twist parameter
$q$. We derive explicit laws of motion for rectangular domains and we
show that the motion of spirals becomes exponentially slow when the
twist parameter exceeds a critical value depending on the size  
of the domain. The oscillation frequency of multiple-spiral patterns is
also analytically obtained.    
\end{abstract}


\section{Introduction}
The complex Ginzburg-Landau equation has a long history in physics. It
arises as
the  amplitude equation in the vicinity of a  
Hopf bifurcation in spatially-extended systems (see for instance \S 2
in \cite{kura84}), and so describes active media
close to the onset of  pattern formation \cite{aranson02,
  hohenberg2015introduction}. 
 The simplest 
examples of such media are chemical oscillations such as the famous
\emph{Belousov-Zhabotinsky} reaction \cite{zaikin}. More complex
examples include 
thermal convection of binary fluids \cite{AransonRef6}, transverse
patterns of high intensity light \cite{moloney90}; more recently, it has also been used to model the interaction of several species in some ecological systems \cite{mowlaei2014}. 

The general cubic complex Ginzburg-Landau equation is given by
\beq
\frac{\dd \Psi}{\dd t} =
\Psi-(1+\ii a) \ |\Psi|^2 \Psi+(1+\ii b) \lap \Psi,
\label{CGL0}
\eeq
where $a$ and $b$ are real parameters and $\Psi$ is a complex field
representing the amplitude and phase of the modulations of the
oscillatory pattern.   

Of particular interest are ``defect'' solutions of \eqref{CGL0} in
$\mathbb{R}^2$. Solutions with a single defect are
 characterised by the fact that $\Psi$
has a single zero around which its phase varies by an
integer multiple of $2\pi$ (that we shall denote as $n$),
known as the winding number.
When
$a=b$ the  isophase lines of such a solution are straight lines
emanating from the zero (see \cite{hagan82, paullet94} for more
details). If $a\neq b$, the 
isophase lines bend to form spirals, left-handed or right-handed
depending on the sign of $n$.
The time dependence of this type of solution
appears as a global oscillation, so that
$\Psi(\bx,t) =\ee^{-\ii\omega  t}\p(\bx)$, where $\omega$ is not free
but needs to be determined as part of the problem. Moreover  
 $\p(\bx) = f(r)\ee^{\ii n\phi+\ii \varphi(r)}$,
with $r$ and $\phi$ the polar radial and azimuthal variables
respectively, where  $f$ and $\varphi$ satisfy  a system of
ordinary differential equations (see \cite{hagan82} for
the derivation and asymptotic properties of these solutions
 and \cite{kopell81} for a result on existence and
uniqueness of solution).

We are concerned here with solutions containing multiple defects
or spirals (we use the terms interchangeably). Such
complex patterns may be 
understood in terms of the position of the centres of the
spirals---if the motion of the 
defects can be determined, much of the dynamics of the solution can be
understood.

Although the time-dependence is now more complicated, it is still
convenient to factor out a global phase oscillation from the 
wavefunction by 
writing
$$\Psi= \ee^{-\ii\omega t}\sqrt{\dfrac{1+\omega b}{1+ab}}\psi,\quad t
= \dfrac{t'}{1+\omega b}, \quad
(x,y)=\sqrt{\dfrac{1+b^2}{1+b\omega}}(x',y')$$ 
in \eqref{CGL0} to give
\beq
\label{pdeCGL2}
(1-\ii b)\dfrac{\partial\psi}{\partial t'}=(1-|\psi|^2)\psi+\ii
q\psi(1-k^2-|\psi|^2)+\nabla^2\psi, 
\eeq
where $q=(a-b)/(1+ab)$ and $k$ is such that 
\beq
\label{disp}
q(1-k^2)=\frac{\omega-b}{1+b\omega}. 
\eeq
The parameters $q$ and $k$ are usually referred to as the \emph{twist
  parameter} and \emph{asymptotic wavenumber} respectively. We note that $q$ is a parameter of the problem, but $k$, like $\omega$,  is not free but determined as part of the solution.

Solutions with finitely-many zeroes evolve in time in such a
way that the spirals preserve their local structure (at least for
$|n|=1$, which is the case we consider here). When the twist
parameter vanishes (that is if $a=b$), multiple-spiral solutions in
$\mathbb{R}^2$ move on a time-scale that is proportional to the
logarithm of the inverse of the typical spiral separation
\cite{neu90}. As $q$ increases the interaction
weakens and eventually becomes exponentially small in the
separation. When $q$ becomes of order one 
numerical simulations reveal that the dynamics becomes ``frozen'', evolving on a very long timescale,  with
 a set of virtually independent spirals separated by shock
lines \cite{bohr97, das2013unlocking}. The singular role of the twist
parameter, as pointed out in \cite{pismen03}, is  to interpolate
between these two very dissimilar behaviours, namely a strong
(algebraic) interaction for small values of $q$ and an exponentially
weak interaction as $q$ approaches the critical value of $q_{c}=\pi/(2
\log d)$, where $d$ is the   spiral separation, as is
shown in \cite{AgChWi08, AgCh09}.  

For a finite set of spirals in the whole of  $\mathbb{R}^2$, the
asymptotic wavenumber $k$ represents the wavenumber of the phase of
$\psi$ at infinity, that is to say, $k=\lim_{r\to\infty}\arg(\p)/r.$
Thus expression \eqref{disp} represents a dispersion relation.
 For small $q$,
on an infinite domain, 
it turns out that $k$
is exponentially small in $q$.

The earliest work on a law of motion for spirals is that of Biktashev \cite{biktashev89}, who derived a law
of motion and the asymptotic wavenumber $k$ in the limit $q \ra 0$ for a pair of spirals separated
by a distance large compared to $\ee^{\pi/2q}$ (or equivalently for a spiral in a half-space, far from the boundary). 
In \cite{pismen92} Pismen \& Nepomnyashchy extend the results of Biktashev to a pair of spirals separated by distances of $O(\ee^{\pi/2q})$. Rather than deriving a law of motion, their main aim was to establish the non-existence of a bound state, that is, a solution in which the spirals move at uniform speed in the direction perpendicular to the line of centres. Unfortunately there are a number of mistakes in \cite{pismen92}, which we elaborate on in Appendix \ref{appA}. 
In Aranson et al. \cite{aranson91b,aranson93} two spirals are again considered, and in the latter a law of motion is derived in the
limit in which the separation is much greater than $\ee^{\pi/2q}$. However \cite{aranson93} does
not require $q$ to be small. On the other hand
\cite{aranson93} seems to assume that the wavenumber $k$ is the same as that of a single spiral.
Brito et al. \cite{brito03} consider the motion of a system of $n$ spirals. They take the equations for
a pair of spirals derived in \cite{aranson93} and sum over all pairs to calculate the motion of each. As
in \cite{aranson93}, they take the wavenumber $k$ to be the same as that for a single spiral, and again the
equations used are valid only when the separation of spirals is much greater than $\ee^{ \pi/2q}$.
The methods in all of these works do not easily generalise to more than two spirals, to spirals in bounded domains, or to spirals not so widely separated.

In our previous work \cite{AgChWi08,
  AgCh09} we used perturbation techniques to determine the asymptotic
wavenumber and to obtain a law of motion for the centres of  an
arbitrary arrangement of spirals
in the whole of $\mathbb{R}^2$. 
In this paper we focus on multiple-spiral  solutions on  a bounded
domain in  $\mathbb{R}^2$ when the twist parameter $q$ is small. We consider
homogeneous Neumann (zero flux)  boundary 
conditions; the extension to periodic boundary conditions
is easy to make, and together these cover the vast majority of  
numerical computations and  physical applications.
 We  extend our results in \cite{AgChWi08,
  AgCh09} to derive laws of motion for spirals 
confined to a general bounded domain  $\Omega$.  
The law of motion we find will be given in terms of the Green's
function for a modified Helmholtz equation on $\Omega$, which encodes how
the shape of the domain affects the motion of defects.
  By way of illustration, we then
focus on rectangular domains where we obtain explicit laws of motion
for a finite set of spirals.

In the limit 
$q \ra 0 $ the interaction of spirals passes from
algebraic to  exponentially small as separation between spirals
increases.
To simulate \eqref{CGL0} numerically
one usually 
assumes that a large rectangular domain will suffice to  approximate
the solution on 
$\mathbb{R}^2$. 
The question then arises as to whether any interesting observed
behaviour,
such as bound states or a change in the direction or sign of the
interaction between spirals, is actually present in $\mathbb{R}^2$ or is an
artifact of truncation.

One of our main results is to show how the size of the domain affects 
the interaction between spirals. In particular,
we find that the  motion of spirals becomes exponentially small 
only when the diameter of the domain approaches $\ee^{\pi/2 q}$, which
gives an indication of the difficulty of approximating the solution on
an infinite domain with that on a truncated domain.

A second important goal of this paper is to describe the role of the
boundaries as a selection mechanism for the oscillation frequency
$\omega$, and hence for the asymptotic wavenumber $k$, which we also
obtain. In this case we find that as the diameter of the domain
approaches $\ee^{\pi/2 q}$, the asymptotic wavenumber $k$ also 
shifts from being algebraic to becoming exponentially small in $q$.  

For ease of exposition we shall take $b=0$  so that, dropping the
primes henceforth, the  equation
 we  consider is 
\begin{equation}
\frac{\dd \p}{\dd t} = \lap \p+(1-|\p|^2)\p+\ii q\p(1-k^2-|\p|^2).
\label{CGL}
\end{equation}
The extension to $b\neq 0$ is briefly analysed in Appendix \ref{secApA}.

The paper is organised as follows. Sections  \ref{canonical} and 
\ref{middle} are devoted to obtaining expressions for the laws of motion
of the centres of the spirals in general bounded domains. We start in
Section  \ref{canonical} by considering what we denote as the
\emph{canonical} or \emph{far-field} scale, which corresponds to
considering domains of diameter $\ee^{\pi/2 q}$. Then, in Section 
\ref{middle}, we consider domains of diameter $\ll\ee^{\pi/2 q}$, which
 provides a new set of
equations for  spiral motion in what we denote as the \emph{near
  field}. 
In
Section  \ref{rectangular} we consider the particular case of
rectangular domains and we obtain  explicit laws of motion
in both the far and near field. In particular we find that the 
interaction between the spirals changes from being exponentially small
and mainly in the azimuthal direction when the parameters are in the
far field regime to becoming algebraic and with a radial component in
the near field. Furthermore, the asymptotic wavenumber of the patterns is exponentially small in the 
far-field scaling but proportional to the square root of $q$ and the
diameter of the domain  in the near field. To reconcile these two regimes,  a composite law of
motion that is valid in both near and far fields  is
proposed. In Section \ref{numerics} this composite law of motion is used to compare the
trajectories of the spirals with direct numerical simulations of the
original system of partial differential equations \eqref{CGL}. Finally,
in Section  \ref{conclusions}, we present our conclusions. 

%



\section{Interaction of spirals in bounded domains at 
the canonical   scale} 
\label{canonical} 
In this section we derive laws of motion for the centres of a finite
set of spirals with unitary winding numbers confined in general
bounded domains with homogeneous Neumann boundary conditions. The law of motion
and the corresponding asymptotic wavenumber, $k$, are given explicitly
in terms of the parameter $q$, which is assumed to be small.

In what
follows we assume that the centres of the spirals are separated from
each other and from the boundaries by distances which are large in
comparison with the  core radius of the spirals. By core radius
we mean the lengthscale over which the modulus of $\psi$ recovers its
equilibrium value close to one (for small $q$) from its value of zero
at the spiral centre. We see from  (\ref{CGL}) that the
core radius is $O(1)$, which means we need the
domain to be large if the spirals are to be well-separated.
We quantify this by introducing the inverse of the domain diameter as 
the  small parameter $\e$, and we suppose that spirals are separated
by distances of $O(1/\e)$.

We therefore consider the system
\begin{equation}
\begin{split}
\p_t &= \p(1 -|\p|^2) + \ii q\,\p(1-k^2-|\p|^2)+\lap\p \quad
\textrm{in}\quad \Omega\label{9}\\ 
\frac{\dd \p}{\dd n}&=0 \quad \textrm{on}\quad \dd \Omega,
\end{split}
\end{equation}
with parameters $0<q\ll1$ and $0<k\ll 1$. As  in
unbounded domains (see \cite{AgChWi08} and \cite{AgCh09}), the
relationship between $\e$, $q$ and $k$ plays a special role  giving
place to different types of interaction. In particular, we shall
show  it is the combination $\al=kq/\e$ that determines the nature of
the interaction between spirals. In this section we shall assume
that $\al$ is an order-one constant, and we shall show that this is
equivalent to assuming that $1/\e$ is of order $\ee^{\pi/(2q)}$. 

\subsection{Outer solution}
\label{sec:2.1}
We follow the same notation as
\cite{AgChWi08} and \cite{AgCh09},  denoting by $\bX=\e \bx$ the
outer space variable
and $T=\m \e^2 t$ the slow time scale on which the spirals 
 interact. At this stage $\m$ is an unknown small
parameter. We will later determine that $\mu = 1/\log(1/\e)$. 

Since in this section we are assuming that $\al=kq/\e=\order(1)$, we
write (\ref{9}) in the outer region as 
\begin{equation}
\e^2\m\p_T =
(1+\ii q)\,\p(1-|\p|^2)-\ii\frac{\e^2\al^2}{q}\p+\e^2\lapX\p,\quad
\textrm{in $\Omega$}\label{10} 
\end{equation}
along with homogeneous Neumann boundary conditions at the domain
boundaries, where $\delX$ now represents the gradient with respect to
$\bX$. We express the solution in amplitude-phase form as $\p=fe^{\ii\ki}$,
giving 
\begin{eqnarray}
\m \e^2 f_T&=& \e^2\delX^2 f-\e^2 f|\delX\ki|^2+f(1-f^2)\label{11},\\
\m\e^2 f^2\ki_T&=&\e^2 \delX\cdot(f^2 \delX
   \ki)+qf^2(1-f^2)-\frac{\e^2\al^2}{q} f^2, \label{12} 
\end{eqnarray}
in $\Omega$, where now the boundary conditions for $f$ and $\ki$ are 
\begin{equation*}
\frac{\dd f}{\dd n}=\frac{\dd \ki}{\dd n}=0\quad
\textrm{on}\quad \dd\Omega. 
\end{equation*}
Expanding in power series in $\e^2$ as 
$$f(\bX,T;\e,q)\sim f_0(\bX,T;q)+\e^2 f_1(\bX,T;q)+ \e^4 f_2(\bX,T;q)+\ldots,$$
$$\ki(\bX,T;\e,q)\sim \ki_0(\bX,T;q)+\e^2 \ki_1(\bX,T;q)+ \e^4 \ki_2(\bX,T;q)+\ldots,$$ 
the leading and first-order terms in (\ref{11})  give
\begin{eqnarray}
 f_0=1,\quad f_1=-\frac{1}{2}|\del \ki_0|^2.\label{P11}
\end{eqnarray}
Substituting (\ref{P11}) into (\ref{12}) gives
\begin{equation}
\begin{split}
\m \frac{\dd\ki_{0}}{\dd T}&=\lap
\ki_0+q|\del\ki_0|^2-\frac{\al^2}{q}\quad\textrm{in}\quad\Omega \label{chi0eqn}\\ 
 \frac{\dd\ki_0}{\dd n}& = 0 \quad\textrm{on}\quad \dd\Omega.
\end{split}
\end{equation}
 We proceed as in \cite{AgCh09} and expand $\ki_0$ in terms of the
 small parameter 
$q$ as
$\ki_0\sim\ki_{00}/q+\ki_{01}+\cdots$ to find, at leading order,
\begin{equation}
\label{22}
\begin{split}
0&=\del^2 \ki_{00}+|\del \ki_{00}|^2-\al^2 \quad\textrm{in}\quad\Omega,\\
\frac{\dd\ki_{00}}{\dd n} &= 0 \quad\textrm{on}\quad \dd\Omega.\\
\end{split}
\end{equation}
Using the Cole-Hopf transformation $\ki_{00}=\log h_0$, equation \eqref{22} is
transformed into the linear problem 
\begin{equation}
\label{23can}
 \begin{split}
  0&=\lap h_0-\al^2 h_0\quad\textrm{in}\quad\Omega,\\
\frac{\dd h_0}{\dd n}& = 0\quad\textrm{on}\quad \dd\Omega,\\
 \end{split}
\end{equation}
Note that although $\chi_0$ is multivalued,  $\chi_{00}$ is single-valued (the $n_j \phi$ terms in the phase
appear in $\chi_{01}$) so that there is no issue with applying the
Cole-Hopf transformation.
If we had not expanded in $q$ but written simply $\chi_0 = (1/q) \log(h)$ as in \cite{pismen92}, then the multivaluedness of $\chi_0$ would induce a  multivaluedness in $\log(h)$ which precludes the superposition of spiral solutions, even though the equation for $h$ is linear. Of course, the multivaluedness and its associated complications have not disappeared, but will appear in the correction term $\chi_{01}$. However, we will find that we can determine the asymptotic law of motion of spirals without calculating $\chi_{01}$.

In order to match to a spiral solution locally near the origin
$h_0$ should have the form \mbox{$h_0 \sim -\beta \log |\bX|$} as  $\bX
\ra {\bf 0}$ for some constant $\beta$ \cite{AgCh09}. Thus, a solution
with $N$ spirals at 
positions $\bX_1, \ldots, \bX_N$ should satisfy (\ref{23can}) along with
 \beq h_0 \sim -\beta_j\log |\bX-\bX_j| \mbox{ as } \bX \ra \bX_j, \mbox{
   for } j = 1,\ldots,N.\label{23can1}
\eeq
The solution to (\ref{23can})-(\ref{23can1}) is therefore 
\begin{equation}
\label{solGc}
h_0 = - 2 \pi \sum_{j=1}^N \beta_j \Gn(\bX;\bX_j) =
\GG\left(\bX;\al(T),\beta_{1}(T),\ldots,\beta_{N}(T), \bX_1(T), \bX_2(T),
  \ldots, \bX_N(T)\right), 
\end{equation}
say, where $\Gn(\bX;\bY)$ is the Neumann
Green's function for the modified Helmholtz equation in $\Omega$, satisfying
\begin{equation}
\label{NeumannGreens}
  \lap \Gn-\al^2 \Gn=\delta(\bX-\bY)\quad\textrm{in}\quad\Omega,\qquad
\frac{\dd \Gn}{\dd n} = 0\quad\textrm{on}\quad \dd\Omega,
\end{equation}
and we have been explicit about the dependence of $\GG$ on the value
of $\alpha$, the weights $\beta_j$, and the position of the spirals
$\bX_j$, all of which may depend on $T$.

\subsection{Inner solution}
\label{innercanonical}
We rescale close to the centre of a spiral
 $\bX_{\ell}$ by writing \mbox{$\bX = \bX_{\ell} + \e \inx $} to give
\begin{eqnarray*}
\e\m \left(\e f_T - \frac{\d \bX_{\ell}}{\d T}\cdot\del f\right) &=& 
\del^2 f -f|\del   \ki|^2+(1-f^2)f,\\     
\e \m f^2\left(\e \ki_T-\frac{\d\bX_{\ell}}{\d T}\cdot\del \ki\right) &=&
           \del  \cdot(f^2\del  \ki)+q(1-f^2)f^2-\frac{\e^2 \al^2f^2}{q},
\end{eqnarray*}
or equivalently
\begin{equation}
\e\m\left(\e\p_T-\frac{\d\bX_{\ell}}{\d T}\cdot\del\p\right) =
\del^2\p+(1+\ii q)(1-|\p|^2)\p-
\ii \frac{\e^2\al^2}{q}\p,\label{innerpsi}
\end{equation}
where $\del$ represents now  the gradient with respect to the inner
variable $\inx$.
Since we assume that the distance between the spiral centre and the boundary is
much greater than the core radius, the inner equations must be
solved on an unbounded domain, with conditions at infinity that come
from matching with the outer solution. Thus the solution in the inner
region mirrors that in \cite{AgCh09}. 

Expanding $f \sim f_0(\inx, T;q,\m)+\e f_1(\inx, T;q,\m)+\e^2
f_2(\inx, T;q,\m)+\ldots$ and $\ki \sim
\ki_0(\inx, T;q,\m)+\e\ki_1(\inx, T;q,\m)+\ldots$, or equivalently
$\p\sim\p_0(\inx, T;q,\m)+\eps \p_1(\inx, T;q,\m)+\ldots$, the leading-order
equation is 
\begin{equation}
0 = \del^2 \p_0 + (1+\ii q)\p_0(1-|\p_0|^2),\label{psi0eqn}
\end{equation}
with solution $f_0 = f_0(r;q)$ and $\ki_0 = n_{\ell}\phi +
\varphi_0(r,T;q)$, where $r$ and $\phi$ are the radial and azimuthal
variables with respect to the spiral's centre, 
$|n_\ell|=1$ is the spiral's winding number, and  $f_0$ and
$\varphi_0$ satisfy ordinary differential equations in $r$ which, as indicated, also
depend on the small parameter $q$. Note that, although (\ref{psi0eqn}) does not depend on $T$, the matching condition with the outer solution causes $\varphi_0$ to depend parametrically on $T$. 
Expanding further in $q$ as  $f_0 \sim f_{00} +f_{01}q+
f_{02}q^2+\cdots$ and $\varphi_0 \sim
\varphi_{00}/q+\varphi_{01}+\varphi_{02}q+\cdots$, 
gives $\varphi_{00} = \varphi_{00}(T)$, $\varphi_{01} =\varphi_{01}(T)$ and also 
\begin{eqnarray}
\label{ki0}
&& f_{00}''+\frac{f_{00}'}{r}-\frac{f_{00}}{r^2}+(1-f_{00}^2)f_{00} =
  0,\\
\label{ki2}
&& \varphi_{02}'(r) = -\frac{1}{r f_{00}^2}\int_{0}^{r} s
  f_{00}^2(1-f_{00}^2) \, ds,
\end{eqnarray}
with boundary conditions $ f_{00}(0)= 0$ and, to match with (\ref{P11}),
$\lim_{r\to \infty}
f_{00}(r) = 1$.
Note that we allow the (constant in space) terms $\varphi_{00}/q$ and
$\varphi_{01}$ in order to enable $\varphi$ to match with the outer
solution $\chi_0 \sim \chi_{00}/q$ (though in fact we will not worry
about these terms further since  we can obtain all
the information we need by matching derivatives of $\varphi$). The
existence of a unique solution for $f_{00}$ has been shown in
\cite{greenberg}. 

At first order in $\e$ we find
\begin{equation}
\label{1orderinner}
-\m \frac{\d\bX_{\ell}}{\d T} \cdot \del \p_0 = \del^2
 \p_1+(1+\ii q)(\p_1(1-2|\p_0|^2)-\p_0^2\p_1^*),
\end{equation}
or equivalently, in terms of $f_1$ and $\ki_1$,
\begin{eqnarray}
- \m \frac{\d\bX_{\ell}}{\d T}\cdot\del f_0 &=& \del^2 f_1
-f_1|\del  \ki_0|^2
-2 f_0 \del  \ki_0 \cdot \del \ki_1+
f_1 - 3 f_0^2 f_1,\label{firstin1}\\
- \m f_0^2 \frac{\d\bX_{\ell}}{\d T}\cdot\del \ki_0 &=& \del \cdot(f_0^2\del
  \ki_1)+\del \cdot(2 f_0 f_1\del
  \ki_0)
+2 q f_0 f_1 -4 q f_0^3 f_1.\label{firstin2}
\end{eqnarray}
Note that we retain the terms proportional to $\mu$ in these
equations since we will later find that
 \mbox{$\mu = O(q) = O(1/|\log \eps|))$}.

\subsection{Inner limit of the outer solution}
We define the regular part of the outer solution $\GG$ near the $\ell$th spiral
by setting 
\begin{equation}
\GG^\ell_{\mathrm{reg}}(\bX)=\GG(\bX) +\beta_{\ell}\log
|\bX-\bX_\ell(T)|.\label{Gell}
\end{equation}
Then, from (\ref{solGc}), as $\bX$ approaches $\bX_\ell$, we find
\begin{equation*}
h_0 \sim -\beta_{\ell}\log |\bX-\bX_\ell| + 
\GG^\ell_{\mathrm{reg}}(\bX_{\ell})+(\bX-\bX_\ell)\cdot\del
\GG^\ell_{\mathrm{reg}}(\bX_{\ell})+\ldots.  
\end{equation*}
Thus, written in terms of the inner variables,
\begin{equation}
\label{40}
\ki_{0} \sim \frac{1}{q}\log h_0
\sim\frac{1}{q}\log\left(-\beta_{\ell}\log(\e r) +
  \GG^\ell_{\mathrm{reg}}(\bX_{\ell})\right)+\frac{\e\inx\cdot\del
  \GG^\ell_{\mathrm{reg}}(\bX_{\ell})}{q\left(-\beta_{\ell}\log (\e r) +
    \GG^\ell_{\mathrm{reg}}(\bX_{\ell})\right)}+\ldots, 
\end{equation}
where $r=R/\e=|\bX-\bX_\ell(T)|/\e$.
\subsection{Outer limit of the inner solution}
\label{sec:outerlimitofinner}
Using  \eqref{ki2} along with the fact that $f_{00}\sim
1-1/r^2$ as $r\to\infty$, it is found that 
\begin{equation}
\label{bc1}
\frac{\dd \varphi_{02}}{\dd r} \sim -q\, \frac{\log r +
  c_{1}}{r} + \ldots,
\end{equation}
as $r \to \infty$, where $c_{1}$ is a constant given by \cite{hagan82}
\[
c_{1} = \lim_{r\to\infty}\left(\int_0^{r}f_0^2(s)\left(1-f_0(s)^2\right)s
\, \d s - \log r\right) \approx -0.098.\]
However, in order to match with the outer expansion we need the outer
limit of the whole expansion in $q$. This can be found to be of the
form
\begin{eqnarray}
\label{innercapouter1}
f_0 &\sim&
1+\frac{1}{r^2}\sum_{i=0}^{\infty}C_i\big(q(\log r +c_1)\big)^{2i}+\cdots,\\
\label{innercapouter2}
\frac{\dd \ki_0}{\dd r} &\sim&
-\frac{1}{r}\sum_{i=0}^{\infty}D_i\big(q(\log r +c_1)\big)^{2i+1}+\cdots,
\end{eqnarray}
where $C_i >0$ and $D_i >0$ are constant values independent of
$q$. The necessity of taking all the terms in $q$ when
matching can be seen, since the expansion in $q$ is valid only when
$q(\log r+c_1)\ll 1$. When $\al=\order(1)$, $q$ turns out to be $O(1/\log
\e)$ and thus all the terms in \eqref{innercapouter1}-\eqref{innercapouter2} are the same order. We can sum all these terms in
the outer limit of the inner expansion using the same method as in
Section 3.3.1 in \cite{AgCh09}. The idea is to rewrite the
leading-order (in $\eps$) inner equations in terms of the outer
variable $R = \e r$ to obtain 
\begin{eqnarray}
0 &=& \e^2(\del^2 f_0 -f_0|\del\ki_0|^2)+(1-f_0^2)f_0,\label{inout1}\\
0 &=& \e^2\del\cdot (f_0^2\del\ki_0)+q(1-f_0^2)f_0^2. \label{inout2}
\end{eqnarray}
We  now  expand again  in powers of $\e$ as $\ki_0 \sim
\widehat{\ki}_{00}(r,\phi;q) 
+\e^2\widehat{\ki}_{01}(r,\phi;q)+\cdots$ and $f_0 \sim \widehat{f}_{00}(r,\phi;q) +\e^2\widehat{f}_{01}(r,\phi;q)+\cdots$.
The leading-order term in this expansion
$\widehat{\ki}_{00}(r,\phi;q)$ is just 
the first term (in $\e)$ in the outer expansion of the leading-order
(in $\eps$) inner solution, including all the terms in $q$. Substituting 
these expansions into (\ref{inout1})--(\ref{inout2}) gives
$\widehat{f}_{00}=1$, $\widehat{f}_{01} = -\frac{1}{2} |\del 
\widehat{\ki}_{00}|^2$ and
\begin{equation*}
0 = \del^2\widehat{\ki}_{00} + q|\del\widehat{\ki}_{00}|^2,
\end{equation*}
that is a Riccati equation which can be linearised
with the change of variable $\widehat{\ki}_{00} = (1/q)\log\widehat{h}_0$
to give $\del^2\widehat{h}_0= 0$.

Since  $\widehat{\ki}_{00}= n_{\ell}\phi + \widehat{\varphi}(R)$ we set
$\widehat{h}_0=
e^{qn_{\ell}\phi}e^{q\widehat{\varphi}(R)} = e^{qn_{\ell}\phi}H_0(R)$ to give
\begin{equation*}
H_0''+\frac{H_0'}{R}+q^2\frac{H_0}{R^2} = 0,
\end{equation*}
with solution
\begin{equation}
\label{solutionH}
H_0 = A_{\ell}(q)\e^{-\ii q n_{\ell}} R^{\ii q n_{\ell}}+B_{\ell}(q)\e^{\ii q n_{\ell}} R^{-\ii q n_{\ell}},
\end{equation}
where $A_{\ell}$ and $B_{\ell}$ are constants that depend on $q$ which may be
different at each vortex, and the factors $\e^{\pm iq n_{\ell}}$ are
included to facilitate their determination by comparison with the solution in the inner variable.
To determine  $A_{\ell}$ and $B_{\ell}$ we need to write $\hat{\ki}_{00}$ in terms
of $r$, expand in powers of $q$, and compare with the limit as $r \ra \infty$ of the expansion in powers of $q$ of $\varphi_0$, that is, with (\ref{innercapouter2}). Since we do not know all the terms $D_i$, only that $D_0=1$ (from (\ref{bc1})), we will only be able to determine the first two terms in the $q$ expansion of $A_{\ell}$ and $B_{\ell}$. However, this will be enough to determine the leading-order law of motion.

Writing the constants in powers of $q$  as $A_{\ell}(q)\sim A_{\ell 0}/q +A_{\ell 1} +q A_{\ell 2} +\cdots$ and $B_{\ell}(q) \sim B_{\ell 0}/q +B_{\ell 1} +q B_{\ell 2} +\cdots$,
and expressing $H_0$ in terms of $r$ we find
\begin{eqnarray*}
H_0(r) & =&  A_{\ell}(q) \ee^{\ii qn_{\ell}\log r} + B_{\ell}(q) \ee^{-\ii qn_{\ell}\log r}\\
&\sim &\frac{A_{\ell 0}+B_{\ell
        0}}{q}+A_{{\ell}1}+B_{{\ell}1}+(A_{{\ell}0}-B_{{\ell}0})\ii
        n_{\ell}\log r\\ 
&& \mbox{ }+q\left(A_{{\ell}2}+B_{{\ell}2}+(A_{{\ell}1}-B_{{\ell}1})\ii n_\ell\log
r-\frac{(A_{{\ell}0}+B_{{\ell}0})}{2}\log^2 r\right)+\cdots, 
\end{eqnarray*}
so that
\begin{eqnarray*}
\frac{\dd \widehat{\ki}_{00}}{\dd r}& =& \frac{H_0'(r)}{qH_0(r)}\sim\frac{n_{\ell}(A_{{\ell}0}-B_{{\ell}0})\ii}{r(A_{{\ell}0}+B_{{\ell}0})}+q\left(\frac{(A_{{\ell}1}-B_{{\ell}1})}{(A_{{\ell}0}+B_{{\ell}0})}\frac{n_{\ell}\ii }{r}-\frac{\log
  r}{r}\right.\\
&+&\left.\frac{(A_{{\ell}0}-B_{{\ell}0})^2}{(A_{{\ell}0}+B_{{\ell}0})^2}\frac{\log
    r}{r} -\left(\frac{\ii
    (A_{{\ell}0}-B_{{\ell}0})(A_{{\ell}1}+B_{{\ell}1})}{(A_{{\ell}0}+B_{{\ell}0})^2}
    \right)\frac{n_{\ell}}{r}\right) + \cdots 
\end{eqnarray*}
Comparing with (\ref{bc1}) (and recalling that $\dd\varphi_{00}/\dd r =  \dd\varphi_{01}/\dd r=0$ and $n_\ell = \pm 1$) 
we see that
\begin{align}
\label{A0}
A_{\ell 0}-B_{\ell 0} =& 0,\\
\label{A1}
\frac{(A_{\ell 1}-B_{\ell 1})}{A_{\ell 0}+B_{\ell 0}}\ii  =& -n_{\ell}c_{1} \quad \textrm{for $\ell =
  1, \ldots, N$},
\end{align}
where $N$ is the total number of spirals.
The remaining equations determining $A_{\ell}$ and $B_{\ell}$ will be fixed when
matching with the outer region.

\paragraph{Outer limit of the first-order inner solution}
We do the same with the first-order (in $\eps$) inner solution. The
details of the 
calculations, which we summarize in what follows, are the same as  in
Section 4.3.4 in \cite{AgCh09}.  
We first write equation (\ref{firstin1})-(\ref{firstin2}) in terms of the outer
variable to give
\begin{align*}
- \e \m \frac{\d\bX_{\ell}}{\dd T}\cdot\del f_0 &=\e^2 \del^2 f_1
- \e^2 f_1|\del  \ki_0|^2
-2 \e^2 f_0 \del  \ki_0 \cdot \del \ki_1+
f_1 - 3 f_0^2 f_1,\\
- \e\m  f_0^2  \frac{\d\bX_{\ell}}{\d T}\cdot\del \ki_0 &=\e^2 \del
 \cdot(f_0^2 \del
  \ki_1)+\e^2\del \cdot(2 f_0 f_1\del
  \ki_0)
+2 q f_0 f_1 -4 q f_0^3 f_1.
\end{align*}
We now expand in powers of $\e$ as $\ki_1 \sim \widehat{\ki}_{10}(r,\phi;q)/\e
+\widehat{\ki}_{11}(r,\phi;q)+\cdots$ and $f_1 \sim \widehat{f}_{10}(r,\phi;q)
+\e\widehat{f}_{11}(r,\phi;q)+\cdots$ 
to give $\widehat{f}_{10} =  0$, $\widehat{f}_{11}= -
\del\widehat{\ki}_{00}\cdot \del 
\widehat{\ki}_{10}$ and 
\begin{equation}
- \m  \frac{\d\bX_{\ell}}{\d T}\cdot\del \widehat{\ki}_{00} =
\del^2 \widehat{\ki}_{10}+ 2 q \del \widehat{\ki}_{00} \cdot \del
\widehat{\ki}_{10}.\label{chi1a}
\end{equation}
Motivated by the transformation we applied to $\widehat{\ki}_{00}$ we
write $ \widehat{\ki}_{10} =\widehat{h}_1/(q \widehat{h}_0) = \widehat{h}_1 e^{- q
    \widehat{\ki}_{00}}/q $ and (\ref{chi1a}) becomes
\[
- \m  \frac{\d\bX_{\ell}}{\d T}\cdot\del \widehat{\ki}_{00}  =
\frac{e^{-q\widehat{\ki}_{00}}}{q}
\del^2 \widehat{h}_1.
\]
Writing $\widehat{\ki}_{00}$ in terms of $\widehat{h}_0$ gives
\beq
- \m  \frac{\d\bX_{\ell}}{\d T}\cdot\del \widehat{h}_0 =
\del^2 \widehat{h}_1.
\label{eqki1}
\eeq
Writing the velocity as
\[ \frac{\d\bX_{\ell}}{\d T} = (V_1,V_2)\]
and recalling that $\widehat{h}_0 = \ee^{qn_{\ell}\phi} H_0(R)$, the left hand
side of \eqref{eqki1} gives
\beqas
\lefteqn{ - \m  \frac{\d\bX_{\ell}}{\d T}\cdot \left(\frac{qn_{\ell} \ee^{qn_{\ell}\phi} H_0(R)}{R}
{\bf e}_{\theta} + H_0'(R) \ee^{qn_{\ell}\phi}{\bf e}_{R}\right)}\hspace{3cm} &&\\
\\ &=&-\frac{\m q n_{\ell} \ee^{qn_{\ell}\phi}}{R}\left(\ee^{ \ii \phi}
R^{ \ii qn_{\ell}} A_{\ell}\e^{-iqn_{\ell}}(V_2+\ii V_1)
-\ee^{-  \ii \phi} R^{- \ii qn_{\ell}}B_{\ell} \e^{iqn_{\ell}} ( V_2  -
\ii V_1)\right), 
\eeqas
since
\beqas
 \frac{H_0(R)}{R} & = &  A_{\ell}(q)\e^{-\ii qn_{\ell}}
 R^{\ii qn_{\ell}-1}+B_{\ell}(q)\e^{\ii qn_{\ell}} R^{-\ii qn_{\ell}-1}, \\
H_0'(R) & = &  \ii  qn_{\ell} A_{\ell}(q)\e^{-\ii qn_{\ell}}
R^{\ii qn_{\ell}-1}- \ii  qn_{\ell} B_{\ell}(q)\e^{\ii qn_{\ell}}
R^{-\ii qn_{\ell}-1}. 
\eeqas
Therefore, writing
\[
\widehat{h}_1 = - \m qn_{\ell} A_{\ell} \e^{-\ii  qn_{\ell}}(V_2+\ii
V_1) g_1(R)\, \ee^{(qn_{\ell}+\ii ) \phi} 
- \m qn_{\ell} B_{\ell} \e^{\ii  qn_{\ell}}(V_2-\ii  V_1) g_2(R)\,
\ee^{(qn_{\ell}-\ii ) \phi} , \] 
yields a system of ordinary differential equations for $g_1$ and $g_2$, whose solution gives  
\beqa
\widehat{h}_1 &=&   -\frac{\m  A_{\ell} \e^{-\ii  qn_{\ell}}(V_1-\ii
  V_2)}{4  }(R^{ \ii qn_{\ell} +1}+\gamma_1R^{1-\ii qn_{\ell}})\,
\ee^{(qn_{\ell}+\ii )\phi} \nonumber \\
&&\mbox{ }- \frac{\m B_{\ell} \e^{\ii  qn_{\ell}}( V_1+ \ii  V_2)}{4  } (R^{
  -\ii qn_{\ell} +1}+\gamma_2R^{1+\ii qn_{\ell}})\, 
\ee^{(qn_{\ell}-\ii )\phi}. \label{h1solution}
\eeqa 
 where $\gamma_1$ and
$\gamma_2$ are unknown constants that will be determined by matching to the inner limit of
the outer solution.
\subsection{Leading-order matching: 
determination of the asymptotic 
  wavenumber} 
Using \eqref{solutionH} and \eqref{A0},  the leading-order (in $\eps$)
outer limit of the leading-order inner solution is found in the limit $q \ra 0$ to be 
\begin{equation*}
\widehat{\ki}_{00}\sim \frac{1}{q}\log H_0 + \order(1)\sim
\frac{1}{q}\log
\left(\frac{A_{0\ell}\ee^{-\ii qn_\ell\log\e}+A_{0\ell}\ee^{\ii
    qn_\ell\log\e}}{q}+\order(1)\right), 
\end{equation*}
while the leading-order inner limit of the leading-order outer solution, according to \eqref{40} reads
\begin{equation}
\label{leadoutin}
\ki_{00}\sim\frac{1}{q}\log\left(-\beta_{\ell}\log (\e r) +
  \GG^\ell_{\mathrm{reg}}(\bX_{\ell}) +\order(\e   r)\right). 
\end{equation}
Hence, in order to match, the order $1/q$ term inside the logarithm in
the outer limit of the inner must vanish, so that
\begin{equation}
\label{epscan}
\ee^{-\ii qn_\ell\log\e}+\ee^{\ii
  qn_\ell\log\e}=\order(q)\quad\textrm{or equivalently}\quad
q|\log\e|=\frac{\pi}{2}+q\nu, 
\end{equation}
where $\nu$ is order one as $q$, $\eps \ra 0$ (recall also that $|n_\ell |=1$). This
expression sets the relative size of the two small parameters $q$
and $\e$ needed  for $\al$ to be an order one
constant. It is  equivalent to assuming that the typical size
domain is $1/\e = \order(\ee^{\pi/2q})$. 

The leading-order outer limit of the leading-order inner solution now reads
\[
\widehat{\ki}_{00}\sim \frac{1}{q}\log \left(-2A_{0\ell}\nu
  +\ii n_\ell(A_{1\ell}-B_{1\ell})-2A_{0\ell}\log R+\ldots\right), 
\]
and matching with \eqref{leadoutin} provides the conditions $A_{0\ell}=\beta_{\ell}/2$ and
\[\GG^\ell_{\mathrm{reg}}(\bX_{\ell})
=-2A_{0\ell}\nu + \ii n_\ell(A_{1\ell}-B_{1\ell}).\] 
Eliminating $A_{1\ell}-B_{1\ell}$ using \eqref{A1} gives
\begin{equation}
\label{eqGreg}
\GG^\ell_{\mathrm{reg}}(\bX_{\ell}) +\beta_{\ell}(c_1+\nu)=0.
\end{equation}
With $\nu$ given by (\ref{epscan}), and for a given set of spiral
positions $\bX_{\ell}$, equation (\ref{eqGreg})  provides a set of $N$ 
equations for the $N+1$ unknowns $\al$ and $\beta_{\ell}$, $\ell = 1,
\ldots,N$ (recall that $\GG^\ell_{\mathrm{reg}}(\bX_{\ell})$, defined
through (\ref{solGc}), (\ref{NeumannGreens}) and (\ref{Gell}),  depends on
$\alpha$ and $\beta_1, \ldots, \beta_N$). 
However, since $\GG^\ell_{\mathrm{reg}}(\bX_{\ell})$ is a
homogeneous, linear function of $\beta_1, \ldots, \beta_N$ (see
(\ref{solGc})), the system 
(\ref{eqGreg})  is a homogeneous linear system of $N$ equations for
$\beta_1, \ldots, \beta_N$. There exists a solution if and only if the
determinant of the system is zero, which provides an equation for 
$\al$. This in turn determines the asymptotic wavenumber, $k = \al
\e/q$, and therefore the oscillation
frequency $\omega$. The coefficients $\beta_1, \ldots, \beta_N$ are
then determined only up to some global scaling (which is equivalent to
adding a constant to $\ki_{00}$).

\subsection{First-order matching: law of motion
 for the centres of the spirals}
\label{sec:2.6}
We now compare one term of the outer $\eps$-expansion with two terms of the
inner $\eps$-expansion (in the notation of Van Dyke \cite{VanDyke}, we equate  (2 terms inner)(1 term outer) with (1 term outer)(2 terms inner)). This
matching will eventually provide a law of motion for the spirals.  

The two-term inner expansion of the one-term outer expansion for $\chi$
is given
in  \eqref{40}. We must compare this with the one-term outer expansion
of the two-term inner expansion $\chi_0 + \e \chi_1$. From \S\ref{sec:outerlimitofinner}  the one-term (in $\e$) outer expansion of this is
\beq
\frac{1}{q}\log(\widehat{h}_0)+\frac{\widehat{h}_1}{q\widehat{h}_0}. 
\label{InnerToOuter2}  
\eeq
Comparing this with \eqref{40} gives the matching condition
\beq
\inx\cdot\del \GG^\ell_{\mathrm{reg}}(\bX_{\ell})=- \frac{\m r \ii
  n_\ell  A_{0\ell}}{4q}
\left(\ee^{\ii\phi}(V_1-\ii V_2)(1+\gamma_1)-\ee^{-\ii\phi}(V_1+\ii
V_2)(1+\gamma_2)\right).
\label{firstmatching}
\eeq
Note that this equation implies that $\mu = O(q)$, as we have been
supposing.
Solving for $\gamma_1$ and $\gamma_2$, substituting into
(\ref{h1solution}), writing $\widehat{\ki}_{10}$ in terms of the inner
variable and expanding in powers of $q$ finally gives, to leading order in $q$,
\begin{equation}
\label{ki10}
\ki_{10}\sim -\frac{\m r}{2q}(V_1 \cos\phi+V_2\sin\phi)+\frac{n_\ell
  r}{\beta_{\ell}}\del \GG^\ell_{\mathrm{reg}}(\bX_{\ell})\cdot {\bf
  e}_{\phi} \qquad \mbox{ as } r \ra \infty.
\end{equation}

\paragraph{Solvability condition and law of motion}
Equation (\ref{ki10}) provides a boundary condition on the first-order
inner equation (\ref{1orderinner}). However, there is a solvability
condition on (\ref{1orderinner}) subject to  (\ref{ki10}), which
determines $V_1$ and $V_2$, thereby providing our law of motion for
the spiral centres. The analysis in this section summarises the
corresponding analysis in \cite{AgCh09}.

Multiplying equation \eqref{1orderinner} by the conjugate $v^*$  of a
solution $v$ 
of the adjoint equation 
\begin{equation*}
 \del^2 v+(1-\ii q)\left(v(1-2|\p_0|^2)-\p_0^2 v^*\right)=0,
\end{equation*}
integrating over a disk $B_{\R}$ of radius $\R$, and using  integration
by parts gives, after some manipulation, 
\begin{equation}
-\int_{B_{\R}} \Re\left\{(1-\ii q)\m v^* \frac{\d\bX_{\ell}}{\d T}\cdot \del
  \p_0 \right\} \d S=\int_{\dd 
  {B_{\R}}}\Re\left\{(1-\ii q)\left(v^*\frac{\dd \p_1}{\dd n}-\frac{\dd v^*}{\dd
    n}\p_1\right)\right\} \d s,
\label{solvability}
\end{equation}
where $\Re$ denotes the real part.
A straightforward calculation shows that  directional derivatives of
$\p_0$ are solutions of the adjoint problem if $q$ is replaced by
$-q$, i.e. $v= {\bf d}\cdot \del\p_0|_{q \rightarrow -q}$, where ${\bf
  d}$ is any vector in $\mathbb{R}^2$. 
To leading order in $q$ and $\mu$ the solvability condition
(\ref{solvability}) is 
\[
0=  \int_{\dd B_{\R}}\Re\left\{\left({\bf d}\cdot \del \p_0^*\right)
  \frac{\dd \p_1}{\dd 
  n}-\frac{\dd( {\bf d}\cdot \del \p_0^*)
  }{\dd n} \p_1\right\}\d s.
\]
Letting the disk radius $\R$ tend to infinity gives
\begin{equation}
\lim_{r \ra \infty} \int_0^{2 \pi}({\bf e}_{\phi} \cdot {\bf d}) \left(
  \frac{\dd \ki_{10}}{\dd r} + \frac{\ki_{10}}{r}\right)\, \d \phi =
0.\label{compatibility} 
\end{equation}
Now using  \eqref{ki10} gives the law of motion, to leading order in $q$,
\begin{equation}
\label{lawCan}
 \frac{\d\bX_{\ell}}{\d T}=-\frac{2 q n_\ell}{\beta_{\ell}\m}\del^{\perp}
 \GG^{\ell}_{\mathrm{reg}}(\bX_{\ell}), 
\end{equation}
where $\nabla^\perp =(-\partial_y, \partial_x)$.

\paragraph{Summary}
The  parameter $\alpha$ and the coefficients $\beta_j$ are determined
(up to a scaling) by the linear system (\ref{eqGreg}), which is
\begin{equation}
 2 \pi \beta_\ell  \Gnreg(\bX_\ell;\bX_\ell)+  2 \pi
 \sum_{j=1,j \not = \ell}^N \beta_j
 \Gn(\bX_\ell;\bX_j)-\beta_{\ell}(c_1+\nu)=0 ,  \label{canalpha} 
\end{equation}
where
\[\Gnreg(\bX;\bY) = \Gn(\bX;\bY)-\frac{1}{2 \pi}\log|\bX-\bY|
,\]
 is the regular part of the Neumann Green's function $\Gn$ for the
 modified Helmholtz equation
\beq
  \lap \Gn-\al^2 \Gn=\delta(\bX-\bY)\quad\textrm{in}\quad\Omega,\qquad
\frac{\dd \Gn}{\dd n} = 0\quad\textrm{on}\quad \dd\Omega,\label{NeumannGreens2}
\eeq
and \mbox{$\nu =
  \log (1/\eps) - \pi/2 q$}. 
The law of motion (\ref{lawCan}) may be written, to leading order in $q$, as
\beqa
 \frac{\d\bX_{\ell}}{\d T}
&=&\frac{4 \pi q n_\ell  }{\beta_{\ell}\m}
\sum_{j=1,j \not = \ell}^N \beta_j \del^{\perp} \Gn(\bX_\ell;\bX_j) 
+\frac{4 \pi q  n_\ell}{\m}
\del^{\perp}
\Gnreg(\bX_\ell;\bX_\ell)  \label{canonicallaw}
\eeqa
As the size of the domain tends to infinity,
\beq
 \Gn(\bX;\bY) \sim  -\frac{1}{2 \pi} K_0(\alpha|\bX-\bY|),\label{freespace}
\eeq
where $K_0$ is the order zero modified Bessel function of second kind,
and equation (\ref{canonicallaw}) agrees with that given in
\cite{AgChWi08} for spirals in an infinite domain.


\section{Interaction of spirals in bounded
 domains in the near-field}
\label{middle}
In the previous section we  assumed the parameter $\al$ is order one
as $\eps \ra 0$, which led to $q$ and $\eps$ being related by
(\ref{epscan}), which implies that the separation of spirals, and therefore
the size of the domain, is exponentially large in $q$. 

We now consider  smaller domains, in which $\al$ will be small. 
In the limit $q$, $\eps \ra 0$ with
$0<q \log(1/\eps) < \pi/2$ we will find that 
$\al=O(q^{1/2})$. This is in contrast to spirals in the near field in the whole of $\mathbb{R}^2$, 
where $\al$ is found to be exponentially small in $q$
\cite{AgChWi08}.

\subsection{Outer region}
\label{jon:outer}
As before we rescale time as  $T=\m \e^2 t$ and use $\bX=\e \bx$ as
the outer variable, to give
\begin{equation*}
\e^2\m\p_T = (1+\ii
q)\,\p(1-|\p|^2)-\ii\frac{\e^2\al^2}{q}\p+\e^2\del^2\p\quad\textrm{in
  $\Omega$}. 
\end{equation*}
Recall that $1/\eps$ is the typical domain diameter in $\bx$, so that
the diameter of the domain is $O(1)$ in terms of $\bX$. Expressing the
solution in amplitude-phase form as $\p=f\ee^{\ii \ki}$ yields 
\begin{eqnarray}
&&\m \e^2 f_T= \e^2\del^2 f-\e^2 f|\del\ki|^2+f(1-f^2),\label{outerf3}\\
&&\m\e^2 f^2\ki_T=\e^2 \del\cdot(f^2 \del 
   \ki)+qf^2(1-f^2)-\frac{\e^2\al^2}{q} f^2,\label{outerchi3} 
\end{eqnarray}
in  $\Omega$, where, as  before, the boundary conditions for $f$ and $\ki$ are
\begin{equation*} 
\frac{\dd f}{\dd n}=\frac{\dd \ki}{\dd n}=0\quad
\textrm{on}\quad \dd\Omega. 
\end{equation*}
Expanding in asymptotic power series in $\e$ as  $f\sim f_0+\e^2
f_1+\ldots$ and $\ki\sim \ki_0+\e^2 \ki_1+\ldots$, the leading- and
first-order terms in $f$ give  
\[ f_0=1,\qquad f_1=-\frac{1}{2}|\del \ki_0|^2.\]
 The equation for the leading-order phase function, $\ki_0$, is
\begin{equation*}
\begin{split}
&\m \frac{\dd \ki_{0}}{\dd T}=\del^2 \ki_0+q|\del\ki_0|^2 - \frac{\al^2}{q}\quad\textrm{in}\quad\Omega,\\
& \frac{\dd\ki_0}{\dd n} = 0 \quad\textrm{on}\quad \dd\Omega.
\end{split}
\end{equation*}
So far the analysis is exactly the same as before. However, we know
that $\alpha$ cannot be $O(1)$ this time, and so must be some lower
order in $q$. The natural assumption is that $\alpha^2 = O(q)$, which
we will verify a posteriori. We thus rescale $\alpha = q^{1/2}
\bar{\alpha}$. We note that  $\al$ being of order $q^{1/2}$ is
consistent with the value of $\al$ that is found in 
\cite{Ag14} for a single spiral in a finite disk with homogeneous Neumann
boundary conditions.

Expanding $\ki_0$ in terms of $q$ as 
$\ki_0\sim\frac{1}{q}(\ki_{00}+q\ki_{01}+\ldots)$ as in \S
\ref{canonical} gives, at  leading
and first order in $q$, 
\begin{eqnarray}
0&=&\del^2 \ki_{00}+|\del \ki_{00}|^2,\label{14}\\
\tilde{\mu}\frac{\dd\ki_{00}}{\dd T}&=&\del^2
 \ki_{01}+2\del\ki_{00}\cdot\del\ki_{01}-\bar{\al}^2,\label{15}
\end{eqnarray}
in $\Omega$, with  homogeneous Neumann boundary
conditions, where $\tilde{\mu} = \mu/q$. 
Integrating (\ref{14}) over $\Omega$ and using the divergence theorem
and the boundary conditions gives
\[
\int_{\Omega} |\del \ki_{00}|^2\,\d S=0,
\]
so that in fact  $\ki_{00}=C_1(T)$.
Now  (\ref{outerf3})-(\ref{outerchi3}) are invariant with
 respect to the transformation
\[\chi \ra \chi - C_1(T)/q, \qquad \alpha^2 \ra \alpha^2 +  \mu C_1'(T),\]
so that we may take $C_1\equiv 0$ without loss of generality. In fact,
if $C_1'(T) \not = 0$ it means we have not factored out all the
global oscillation when making the change of variables which leads to
(\ref{pdeCGL2}). However, we must be careful when matching with the
inner region near each spiral, since changing $C_1$ is equivalent to scaling
$A_{\ell}$ in the inner region. With $C_1=0$ we will find that the 
inner expansions for $A_{\ell}$ and $B_{\ell}$  start at $O(1)$ rather
than $O(1/q)$ as they did in \S \ref{sec:outerlimitofinner}.

The first-order equation (\ref{15})  becomes
\begin{equation}
 \begin{split}
  \del^2\ki_{01}& =  \bar{\al}^2, \quad\textrm{ in }\Omega,\\
\frac{\dd \ki_{01}}{\dd{ n}}&=0\quad\textrm{ on }\dd\Omega,\\
 \ki_{01}&\sim C_{2j}(T)\log
 R_j +n_j\phi_j, \quad \textrm{ as
   $R_j\ra 0$, \quad for $j = 1,\ldots,N$},  
 \end{split}
\label{mid}
\end{equation}
where $R_j = |\bX-\bX_j(T)|$ and $\phi_j$ are polar coordinates
centred on the $j$th spiral, and we have assumed that the
singularities due to the spirals are locally of the same form as the
corresponding singularities when $\Omega=\mathbb{R}^2$ \cite{AgChWi08}. 
 We thus have a set of unknown slow-time-dependent parameters,
 $C_{2j}(T)$, one for each spiral, which are 
determined by matching at each spiral core. 

To determine $\bar{\al}$ we integrate  equation \eqref{mid} over the
domain $V_{\delta}=\Omega\backslash\sum_{j=1}^N B_{\delta}(\bX_j(T))$, which is
the domain that is left after removing disks of radius $\delta$ centred at
each spiral. Applying the divergence Theorem on this domain (on which
solutions are regular), and then taking the limit $\delta \ra 0$, gives 
\beq
\bar{\al}^2|\bar{\Omega}|=\lim_{\delta \ra 0}\int_{\dd V_{\delta}} \frac{\dd \ki_{01}}{\dd
  { n}}\,\d s=\int_{\dd \Omega} \frac{\dd \ki_{01}}{\dd {
    n}}\,\d s+\sum_{j=1}^N \lim_{\delta \ra 0}\int_{\dd B_{\delta}(\bX_j(T))} \frac{\dd
  \ki_{01}}{\dd { n}}\,\d s= - 2\pi \sum_{j=1}^N
C_{2j}, \label{alfa}
\eeq
where
\[ |\bar{\Omega}| = \int_\Omega \d \bX = \eps^2\int_\Omega \d
\bx=\eps^2|\Omega|,\]
is the area of the domain in terms of the outer variable $\bX$.

\subsection{Inner region}

 The inner region is exactly the same as in \S\ref{innercanonical}. 
\subsection{Inner limit of the outer} 
The solution to \eqref{mid} may be written as
\[ \ki_{01}=2 \pi \sum_{j=1}^N C_{2j}(T) \oGn(\bX;\bX_{j}) + 2
\pi\sum_{j=1}^N n_j \oH(\bX;\bX_{j}) = \oGG,\] 
say, where $\oGn(\bX;\bY)$ is the Neumann Green's function for
Laplace's equation in $\Omega$, satisfying
\beq
  \del^2 \oGn  =  \delta(\bX-\bY) - \frac{1}{|\bar{\Omega}|} \quad\textrm{
                  in }\Omega,\qquad\quad
\frac{\dd \oGn}{\dd{ n}}=0\quad\textrm{ on
}\dd\Omega,\label{LaplaceNeumannGreen} 
\eeq
and $\oH$ satisfies
\[
  \del^2 \oH  =  0 \quad\textrm{
                  in }\Omega\backslash\{\bY\},\qquad\quad
\frac{\dd \oH}{\dd{ n}}=0\quad\textrm{ on }\dd\Omega,\qquad\quad
\oH \sim \frac{\phi}{2 \pi} \mbox{ as } \bX \ra \bY,
\]
where $\phi$ is the azimuthal angle centred at $\bY$.
If  $\oGd(\bX;\bY)$ is the  Dirichlet Green's function,
satisfying
\[
  \del^2 \oGd  =  \delta(\bX-\bY) \quad\textrm{
                  in }\Omega,\qquad\quad
\oGd=0\quad\textrm{ on }\dd\Omega,
\]
then $\oH$ is its harmonic conjugate, so that, with $\bX = (X,Y)$, 
\[ \frac{\dd \oH}{\dd X} = -\frac{\dd \oGd}{\dd Y}, \qquad
\frac{\dd \oH}{\dd Y} = \frac{\dd \oGd}{\dd X}.\]
Defining the regular part of $\oGn$, $\oH$ and $\oGd$ as
\beqas
\oGn(\bX;\bY) &=& \frac{1}{2 \pi}\log|\bX-\bY| + \oGnreg(\bX;\bY),\\
\oH(\bX;\bY) &=& \frac{\phi}{2 \pi} +\oHreg(\bX;\bY),\\
 \oGd(\bX;\bY) &=& \frac{1}{2
  \pi}\log|\bX-\bY| + \oGdreg(\bX;\bY),
 \eeqas
 and
 \beqa
 \oGG^\ell_{\mathrm{reg}}&=&2
\pi C_{2\ell}(T)\oGnreg(\bX;\bX_{\ell})+ 2
\pi n_{\ell}\oHreg(\bX;\bX_{\ell}) \nonumber
\\ && \mbox{ }+2 \pi \sum_{j=1,j \not = \ell}^N C_{2j}(T) \oGn(\bX;\bX_{j}) + 2
\pi\sum_{j=1,j \not = \ell}^N n_j \oH(\bX;\bX_{j})   ,\label{oGGregdefn}
\eeqa
we find that as $\bX\ra\bX_\ell(T)$,
\beqa
\ki_{0} &\sim&  n_\ell\phi_\ell +C_{2\ell}\log |\bX-\bX_\ell(T)| +
\oGG^\ell_{\mathrm{reg}}(\bX_{\ell})+ \left(\bX-\bX_\ell(T)\right)\cdot\del
\oGG^\ell_{\mathrm{reg}}(\bX_{\ell}) +\cdots
\eeqa
Written in terms of the inner variable $\eps \inx = \bX-\bX_\ell(T)$
this is
\begin{equation}
\label{kimid}
\ki_{0} \sim  n_\ell\phi +C_{2\ell}\log (\eps r) +
\oGG^\ell_{\mathrm{reg}}(\bX_{\ell})+ \eps \inx\cdot\del
\oGG^\ell_{\mathrm{reg}}(\bX_{\ell}) +\cdots,
\end{equation}
where $r$ and $\phi$ are the polar representation of $\inx$.

\subsection{Outer limit of the inner solution}
We sum the $q$-expansion of the outer limit of the inner solution in exactly
the same was as in \S\ref{sec:outerlimitofinner} to give
$\widehat{\ki}_{00} = n_{\ell}\phi 
+ (1/q)\log H_0$ with 
\begin{equation*}
H_0 = A_{\ell}(q)\e^{-\ii q n_{\ell}} R^{\ii q
  n_{\ell}}+B_{\ell}(q)\e^{\ii q n_{\ell}} R^{-\ii q n_{\ell}}. 
\end{equation*}
To determine  $A_{\ell}$ and $B_{\ell}$ we need to write $\hat{\ki}_{00}$ in terms
of $r$, expand in powers of $q$, and compare with (\ref{bc1}).
Crucially though, as mentioned in \S \ref{jon:outer}, and in contrast to
\S\ref{sec:outerlimitofinner}, the expansions for 
 $A_{\ell}$ and $B_{\ell}$ proceed now as
 $A_{\ell}(q)\sim A_{\ell 0} +q A_{\ell 1} +\cdots$ and $B_{\ell}(q)
 \sim B_{\ell 0} +q B_{\ell 1}  +\cdots$.
Expressing $H_0$ in terms of $r$ we find
\begin{eqnarray}
H_0(r)&\sim &A_{\ell 0}+B_{\ell
        0}+q(A_{{\ell}1}+B_{{\ell}1})+q(A_{{\ell}0}-B_{{\ell}0})\ii
        n_{\ell}\log r\nonumber\\ 
&& \mbox{ }+q^2\left(A_{{\ell}2}+B_{{\ell}2}+(A_{{\ell}1}-B_{{\ell}1})\ii n_k\log
r-\frac{(A_{{\ell}0}+B_{{\ell}0})}{2}\log^2 r\right)+\cdots, \label{jon2}
\end{eqnarray}
so that
\begin{eqnarray*}
\frac{\dd \widehat{\ki}_{00}}{\dd r} = \frac{H_0'(r)}{qH_0(r)}
& \sim& \frac{n_{\ell}(A_{{\ell}0}-B_{{\ell}0})\ii}{r(A_{{\ell}0}+B_{{\ell}0})}+q\left(\frac{(A_{{\ell}1}-B_{{\ell}1})}{(A_{{\ell}0}+B_{{\ell}0})}\frac{n_{\ell}\ii }{r}-\frac{\log
  r}{r}\right.\\
&&\mbox{ }\qquad+\left.\frac{(A_{{\ell}0}-B_{{\ell}0})^2}{(A_{{\ell}0}+B_{{\ell}0})^2}\frac{\log
    r}{r} -\frac{\ii
    (A_{{\ell}0}-B_{{\ell}0})(A_{{\ell}1}+B_{{\ell}1})}{(A_{{\ell}0}+B_{{\ell}0})^2}
   \frac{n_{\ell}}{r}\right) + \cdots 
\end{eqnarray*}
Comparing with (\ref{bc1}) (and recalling that $n_\ell = \pm 1$) 
we see that
\begin{align}
\label{A0a}
A_{\ell 0}-B_{\ell 0} =& 0,\\
\label{A1a}
\frac{(A_{\ell 1}-B_{\ell 1})}{A_{\ell 0}+B_{\ell 0}}\ii  =& -n_{\ell}c_{1} \quad \textrm{for $\ell =
  1, \ldots, N$}.
\end{align}
The remaining equations determining $A_{\ell}$ and $B_{\ell}$ will be fixed when
matching with the outer region.

Using (\ref{A0a}) we now find that (\ref{jon2}) gives
the outer limit of 
the leading-order inner expansion as
\begin{equation}
\label{midkiin}
\begin{split}
\widehat{\ki}_{00}&\sim \frac{1}{q}\log\left(
A_{0\ell}(\ee^{-\ii qn_\ell\log\e}+\ee^{iqn_\ell\log\e})\right)+
n_\ell\phi+\frac{A_{1\ell}\ee^{-\ii qn_\ell\log\e}+B_{1\ell}\ee^{iqn_\ell\log\e} 
}{A_{0\ell}(\ee^{-\ii qn_\ell\log\e}+\ee^{\ii qn_\ell\log\e})}\\ 
&+\ii n\frac{(\ee^{-\ii qn_\ell\log\e}-\ee^{\ii qn_\ell\log\e}) }
{(\ee^{-\ii qn_\ell\log\e}+\ee^{\ii qn_\ell\log\e})}\log R +\order(q).
\end{split}
\end{equation}
Similarly, the leading-order outer limit of the first correction to
the inner expansion $\widehat{\ki}_{10}$ is now
\beq
\widehat{\ki}_{10}\sim-\frac{\m}{4q}R\left(\frac{\ee^{-\ii 
    qn_\ell\log\e}(V_{1}-\ii V_{2})(1+\gamma_{1})\ee^{\ii\phi}}{\ee^{-\ii
    qn_\ell\log\e}+\ee^{\ii qn_\ell\log\e}} +\frac{\ee^{\ii qn_\ell\log\e}(V_{1}+\ii
  V_{2})(1+\gamma_{2})\ee^{-\ii\phi}}{\ee^{-\ii
    qn_\ell\log\e}+\ee^{\ii qn_\ell\log\e}}\right).
\label{nearfieldchi1match}
\eeq

\subsection{Leading-order matching: determination of the asymptotic wavenumber}
Matching  \eqref{kimid} with \eqref{midkiin} gives
\begin{eqnarray}
0&=&\log\Big( A_{0\ell}(\ee^{-\ii qn_\ell\log\e}+\ee^{\ii
       qn_\ell\log\e})\Big),\label{jon3}\\ 
C_{2\ell}&=&\ii n_\ell\frac{(\ee^{-\ii qn_\ell\log\e}-\ee^{\ii qn_\ell\log\e}) }
{(\ee^{-\ii qn_\ell\log\e}+\ee^{\ii
             qn_\ell\log\e})}=n_\ell\tan(qn_\ell\log\e),\label{C2j}\\ 
\oGG_{\mathrm{reg}}(\bX_{\ell})&=&
\frac{A_{1\ell}\ee^{-\ii qn_\ell\log\e}+B_{1\ell}\ee^{\ii qn_\ell\log\e}
}{A_{0\ell}(\ee^{-\ii qn_\ell\log\e}+\ee^{\ii qn_\ell\log\e})}. 
\end{eqnarray}
Equation (\ref{jon3}) gives $2A_{0\ell} = \mbox{cosec}(qn_\ell\log\e)$.
When $|n_j|=1$ equation \eqref{C2j} implies the constants $C_{2j}$ are
 all equal and given by
$$C_{2j}=-\tan(q\log(1/\e)) \qquad \forall j.$$
Equations \eqref{alfa} and \eqref{C2j} together determine $\bar{\al}$ via
\begin{equation}
\bar{\al}^2 =  \frac{2\pi
  }{|\bar{\Omega}|}\sum_{j=1}^{N} n_j\tan(q n_j \log(1/\e)) =  \frac{2\pi N
  }{|\bar{\Omega}|} \tan(q  \log(1/\e)).
\label{alfamid}
\end{equation}
The asymptotic wavenumber is related
 to $\al$ by $k=\al \e/q$ and so, since $\al=
 q^{1/2}\bar{\al}$, 
\begin{equation}
\label{kmid} 
k = \frac{\e \bar{\al}}{q^{1/2}} = \frac{\e}{q^{1/2}}
\left( \frac{2\pi 
  N}{|\bar{\Omega}|} \tan(q \log(1/\e))\right)^{1/2}= 
\left( \frac{2\pi 
  N}{q \areax} \tan(q \log(1/\e))\right)^{1/2}. 
\end{equation}
As $q \log(1/\eps) \ra \pi/2$ this expression matches smoothly into
that given by (\ref{eqGreg}); we demonstrate this  in Section
\ref{sec:composite} when we develop a uniform composite approximation.

\subsection{First-order matching: law of motion for the spirals}
\label{sec:3.6}
Matching (\ref{kimid}) with (\ref{nearfieldchi1match}) gives
\beqas
\inx\cdot\del \oGG^\ell_{\mathrm{reg}}(\bX_{\ell})&\sim&-\frac{\m}{4q}\left(\frac{\ee^{-\ii 
    qn_\ell\log\e}(V_{1}-\ii V_{2})(1+\gamma_{1})r\ee^{\ii\phi}}{\ee^{-\ii
    qn_\ell\log\e}+\ee^{\ii qn_\ell\log\e}} +\frac{\ee^{\ii qn_\ell\log\e}(V_{1}+\ii
  V_{2})(1+\gamma_{2})r\ee^{-\ii\phi}}{\ee^{-\ii
    qn_\ell\log\e}+\ee^{\ii qn_\ell\log\e}}\right)
.
\eeqas
Solving for $\gamma_1$ and $\gamma_2$ and substituting into
(\ref{InnerToOuter2}) using (\ref{h1solution}) gives, finally,
\begin{eqnarray}
\chi_{10} & \sim&-\frac{\tilde{\m}r}{4}\left(V_{1}\cos \phi +
                  V_{2}\sin\phi\right) 
+\frac{\tilde{\m} 
  r}{4}\left(V_{1}\cos(\phi-2qn_{\ell}\log\e)+V_{2}
\sin(\phi-2qn_{\ell}\log\e)\right)\nonumber\\  
&&+r\cos(qn_{\ell}\log\e)\left(\frac{\dd \oGG^\ell_{\mathrm{reg}}}{\dd
   X}({\bf
  X}_{\ell})\cos(\phi-qn_{\ell}\log 
    \e)+\frac{\dd \oGG^\ell_{\mathrm{reg}}}{\dd Y}({\bf 
    X}_{\ell})\sin(\phi-qn_{\ell}\log \e)\right),
\label{ji1r}
\end{eqnarray}
as $r \ra \infty$.
The compatibility condition (\ref{compatibility}) then gives the law
of motion as
\begin{equation}
\label{lawmid}
\frac{\d\bX_{\ell}}{\d T} = 
\frac{2}{\tilde{\m}}\cot(qn_\ell\log\e)\del^{\perp} 
\oGG^\ell_{\mathrm{reg}}(\bX_{\ell}).
\end{equation}
Using (\ref{oGGregdefn}) and (\ref{C2j}) we may write this as
\beqa
\frac{\tilde{\m}}{2}\tan(qn_\ell\log\e)\frac{\d\bX_{\ell}}{\d T} & = & 
2
\pi( n_\ell\tan(qn_\ell\log\e))\del^{\perp} \oGnreg(\bX_{\ell};\bX_{\ell})+ 2
\pi n_{\ell}\del^{\perp}\oHreg(\bX_{\ell};\bX_{\ell}) \nonumber
\\ && \mbox{ }+2 \pi \sum_{j=1,j \not = \ell}^N (
n_j\tan(qn_j\log\e))  \del^{\perp}\oGn(\bX_{\ell};\bX_{j})  \nonumber
\\ && \mbox{ }+ 2
\pi\sum_{j=1,j \not = \ell}^N n_j\del^{\perp} \oH(\bX_{\ell};\bX_{j})
\nonumber \\& = & 
2
\pi n_\ell\tan(qn_\ell\log\e)\del^{\perp} \oGnreg(\bX_{\ell};\bX_{\ell})- 2
\pi n_{\ell}\del\oGdreg(\bX_{\ell};\bX_{\ell}) \nonumber
\\ && \mbox{ }+2 \pi \sum_{j=1,j \not = \ell}^N 
n_j\tan(qn_j\log\e)  \del^{\perp}\oGn(\bX_{\ell};\bX_{j})  \nonumber
\\ && \mbox{ }- 2
\pi\sum_{j=1,j \not = \ell}^N n_j\del \oGd(\bX_{\ell};\bX_{j}).\label{boundedlaw}
\eeqa
Thus we see the motion due to each spiral is a combination of the gradient of
the Dirichlet 
Green's function and the perpendicular gradient of the Neumann Green's
function.

Since we are considering only the case that $|n_j|=1$ for all $j$ we may
simplify to 
\beqa
n_\ell\frac{\m}{2q}\tan(q\log\e)\frac{\d\bX_{\ell}}{\d T} & = & 
 2
\pi\tan(q\log\e)\del^{\perp} \oGnreg(\bX_{\ell};\bX_{\ell})- 2
\pi n_{\ell}\del\oGdreg(\bX_{\ell};\bX_{\ell}) \nonumber
\\ && \mbox{ }+2 \pi \tan(q\log\e)\sum_{j=1,j \not = \ell}^N 
  \del^{\perp}\oGn(\bX_{\ell};\bX_{j})  \nonumber
\\ && \mbox{ }- 2
\pi\sum_{j=1,j \not = \ell}^N n_j\del
\oGd(\bX_{\ell};\bX_{j})\label{boundedlaw1} 
\eeqa

As the size of the domain tends to infinity both the Neumann and
Dirichlet Green's functions tend to 
\[ \frac{1}{2 \pi} \log |\bX - \bY|.\]
Equation (\ref{boundedlaw}) then becomes
\beqas
\frac{\tilde{\m}}{2}\tan(qn_\ell\log\e)\frac{\d\bX_{\ell}}{\d T} & = & 
 \sum_{j=1,j \not = \ell}^N 
  \frac{n_j\tan(qn_j\log\e)}{|\bX_{\ell} -
  \bX_j|} {\bf e}_{\phi_{j}}  
+ \sum_{j=1,j \not = \ell}^N  \frac{n_j}{|\bX_{\ell} -
  \bX_j|} {\bf e}_{r_{j}}
\eeqas
in agreement with \cite{AgChWi08}.


\section{Rectangular domains}
\label{rectangular}
In this section we evaluate our results for  a rectangular domain with sides of length $\lx$ and 
$\ly$, in preparation for a
comparison with direct numerical
simulations in \S\ref{numerics}.
As we have shown in the previous sections, we find two different
laws of motion for spirals depending on the relative sizes of the
domain and the parameter $q$. 
We first evaluate these two laws of motion for the case of a
rectangle, before formulating a uniform approximation valid in both regimes.

\subsection{Canonical scale} 
\label{sec:numcan}
For spirals in a rectangular domain in which  $\lx,\ly \sim 1/\eps
\sim \ee^{\pi/2q}$ 
 the motion takes place in
the canonical scaling.
Recalling that the outer variable is defined as $\bX = \eps \bx$,
equation (\ref{NeumannGreens}) for the Neumann
Green's function $\Gn(\bX;\hat{\bX})$ for the modified Helmholtz equation is,
in this case   
\beqas
  \lap \Gn-\al^2
  \Gn&=&\delta(\bX-\hat{\bX})\quad\textrm{in}\quad[0,\eps
  \lx]\times[0,\eps \ly],\\ 
\frac{\dd \Gn}{\dd X} &=& 0\quad\textrm{on}\quad X =0 \mbox{ and }X=\eps
\lx,\\
\frac{\dd \Gn}{\dd Y} &=& 0\quad\textrm{on}\quad Y =0 \mbox{ and }Y=\eps
\ly,
\eeqas
where $\bX = (X,Y)$ and $\hat{\bX} = (\hat{X},\hat{Y})$.
Using the method of images, and noting that the free space Green's
function is given by (\ref{freespace}), the solution is
\beqas
 \Gn(\bX;\hat{\bX}) &=&  -\frac{1}{2 \pi}\sum_{m,n=-\infty}^\infty
K_0\left(\alpha\left((X-\hat{X} + 2 n  \eps \lx)^2 + (Y-\hat{Y} + 2 m
    \eps \ly)^2\right)^{1/2}\right) 
\\
&&\mbox{ }-  \frac{1}{2 \pi}\sum_{m,n=-\infty}^\infty
K_0\left(\alpha\left((X+\hat{X} + 2 n \eps  \lx)^2 + (Y-\hat{Y} + 2 m  \eps \ly)^2\right)^{1/2}\right) 
\\
&&\mbox{ }-  \frac{1}{2 \pi}\sum_{m,n=-\infty}^\infty
K_0\left(\alpha\left((X-\hat{X} + 2 n \eps  \lx)^2 + (Y+\hat{Y} + 2 m  \eps \ly)^2\right)^{1/2}\right) 
\\
&&\mbox{ }-  \frac{1}{2 \pi}\sum_{m,n=-\infty}^\infty
K_0\left(\alpha\left((X+\hat{X} + 2 n  \eps \lx)^2 + (Y+\hat{Y} + 2 m \eps  \ly)^2\right)^{1/2}\right).
\eeqas
The series are rapidly convergent since  $K_{0}(z)$
decays exponentially for large $z$. We also defined the regular part
of the Green's function by
\[\Gnreg(\bX;\hat{\bX}) =\Gn(\bX;\hat{\bX})- \frac{1}{2
  \pi}\log|\bX-\hat{\bX}|  
.\]
In order to compare with direct numerical
simulation, we rewrite $\Gn$ in terms of the original variable $\bx$ by
setting 
\beqas
 \Gpn(\bx;{\boldsymbol \xi}) = \Gn(\eps \bx;\eps {\boldsymbol \xi}) &=& - \frac{1}{2 \pi}\sum_{m,n=-\infty}^\infty
K_0\left(q k\left((x-\xi + 2 n   \lx)^2 + (y-\eta + 2 m \ly)^2\right)^{1/2}\right)
\nonumber \\
&&\mbox{ }-  \frac{1}{2 \pi}\sum_{m,n=-\infty}^\infty
K_0\left(q k\left((x+\xi + 2 n   \lx)^2 + (y-\eta + 2 m  \ly)^2\right)^{1/2}\right) 
\nonumber \\
&&\mbox{ }-  \frac{1}{2 \pi}\sum_{m,n=-\infty}^\infty
K_0\left(q k\left((x-\xi + 2 n   \lx)^2 + (y+\eta + 2 m   \ly)^2\right)^{1/2}\right) 
\nonumber \\
&&\mbox{ }-  \frac{1}{2 \pi}\sum_{m,n=-\infty}^\infty
K_0\left(q k\left((x+\xi + 2 n   \lx)^2 + (y+\eta + 2 m
    \ly)^2\right)^{1/2}\right), \qquad
\eeqas
where $\hat{\bX} = \eps {\boldsymbol \xi} = \eps(\xi,\eta)$, and we have written $\eps
\alpha = q k$. Then
\[
\Gpnreg(\bx;{\boldsymbol \xi}) =\Gpn(\bx;{\boldsymbol \xi})
-
\frac{1}{2
  \pi}\log|\bx-{\boldsymbol \xi}| 
 = 
\Gnreg(\eps \bx;\eps {\boldsymbol \xi}) + \frac{1}{2 
  \pi}\log \eps .
\]

\paragraph{With a single spiral.} In the particular case where there
is only one spiral at position $\bX_1$ with unitary winding
number $n_1$, the law of motion (\ref{canonicallaw})   simply reads 
\beqa
 \frac{\d\bX_{1}}{\d T}
&=&\frac{4 \pi q  n_1}{\m}
\del^{\perp}
\Gnreg(\bX_1;\bX_1),  \label{canonicallawEX1}
\eeqa
and $\alpha$ is given by
\begin{equation}
 -2 \pi   \Gnreg(\bX_1;\bX_1)+c_1+\log (1/\eps) - \pi/2 q=0. 
 \label{alEX1} 
\end{equation}
Written in terms of the original variables $\bx$, $t$ and $k$ equation
(\ref{canonicallawEX1}) becomes
\beqa
 \frac{\d\bx_{1}}{\d t}
&=&4 \pi q  n_1
\del^{\perp}
\Gpnreg(\bx_1;\bx_1)  \label{canonicallawEX1a}
\eeqa
where $\del$ now represents the gradient with respect to $\bx$. 
Equation (\ref{alEX1}) becomes
\begin{equation*}
 -2 \pi   \Gpnreg(\bx_1;\bx_1)+c_1 - \pi/2 q=0.  
\end{equation*}
Note that neither of these equations depends on the scaling parameters
$\eps$ or $\mu$, as expected.

\paragraph{With two spirals}  Written in terms of the original coordinate
$\bx$, with  spirals at positions
$\bx_1$ and $\bx_2$, (\ref{canalpha}) gives
\beqas
 2 \pi \beta_1  \Gpnreg(\bx_1;\bx_1)+  2 \pi
 \beta_2
 \Gpn(\bx_1;\bx_2)-\beta_{1}(c_1 - \pi/2 q)&=&0 , \\  
 2 \pi \beta_2  \Gpnreg(\bx_2;\bx_2)+  2 \pi
  \beta_1
 \Gpn(\bx_2;\bx_1)-\beta_{2}(c_1 - \pi/2 q)&=&0 .
\eeqas
The equation for $k$ is thus
\[
\left(-2 \pi \Gpnreg(\bx_1;\bx_1)+c_1 - \pi/2 q\right)
\left(- 2 \pi \Gpnreg(\bx_2;\bx_2)+c_1 - \pi/2 q \right) =   
4  \pi^2  \Gpn(\bx_2;\bx_1)\Gpn(\bx_1;\bx_2),
\]
while
\[ \frac{\beta_2}{\beta_1} = \frac{2 \pi
  \Gpnreg(\bx_1;\bx_1)-c_1 + \pi/2 q}{2 \pi\Gpn(\bx_1;\bx_2)
} =  \frac{2 \pi\Gpn(\bx_2;\bx_1)}{2 \pi
  \Gpnreg(\bx_2;\bx_2)-c_1 + \pi/2 q}.\]
Note that $\Gpn(\bx_2;\bx_1) = \Gpn(\bx_1;\bx_2)$.

Written in terms of the original variables $\bx$ and $t$ the law of
motion 
(\ref{canonicallaw}) for two spirals is
\beqas
 \frac{\d\bx_{1}}{\d t}
&=&4 \pi q n_1 \frac{\beta_2 }{\beta_{1}}
 \del^{\perp} \Gpn(\bx_1;\bx_2) 
+4 \pi q  n_1
\del^{\perp}
\Gpnreg(\bx_1;\bx_1)\\\
 \frac{\d\bx_{2}}{\d t}
&=&4 \pi q n_2  \frac{\beta_1}{\beta_{2}}
  \del^{\perp} \Gpn(\bx_2;\bx_1) 
+4 \pi q  n_2
\del^{\perp}
\Gpnreg(\bx_2;\bx_2).
\eeqas

\begin{rem}
\label{remCan}
We note that if initially $\bx_1 +\bx_2 = (\lx,\ly)$, so that the spirals
are placed symmetrically with respect to the centre of the domain,
then if $n_1=n_2$ they keep this symmetry during the motion.
In this case $  \Gpnreg(\bx_1;\bx_1) =
\Gpnreg(\bx_2;\bx_2)$ so that $\beta_2/\beta_1=1$. 
\end{rem}

\subsection{Near-field scale}
\label{sec:numnear}
In the near field scaling the relevant Green's functions are the
Neumann and Dirichlet Green's functions for Laplace's
equation. We rewrite these in the original variables as $\oGpn(\bx;\bxi)
= \oGn(\eps \bx;\eps \bxi)$, $\oGpd(\bx;\bxi)= \oGd(\eps \bx;\eps
\bxi)$.
As before, we evaluate the Green's functions by the method of
images. However, we must be a little careful, because the sums over 
images for the Green's functions themselves do not
converge. However, the corresponding sums over images for the derivatives of the
Green's functions do converge, and these are what we need for the law
of motion.
Defining
\beqas
 \Vx(\bx;\xi,\eta) &=& \frac{1}{2\pi}\sum_{n,m=-\infty}^\infty \frac{x - \xi + 2 \lx n}{(x-\xi + 2 n   \lx)^2 + (y-\eta + 2 m
    \ly)^2}\\
& =& \frac{1}{2\pi}\sum_{m=-\infty}^\infty
\frac{\pi\sin(\pi (x-\xi)/\lx)}{2\lx(\cosh(\pi((y-\eta)+2\ly m)/\lx)-\cos(\pi
  (x-\xi)/\lx))} 
,\\
 \Vy(\bx;\xi,\eta) &=& \frac{1}{2\pi}\sum_{n,m=-\infty}^\infty \frac{y - \eta + 2 \ly m}{(x-\xi + 2 n   \lx)^2 + (y-\eta + 2 m
    \ly)^2}\\
& =& \frac{1}{2\pi}\sum_{n=-\infty}^\infty
\frac{\pi\sin(\pi (y-\eta)/\ly)}{2\ly(\cosh(\pi((x-\xi)+2\lx n)/\ly)-\cos(\pi
  (y-\eta)/\ly))} 
,
\eeqas
then
\beqas
\frac{\dd \oGpn}{\dd x}(\bx;\bxi) & = & \Vx(\bx;\xi,\eta) + \Vx(\bx;-\xi,\eta) + \Vx(\bx;\xi,-\eta) + \Vx(\bx;-\xi,-\eta) ,
\\
\frac{\dd\oGpn}{\dd y}(\bx;\bxi) & = & \Vy(\bx;\xi,\eta) + \Vy(\bx;-\xi,\eta) + \Vy(\bx;\xi,-\eta) + \Vy(\bx;-\xi,-\eta) ,
\\
\frac{\dd\oGpd}{\dd x}(\bx;\bxi) & = & \Vx(\bx;\xi,\eta) - \Vx(\bx;-\xi,\eta) - \Vx(\bx;\xi,-\eta) + \Vx(\bx;-\xi,-\eta) ,
\\
\frac{\dd\oGpd}{\dd y}(\bx;\bxi) & = & \Vy(\bx;\xi,\eta) - \Vy(\bx;-\xi,\eta) - \Vy(\bx;\xi,-\eta) + \Vy(\bx;-\xi,-\eta).
\eeqas
Note that the final sums above again converge exponentially quickly.
In terms of $\bx$ and $t$ the law of motion (\ref{boundedlaw1}) is 
\beqa
\frac{n_\ell}{2q}\tan(q\log\e)\frac{\d\bx_{\ell}}{\d t} & = & 
 2
\pi\tan(q\log\e)\del^{\perp} \oGpnreg(\bx_{\ell};\bx_{\ell})- 2
\pi n_{\ell}\del\oGpdreg(\bx_{\ell};\bx_{\ell}) \nonumber
\\ && \mbox{ }+2 \pi \tan(q\log\e)\sum_{j=1,j \not = \ell}^N 
  \del^{\perp}\oGpn(\bx_{\ell};\bx_{j})  
- 2\pi\sum_{j=1,j \not = \ell}^N n_j\del
\oGpd(\bx_{\ell};\bx_{j}).\qquad \label{nearlawx}
\eeqa
Recall also that 
\begin{equation*}
 k = 
\left( \frac{2\pi 
  N}{q\areax} \tan(q \log(1/\e))\right)^{1/2},
\end{equation*}
where $\areax$ is the area of $\Omega$ in the
original variable $\bx$.

\paragraph{With a single spiral} Written out in component form, the
law of motion \eqref{nearlawx} for a single spiral 
at $\bx_1$ with winding number $|n_1|=1$ is
\beqas
\frac{\d x_{1}}{\d t} 
& = & 
 -4\pi q  n_1\frac{\dd  \oGpnreg(\bx_{1};\bx_{1})}{\dd
     y} - 4
\pi q\cot(q\log\e)\frac{\dd  \oGpdreg(\bx_{1};\bx_{1})}{\dd
     x} ,\\
\frac{\d y_{1}}{\d t} 
& = & 
 4\pi q n_1\frac{\dd  \oGpnreg(\bx_{1};\bx_{1})}{\dd
     x}- 4
\pi q\cot(q\log\e)\frac{\dd  \oGpdreg(\bx_{1};\bx_{1})}{\dd
     y}.
\eeqas

\paragraph*{With two spirals} Written out in component form, the
law of motion  \eqref{nearlawx} for spirals at positions $\bx_1$ and $\bx_2$ with
winding numbers $|n_1|=|n_2|=1$ is

\beqa
\frac{\d x_{1}}{\d t} 
& = & 
- 4\pi q n_1 \frac{\dd  \oGpnreg(\bx_{1};\bx_{1})}{\dd
     y} - 4
\pi q \cot(q\log\e)\frac{\dd  \oGpdreg(\bx_{1};\bx_{1})}{\dd
     x} \nonumber
\\ && \mbox{ }-4 \pi q n_1 
 \frac{\dd  \oGpn(\bx_{1};\bx_{2})}{\dd
     y}  \nonumber
- 4
\pi q n_2 n_1 \cot(q\log\e)\frac{\dd  \oGpd(\bx_{1};\bx_{2})}{\dd
    x},\\
\frac{\d y_{1}}{\d t} 
& = & 
 4\pi q n_1 \frac{\dd  \oGpnreg(\bx_{1};\bx_{1})}{\dd
     x} - 4
\pi q \cot(q\log\e) \frac{\dd  \oGpdreg(\bx_{1};\bx_{1})}{\dd
     y} \nonumber
\\ && \mbox{ }+4 \pi n_{1} q 
 \frac{\dd  \oGpn(\bx_{1};\bx_{2})}{\dd
     x}  \nonumber
- 4
\pi q n_{1} n_2\cot(q\log\e)\frac{\dd  \oGpd(\bx_{1};\bx_{2})}{\dd
     y},\\
\frac{\d x_{2}}{\d t} 
& = & -
4\pi n_2 q\frac{\dd  \oGpnreg(\bx_{2};\bx_{2})}{\dd
     y} - 4
\pi q\cot(q\log\e) \frac{\dd  \oGpdreg(\bx_{2};\bx_{2})}{\dd
     x} \nonumber
\\ && \mbox{ }-4 \pi q  n_{2} 
  \frac{\dd  \oGpn(\bx_{2};\bx_{1})}{\dd
     y}  \nonumber
- 4
\pi q n_1 n_{2} \cot(q\log\e) \frac{\dd  \oGpd(\bx_{2};\bx_{1})}{\dd
    x},\\
\frac{\d y_{2}}{\d t} 
& = & 
 4\pi q n_2\frac{\dd  \oGpnreg(\bx_{2};\bx_{2})}{\dd
     x} - 4
\pi q\cot(q\log\e)\frac{\dd  \oGpdreg(\bx_{2};\bx_{2})}{\dd
     y} \nonumber
\\ && \mbox{ }+4 \pi q  n_{2} 
  \frac{\dd  \oGpn(\bx_{2};\bx_{1})}{\dd
     x} \nonumber
- 4
\pi q n_1 n_{2}\cot(q\log\e)  \frac{\dd  \oGpd(\bx_{2};\bx_{1})}{\dd
     y}.
\eeqa


\subsection{A uniform composite expansion}
\label{sec:composite}
To compare with direct numerical simlulations  we combine the
expansions of Sections \ref{sec:numcan} and \ref{sec:numnear} into a
single composite expansion valid in both regions. We first consider
the asymptotic wavenumber.
As $\alpha \ra 0$ in (\ref{NeumannGreens2}) we find
\[ \Gn(\bX;\bY) \sim -\frac{1}{|\bar{\Omega}|\alpha^2} + \oGn(\bX;\bY)
  + \cdots,\] 
where $\oGn(\bX;\bY)$ is the Neumann Green's function for Laplace's
equation given by (\ref{LaplaceNeumannGreen}).
Thus (\ref{canalpha}) implies that the $\beta_{\ell}$ are all equal to
leading order and $\alpha$ is
given by
\[ \alpha^2 \sim \frac{2 \pi Nq}{|\bar{\Omega}|(\pi/2 - q|\log \eps|)}.\]
We see that this matches smoothly into the near-field $\alpha$ we
found in (\ref{alfamid}), since
\[
\alpha^2 = q \bar{\al}^2  =  \frac{2\pi q N
  }{|\bar{\Omega}|} \tan(q  \log(1/\e)) \sim  \frac{2 \pi
    Nq}{|\bar{\Omega}|(\pi/2 - q|\log \eps|)}
\]
as $q |\log \eps| \ra \pi/2$.
We may generate a uniform approximation to $\alpha$ by taking
\[ \alpha^2 = \alpha^2_{\mathrm{canonical}} + \alpha^2_{\mathrm{near}}
-  \frac{2 \pi
    Nq}{|\bar{\Omega}|(\pi/2 - q|\log \eps|)}
.\]
The corresponding uniform approximation to $k$ is given by
\beq
k^2 = k^2_{\mathrm{canonical}} +  \frac{2\pi  N
  }{q|\Omega|} \tan(q  \log(1/\e))
-  \frac{2 \pi
    N}{q|\Omega|(\pi/2 - q|\log \eps|)}. \label{uniformk}
\eeq
For the law of motion the simplest uniformly valid  composite expansion
is
\beqa
\frac{\d\bx_{\ell}}{\d t} & = & 
\frac{4 \pi q n_\ell}{\beta_\ell} \sum_{j=1,j \not = \ell}^N \beta_j
  \del^{\perp}\Gpn(\bx_{\ell};\bx_{j})  + 4\pi q n_\ell\del^{\perp} \Gpnreg(\bx_{\ell};\bx_{\ell})\nonumber \\ && \mbox{ }-4 \pi q \cot(q\log\e) \del\Gpdreg(\bx_{\ell};\bx_{\ell}) 
- \frac{4 \pi q\cot(q\log\e)}{n_\ell}\sum_{j=1,j \not = \ell}^N n_j\del
\Gpd(\bx_{\ell};\bx_{j}),\qquad \label{uniformv}
\eeqa
where $\Gpd$ is the Dirichlet Green's function for the modified Helmholtz equation given by 
\begin{align}
  \lap \Gpd-q^2 k^2 \Gpd&=\delta(\bx-\by)\quad\textrm{in}\quad\Omega,&\qquad
\Gpd &= 0\quad\textrm{on}\quad \dd\Omega,
  \end{align}
with
\[
\Gpdreg(\bx;\by) =\Gpd(\bx;\by)-
\frac{1}{2
  \pi}\log|\bx-\by|,\]
and $\beta_\ell$ and $k$ are given by the canonical approximation in \S\ref{sec:numcan}.


\subsection{Choice of $\eps$}
\label{sec:eps}

In order to plot the trajectories obtained from the uniformly valid asymptotic approximation to the law of motion  we need to make one final 
choice as to the value of $\eps$, which is the inverse of the typical
separation between spirals (and their images). In principle, the full asymptotic expansion is independent of $\eps$ when written in the original coordinates  (note that $\eps$ disappears 
from the approximation for $k$ in the canonical region, for example,
when it is rewritten in the original variables)---this is reflected in
the law of motion by 
the fact that $\eps$ only appears in (\ref{uniformk}) and (\ref{uniformv}) as $\log \eps$: multiplying $\eps$
by any factor does not change the law of motion
asymptotically.
However, $\eps$ will only disappear from the near-field (and uniform)
law of motion if we include the full expansion to all powers of $\log
\eps$ (i.e. all powers of $q$).
 Since this is not possible,
we must choose an appropriate lengthscale to use for $\eps$.
In principle any choice will do (all lead to the same law of motion at
leading order).

In our numerical comparisons we consider two natural choices for $\epsilon$. The first is simply to choose $\eps$ to be a constant proportional  to the inverse of  the domain diameter---we take $\eps = 4/(\lx +\ly)$, which is 0.01 for the square domain of side length 200 we consider in \S\ref{numerics}.
The second natural choice is to take  $\eps$ to be 
 proportional to the inverse distance from a spiral to the  boundary or between spirals.
For a single spiral at $(x,y)$  we approximate this by setting
\beq
\eps = \left(\frac{1}{x^2} + \frac{1}{|\lx-x|^2} +
\frac{1}{y^2} + \frac{1}{|\ly-y|^2}\right)^{1/2}.\label{epseqn}
\eeq
For two spirals at $(x,y)$ and $(\lx-x,\ly-y)$ we take
\beq
\eps = \left(\frac{1}{x^2} + \frac{1}{|\lx-x|^2} +
  \frac{1}{y^2} + \frac{1}{|\ly-y|^2}
+
\frac{1}{|\lx/2-x|^2+|\ly/2-y|^2}\right)^{1/2}.\label{epseqn2} 
\eeq
In this case $\eps$ evolves slowly as the spirals move.

\section{Comparison with direct numerical simulations}
\label{numerics}
To test the accuracy of our results, numerical simulations were
carried out for the Neumann problem.
Letting $\psi(x,y,t)= e^{-ik^2qt} \ph(x,y,t)$, equation
\eqref{9} becomes
\begin{equation}
\ph_t = (1+iq)(1 -|\ph|^2)\ph  +\lap\ph \quad
\textrm{in}\quad \Omega,\qquad
\frac{\dd \ph}{\dd n}=0 \quad \textrm{on}\quad \dd \Omega,
\label{eq70}
\end{equation}
Numerical simulations of equation
\eqref{eq70} were carried out using finite differences
applied to the coupled reaction-diffusion equations for the real and
imaginary parts of $\ph$ subject to homogeneous Neumann boundary conditions
on a large square domain. A uniform spatial discretization was used with
$\Delta y=\Delta x$.
Following the approach described in \cite{Barkley1997,Barkley1990}, a
nine-point stencil for the Laplacian operator
was used to obtain more accurate approximations of the
oscillating solutions.
Explicit timestepping using Euler's method
with a small timestep, $\Delta t= (\Delta x)^2/20$, was found to be stable and
computationally efficient.

Initial conditions were chosen to have zeros with a unit winding
number at the desired initial location of the spirals.
In particular,
for a single
spiral, initial data at $t=0$ was chosen as
$$\ph_0({\bf x})= \finit(r_1) \ee^{\ii\chiinit(r_1,\theta_1)}$$
where $(r_1,\theta_1)$ are polar coordinates with respect to the intended
starting position, ${\bf x}_1$, for the centre
of the spiral, $\finit(r)=\tanh(Ar)$
where $A=0.583189$ was chosen to match $\finit'(0)$ with the solution for a steady spiral in an infinite domain \cite{hagan82},
and the phase
$\chiinit$ varies by $2 \pi$ as  $\bx_1$ is circled anticlockwise. Since the leading-order equation for the phase in the outer region ((\ref{22}) and (\ref{mid})  in the canonical and near-field scalings respectively) is quasi-steady, in principle any initial condition $\chiinit$ will do, since $\chi$ will equilibriate over a short timescale. However, since the timescale for evolution of $\chi$ is logarithmically smaller (i.e. $O(1/|\log \eps|)$) than that for the motion of spirals, in practice initial transients in $\chi$ can significantly perturb the motion. To eliminate this as much as possible we choose the initial $\chiinit$ to be given by either the near-field  $\chi_{01}$ corresponding to the initial position of the spiral (the solution of (\ref{mid})) or the canonical  $\theta_1 + \chi_{00}/q$ corresponding the initial position of the spiral (where $\chi_{00}$ is the solution of (\ref{22})). In the near-field this requires us to choose a value of $\eps$; we take $\eps = 0.01$ for simplicity.

\begin{figure}[p]
\begin{center}
  \begin{subfigure}[t]{0.31\textwidth}
    \includegraphics[width=\textwidth]{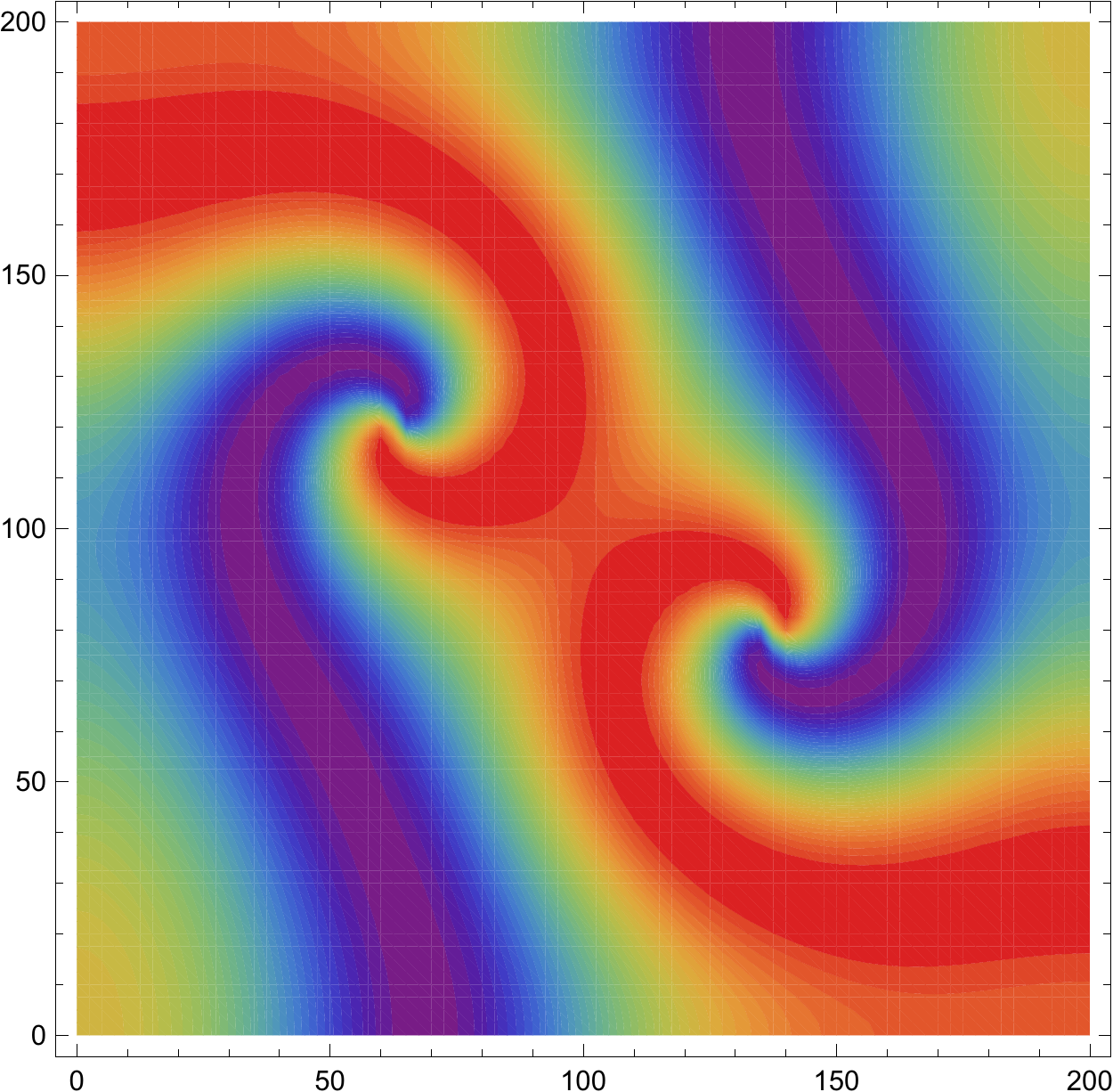}
    \caption{$\re(\ph)$}
        \label{repsi}
    \end{subfigure}
   \begin{subfigure}[t]{0.31\textwidth}
    \includegraphics[width=\textwidth]{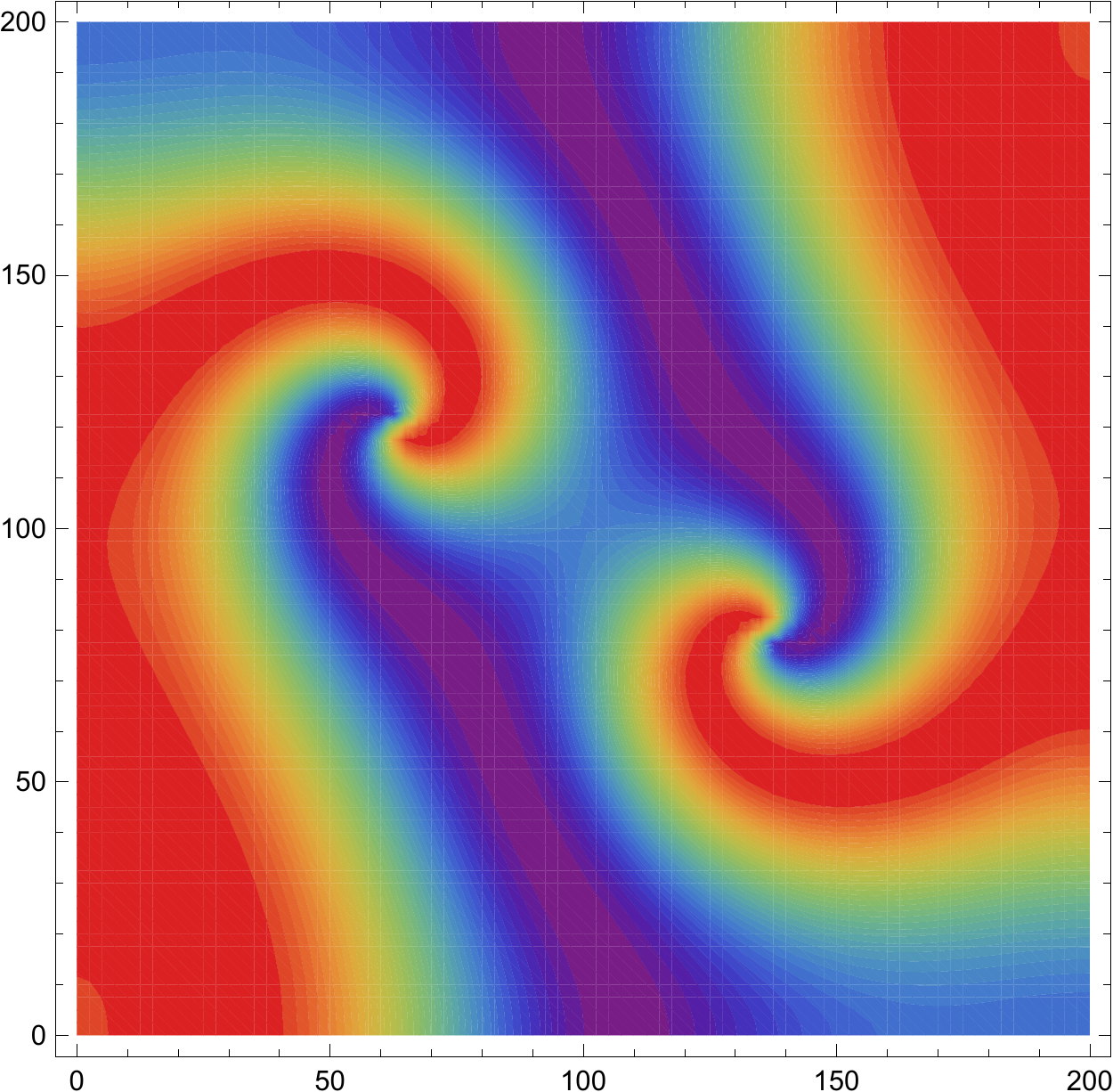}
    \caption{$\im(\ph)$}
        \label{impsi}
    \end{subfigure}
  \begin{subfigure}[t]{0.31\textwidth}
    \includegraphics[width=\textwidth]{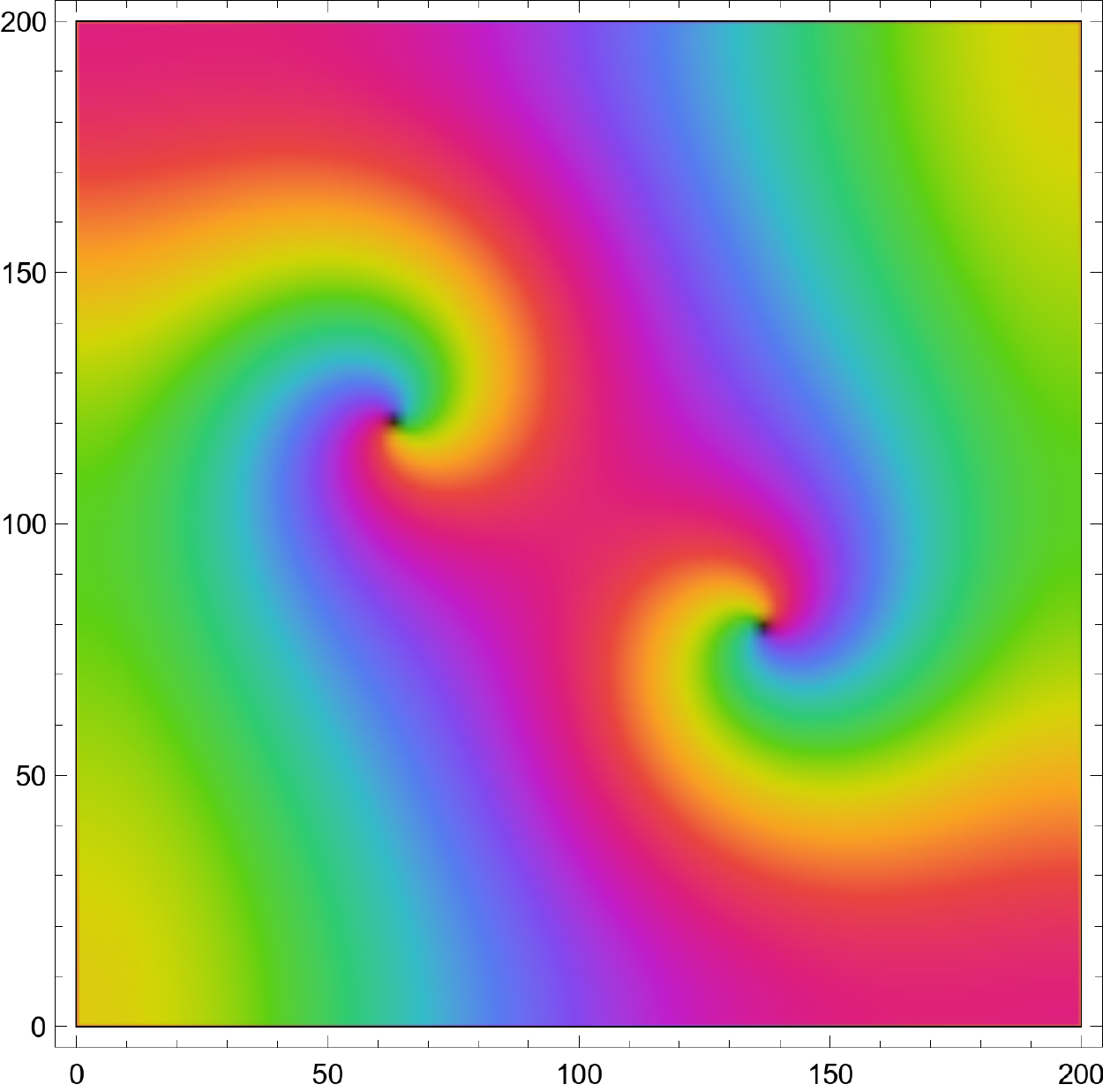}
    \caption{$\arg(\ph)$}
        \label{chi}
  \end{subfigure}
  \end{center}
\caption{A snapshot of a simulation with two spirals for $q=0.1$.}
\label{fig:ph9000}
\end{figure}

For pairs of spirals starting at ${\bf x}_1$ and ${\bf x}_2$, 
the initial condition was given as
\[
\ph_0({\bf x})=
\finit(r_1)\finit(r_2)\exp(\ii[\chiinit(r_1,\theta_1)+\chiinit(r_2,\theta_2)]).
\]
It was observed that this choice of initial data led to  brief
transients after which $\ph$ was smooth, slowly-evolving and satisfied the
boundary conditions. For the most part 
the transients caused only relatively small
changes to the starting positions ${\bf x}_i$ of the spirals. However, when the near-field initial condition for $\chi$ was used with too large a value of $q$ (in which $q |\log \eps|$ is too close to $\pi/2$) it was observed that many more zeros of $\psi$ were generated locally during the initial transient, and that these additional spirals did not always annihilate with each other. We used this behaviour to determine when to switch from the near-field initial condition to the canonical initial condition: for a single spiral we use the near-field initial condition for $q\leq 0.3$ and the canonical initial condition for $q \geq 0.35$; for two spirals we use the near-field initial condition for $q \leq 0.2$   canonical initial condition for $q \geq 0.25$.
Figure \ref{fig:ph9000} shows a snapshot of the real and imagainary parts and the phase of $\ph$ for an example with two spirals.

\begin{figure}
  \begin{center}
  \begin{subfigure}[t]{0.31\textwidth}
    \includegraphics[width=\textwidth]{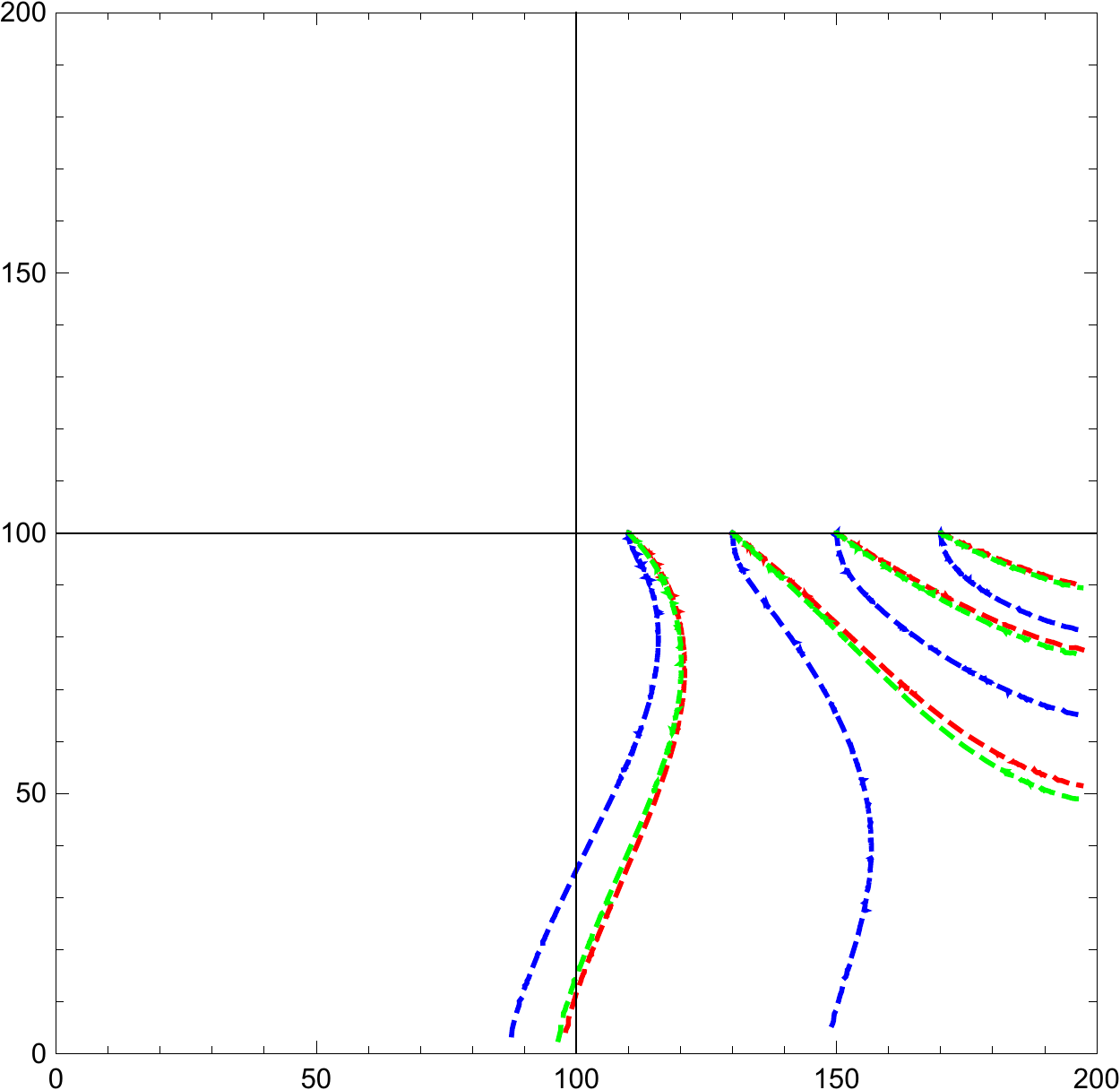}
    \caption{$q=0.1$}
        \label{fig:q100num}
    \end{subfigure}
 \begin{subfigure}[t]{0.31\textwidth}
    \includegraphics[width=\textwidth]{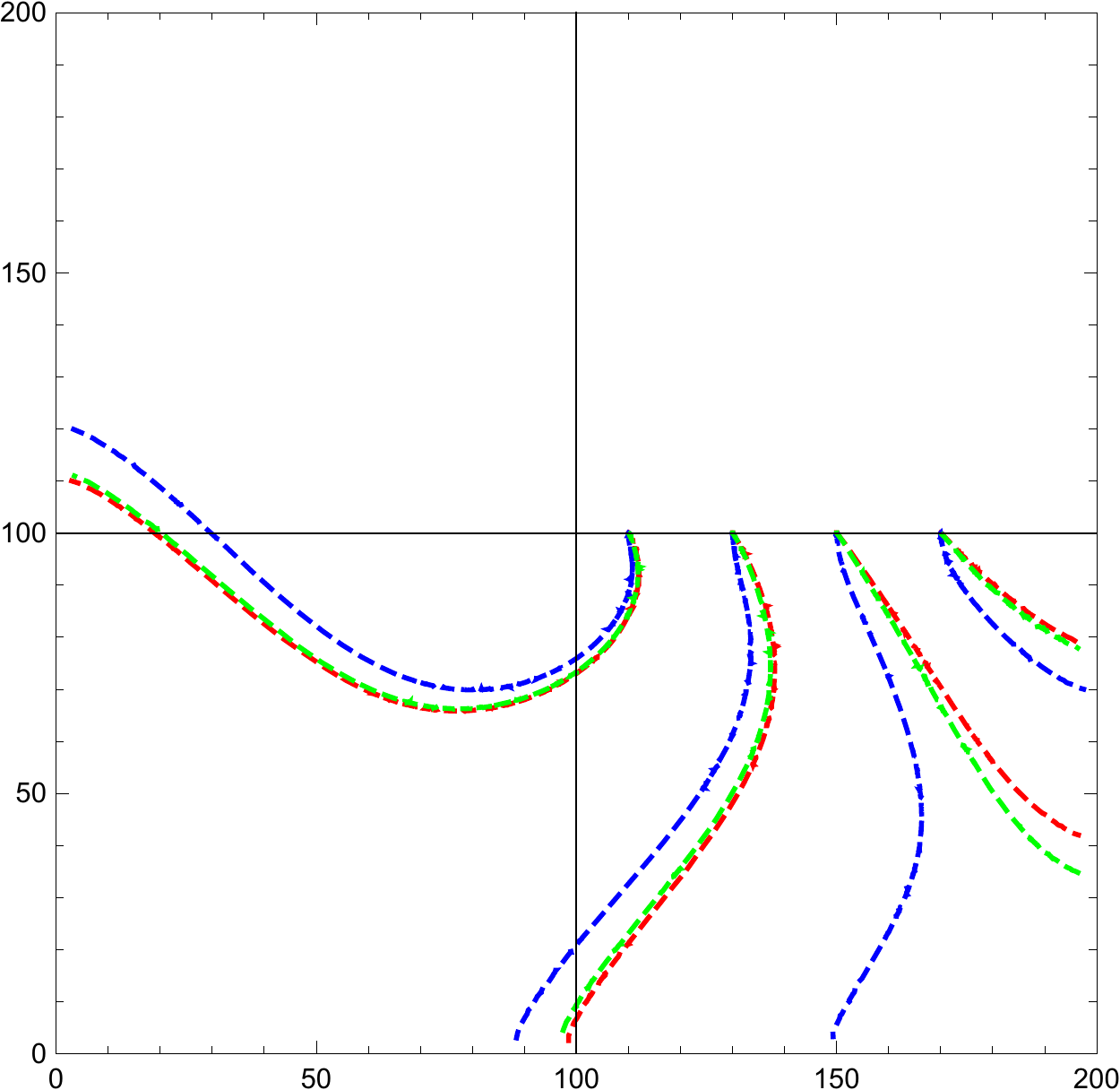}
    \caption{$q=0.2$}
        \label{fig:q200num}
    \end{subfigure}
  \end{center}
  \caption{Numerical trajectories for a single spiral starting at positions $y=100$ and $x = 110$, 130, 150 and 170. The different colours correspond to different initial conditions for the phase $\chi$: blue is the canonical initial condition, red is the near-field initial condition with $\eps=0.01$, and green is the near-field initial condition with $\eps=0.005$.}
\label{fig:numtraj}
\end{figure}

In order to compare the simulations with the asymptotic predictions in Section \ref{sec:composite}
we need to calculate the trajectories of the spiral centres,
${\bf x}_i(t)$. From the simulations, at
regularly spaced times, $t_j$,  the positions of local minima of $|\psi|$ were 
interpolated to sub-grid resolution by fitting computed values at
grid points surrounding
the discrete minimum to a paraboloid.
In Figure \ref{fig:numtraj} we show some  examples of the trajectories obtained from this procedure, for a single $+1$ spiral with various starting positions. This figure illustrates the effect that the initial condition on the phase $\chi$ can have on the trajectory of the spiral. It also illustrates a difficulty we will have when comparing our asymptotic trajectories to the numerically determined trajectories: because trajectories from nearby initial conditions are diverging, any small differences in the velocities will be compounded over time so that small errors in velocity may lead to large errors in position and quite different paths.

\begin{figure}[p]
  \begin{center}
  \begin{subfigure}[b]{0.48\textwidth}
    \begin{overpic}[width=0.49\textwidth]{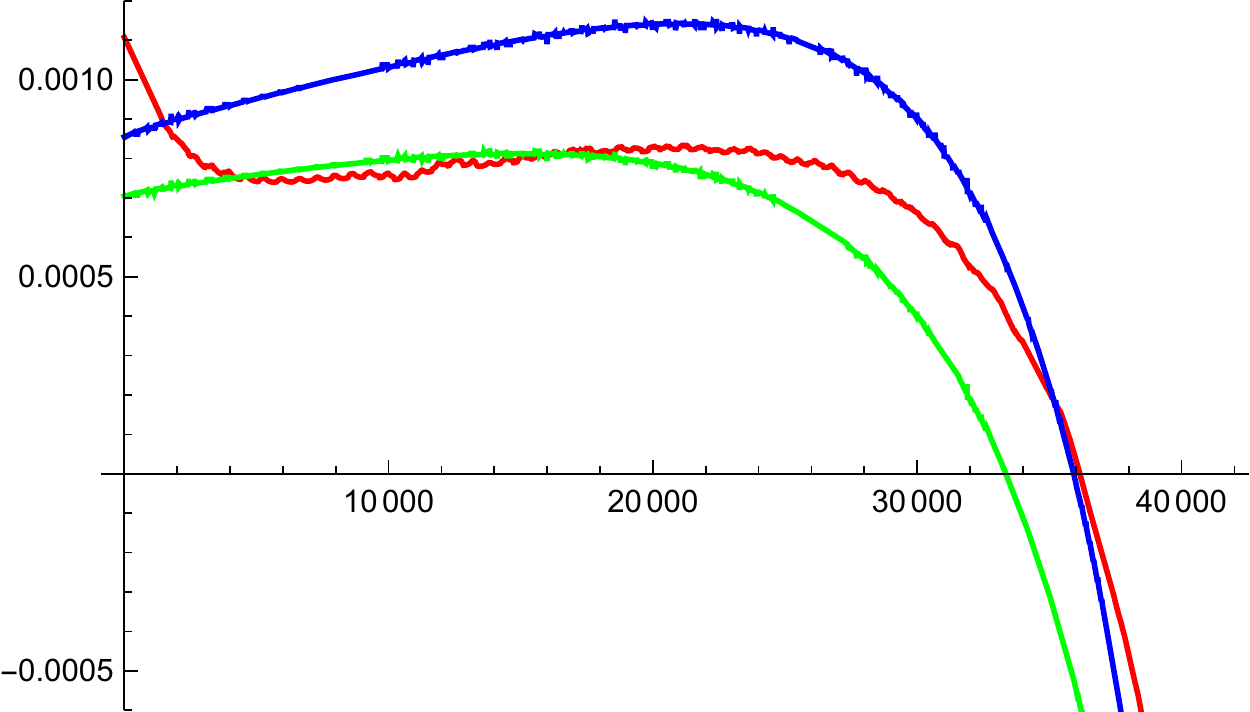}
      \put(0,21){{\tiny $\dot{x}_1$}}
      \put(93,11){{\tiny $t$}}
\end{overpic}
    \begin{overpic}[width=0.49\textwidth]{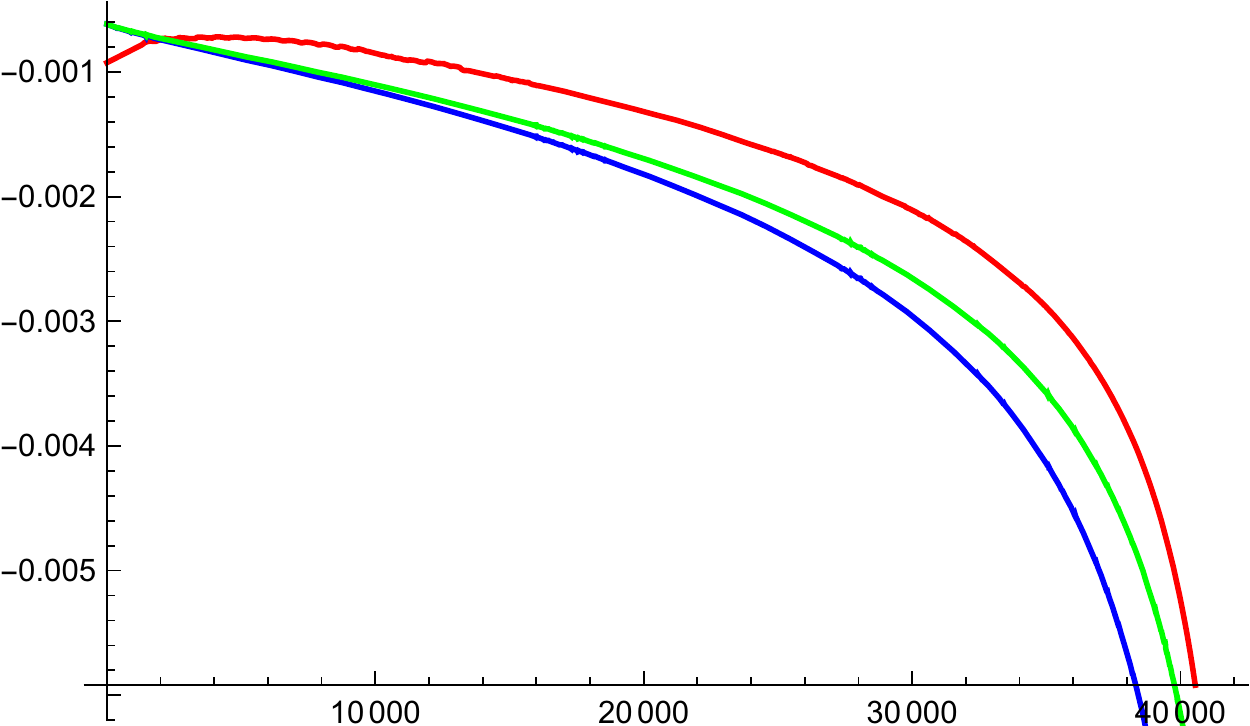}
 \put(0,55){{\tiny $\dot{y}_1$}}
 \put(92,-5){{\tiny $t$}}
     \end{overpic}
     \caption{\footnotesize $q=0.1$, $x_1(0) = 120$}
        \label{fig:q100L}
      \end{subfigure}
  \begin{subfigure}[b]{0.48\textwidth}
    \begin{overpic}[width=0.49\textwidth]{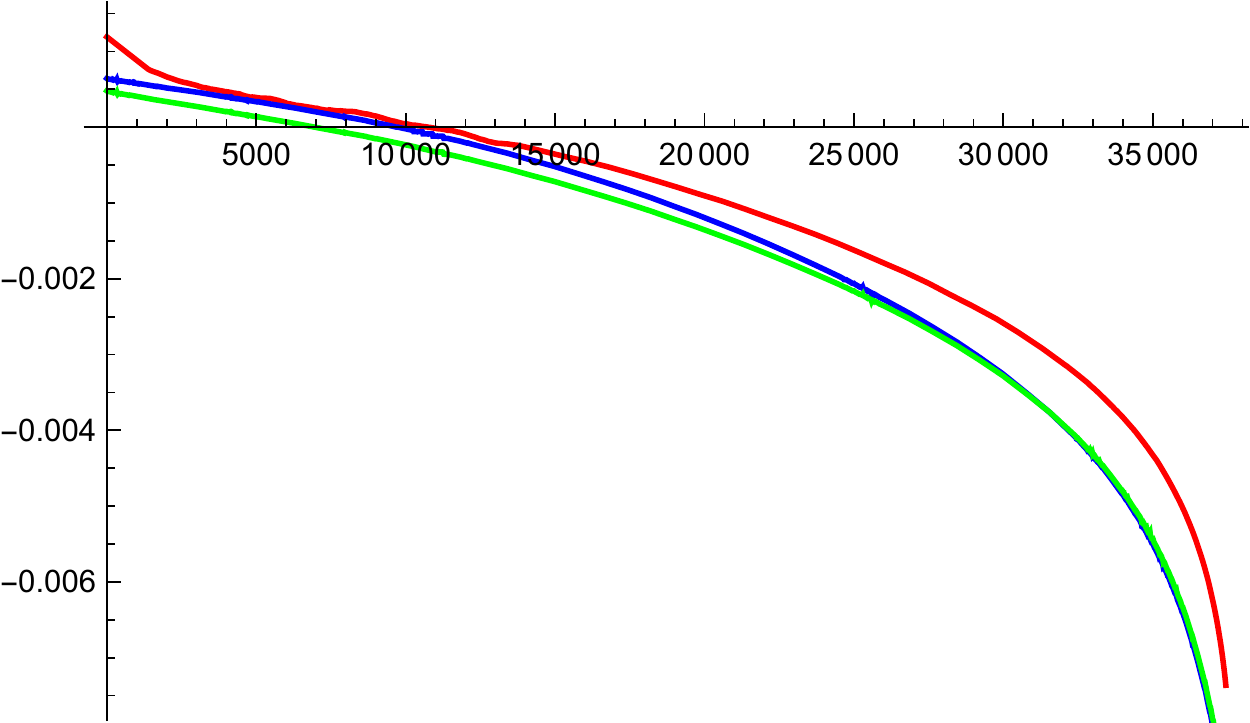}
      \put(0,0){{\tiny $\dot{x}_1$}}
      \put(93,40){{\tiny $t$}}
      \end{overpic}
    \begin{overpic}[width=0.49\textwidth]{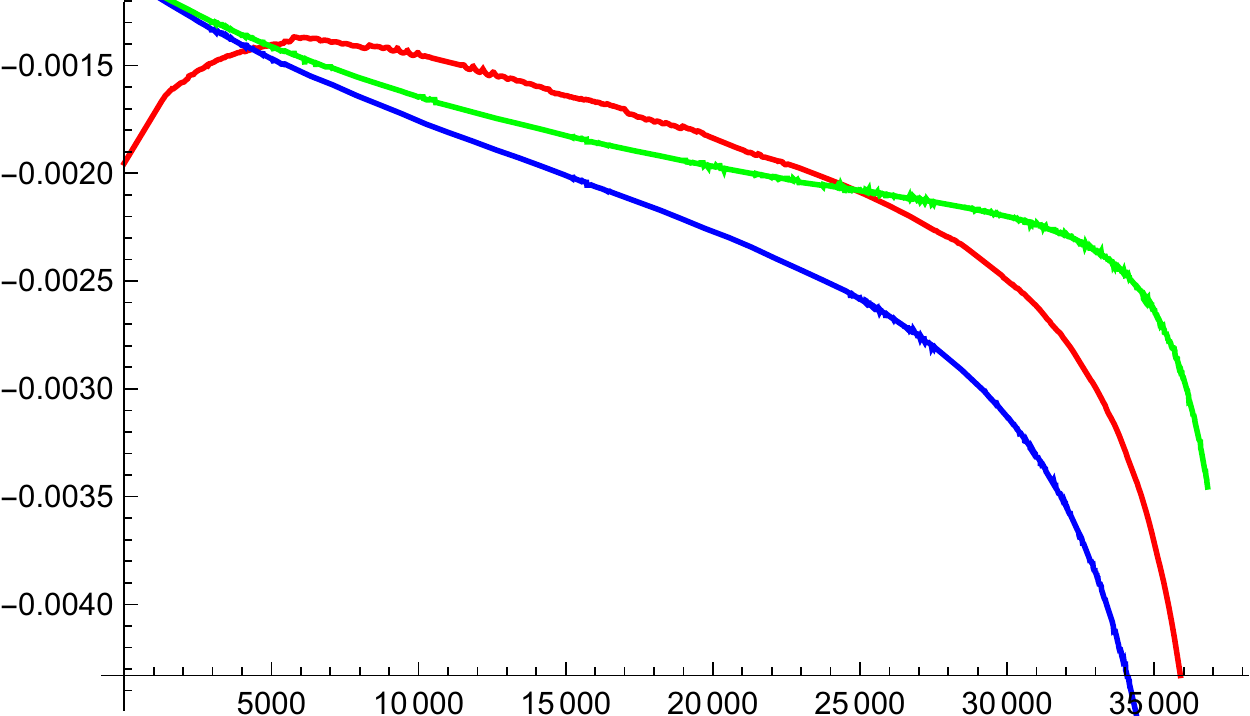}
\put(0,0){{\tiny $\dot{y}_1$}}
\put(92,-5){{\tiny $t$}}
       \end{overpic}
    \caption{\footnotesize $q=0.2$, $x_1(0) = 120$}
        \label{fig:q200L}
    \end{subfigure}\\[2mm]

  \begin{subfigure}[b]{0.48\textwidth}
    \begin{overpic}[width=0.49\textwidth]{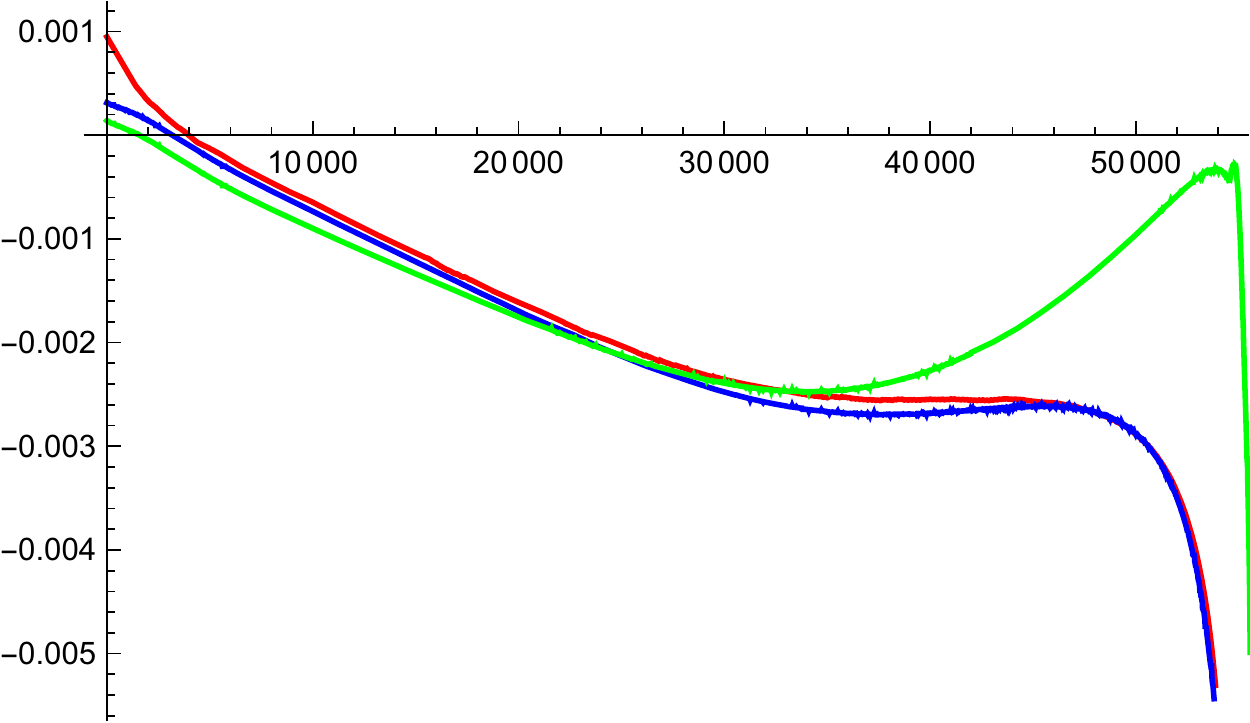}
      \put(0,0){{\tiny $\dot{x}_1$}}
     \put(93,41){{\tiny $t$}}
       \end{overpic}
    \begin{overpic}[width=0.49\textwidth]{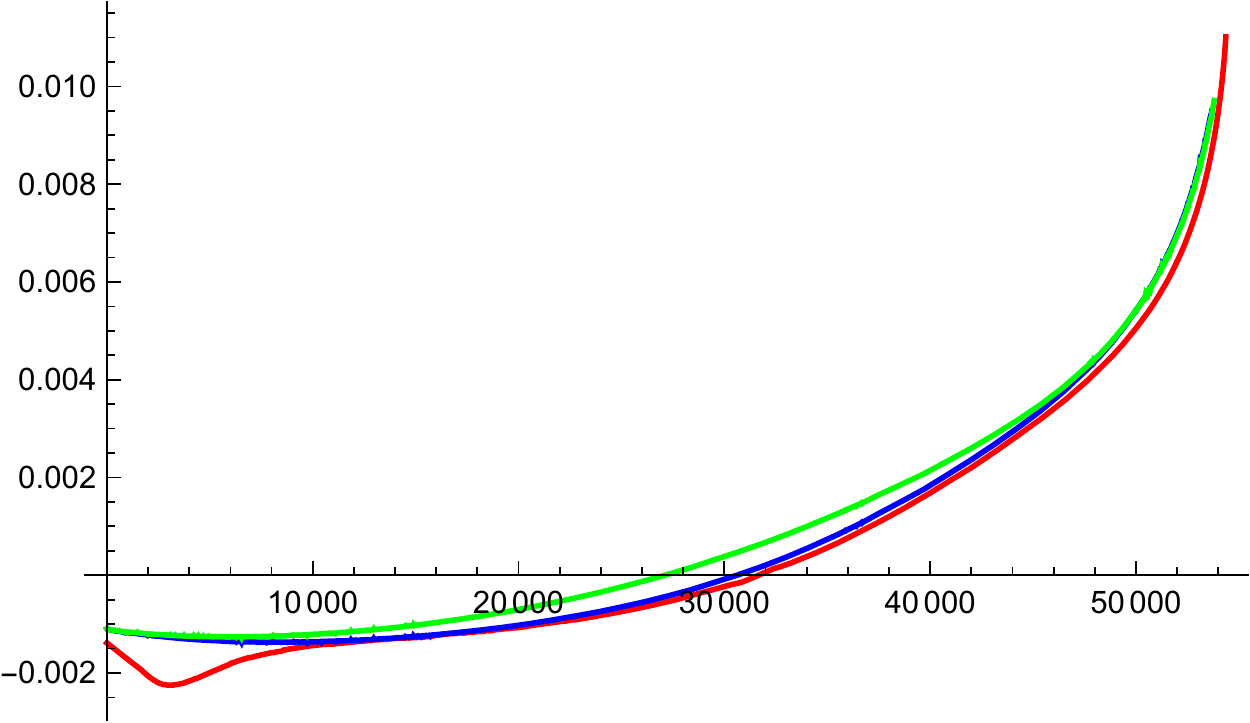}
 \put(0,010){{\tiny $\dot{y}_1$}}
 \put(92,02){{\tiny $t$}}
       \end{overpic}
    \caption{\footnotesize $q=0.3$, $x_1(0) = 120$}
        \label{fig:q300L}
      \end{subfigure}
  \begin{subfigure}[b]{0.48\textwidth}
    \begin{overpic}[width=0.49\textwidth]{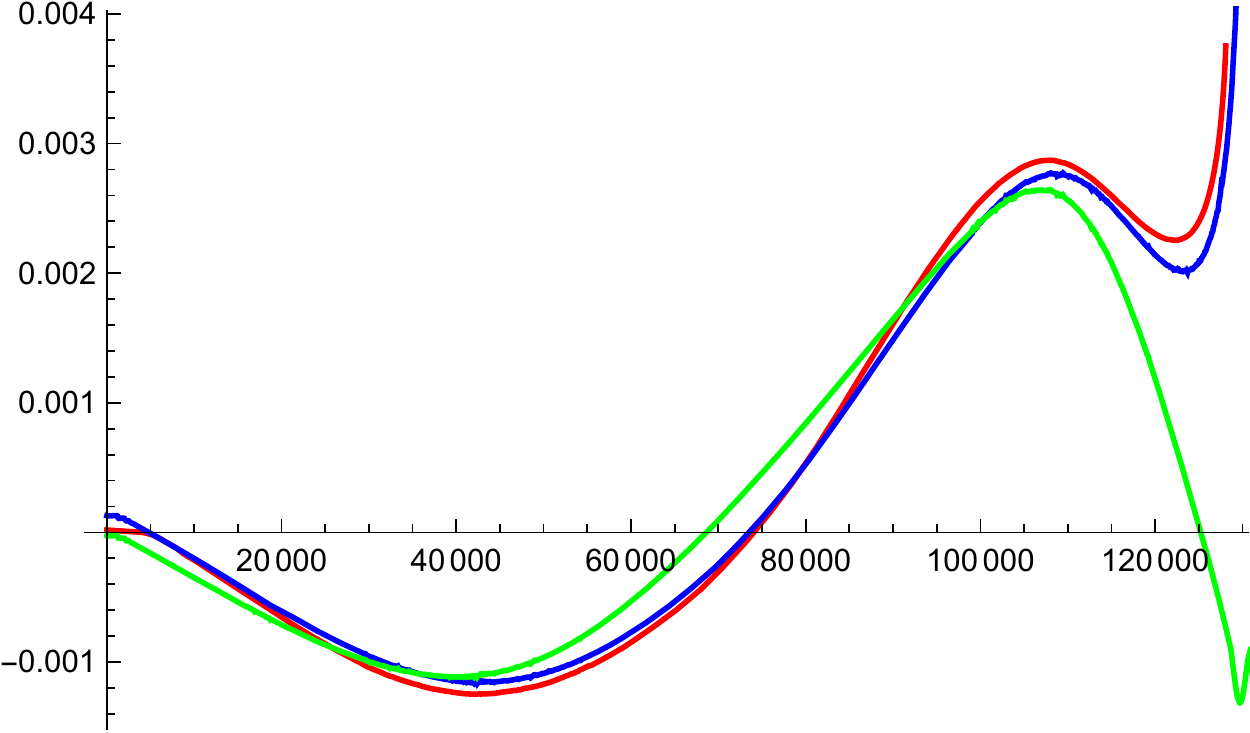}
      \put(0,04){{\tiny $\dot{x}_1$}}
      \put(93,04){{\tiny $t$}}
       \end{overpic}
    \begin{overpic}[width=0.49\textwidth]{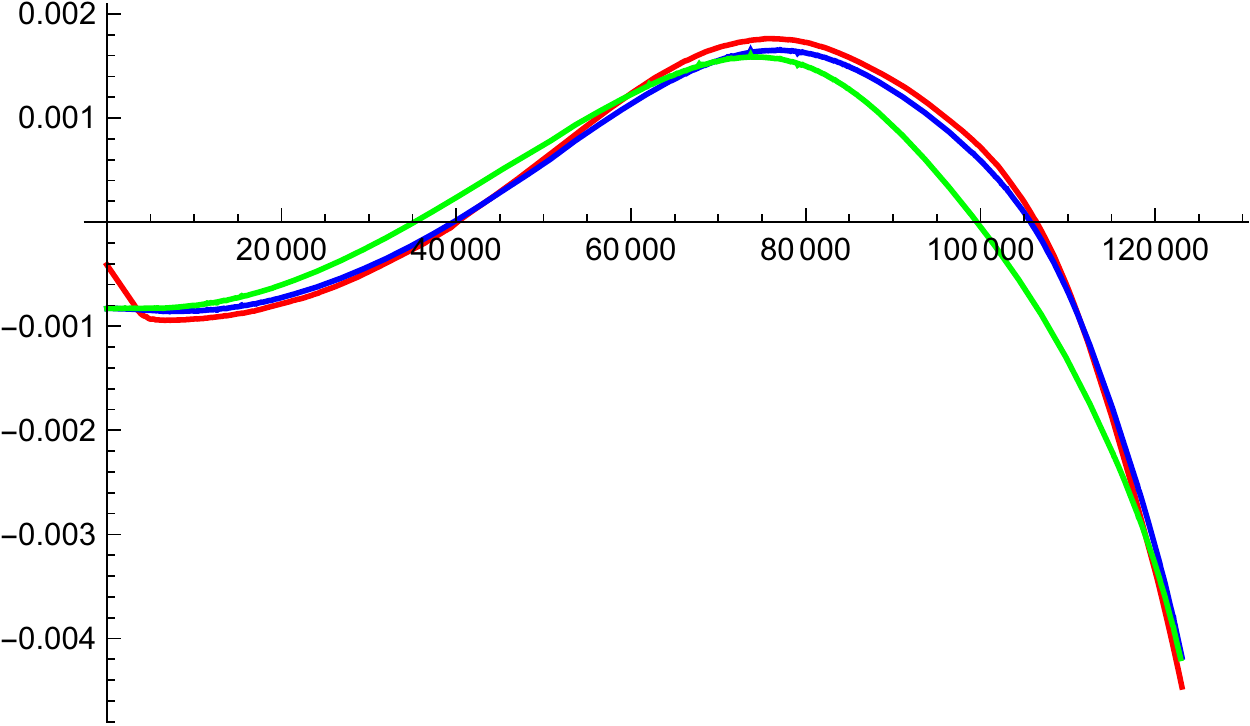}
  \put(0,40){{\tiny $\dot{y}_1$}}
  \put(92,30){{\tiny $t$}}
       \end{overpic}
    \caption{\footnotesize$q=0.35$, $x_1(0) = 120$}
        \label{fig:q350L}
      \end{subfigure}\\[2mm]

  \begin{subfigure}[b]{0.48\textwidth}
    \begin{overpic}[width=0.49\textwidth]{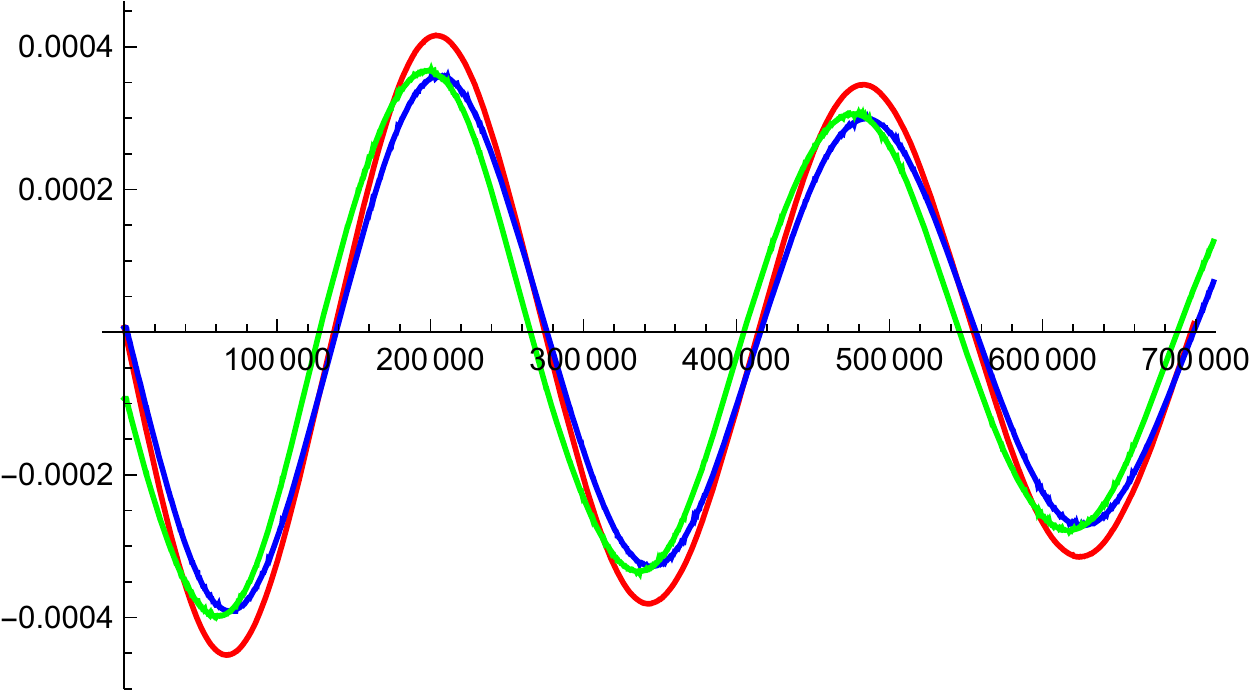}
     \put(0,28){{\tiny $\dot{x}_1$}}
     \put(95,20){{\tiny $t$}}
       \end{overpic}
    \begin{overpic}[width=0.49\textwidth]{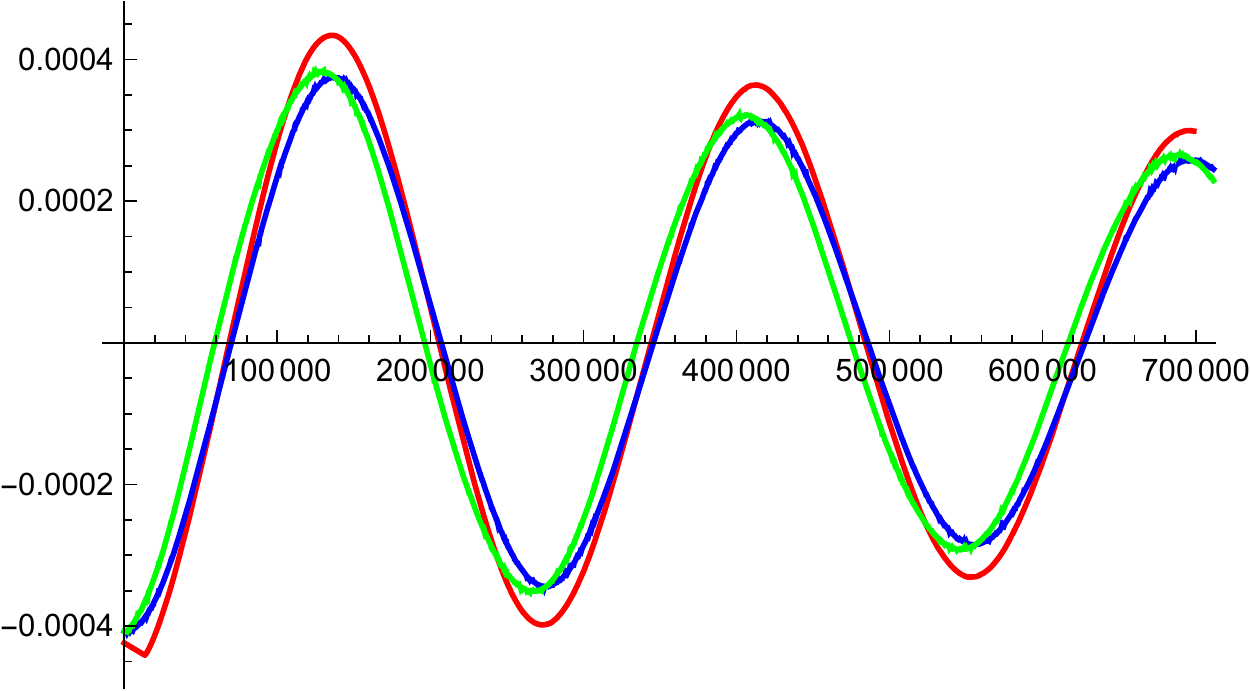}
  \put(0,28){{\tiny $\dot{y}_1$}}
  \put(92,20){{\tiny $t$}}
        \end{overpic}
   \caption{\footnotesize$q=0.4$, $x_1(0) = 120$}
        \label{fig:q400L}
      \end{subfigure}
 \begin{subfigure}[b]{0.48\textwidth}
    \begin{overpic}[width=0.49\textwidth]{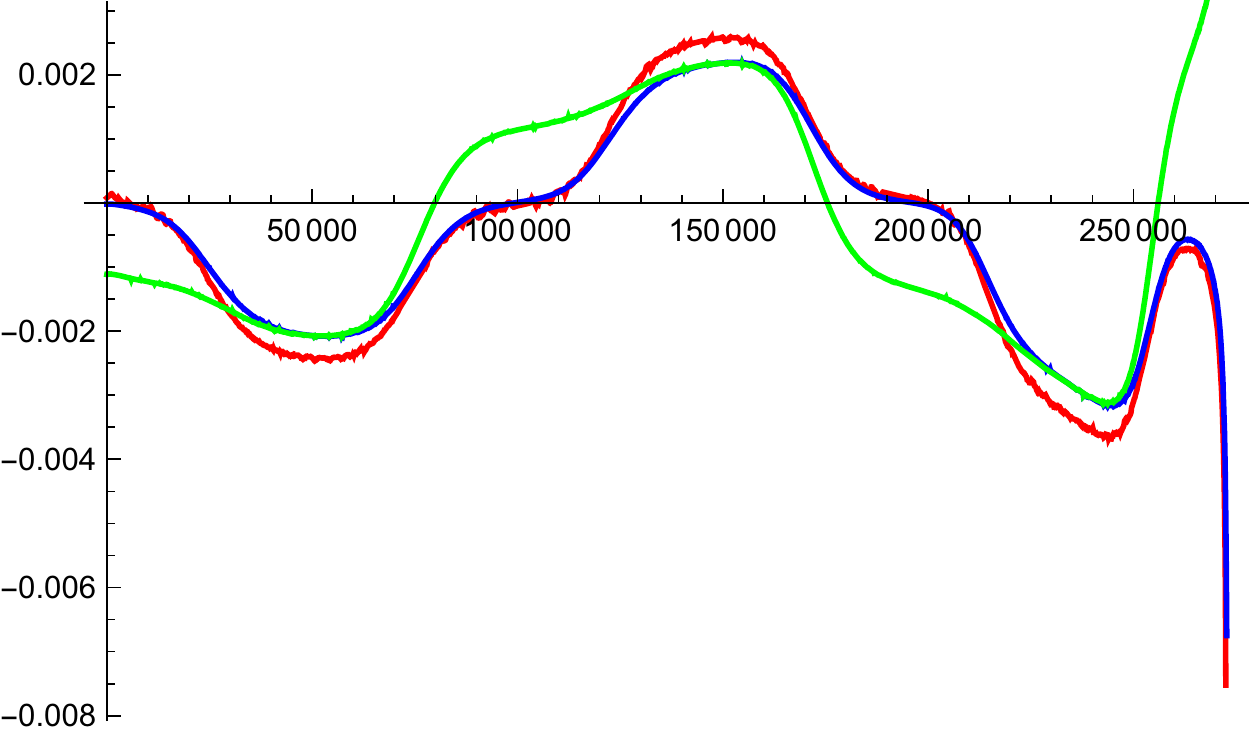}
     \put(0,40){{\tiny $\dot{x}_1$}}
     \put(95,43){{\tiny $t$}}
        \end{overpic}
    \begin{overpic}[width=0.49\textwidth]{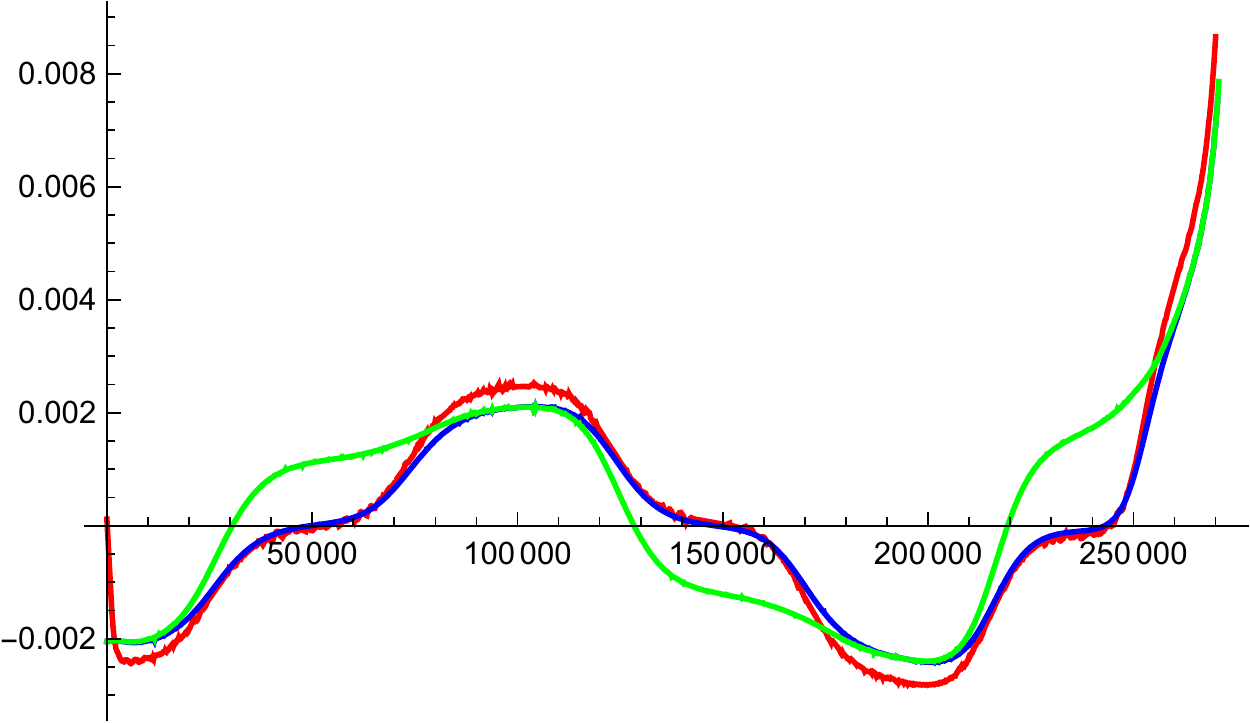}
  \put(0,18){{\tiny $\dot{y}_1$}}
  \put(92,08){{\tiny $t$}}
        \end{overpic}
   \caption{\footnotesize $q=0.45$, $x_1(0) = 161$}
        \label{fig:q450R}
    \end{subfigure}
    \caption{A comparison between the numerically-determined velocity (red), the predicted asymptotic velocity with $\eps=0.01$ (green), and the predicted asymptotic velocity with $\eps$ given by (\ref{epseqn}) (blue), as a function of time along the numerically-determined spiral trajectory, for a single spiral in the square domain $[0,200]\times[0,200]$. The starting $y$-value for each trajectory is $y_1(0) =100$. The numerical results have been locally averaged to reduce some of the noise. The trajectories themselves may be seen in Figure \ref{singnew}.}    
   \label{fig:velocity}
\end{center}
\end{figure}

In Figure \ref{fig:velocity}
we show a comparison between the numerically determined velocity (by finite differencing and smoothing \cite{anderssen} the numerically determined path in time) and the   velocity predicted by the uniform asymptotic approximation described above, as a function of time along the numerically determined spiral trajectory. The numerical solutions are for a single $+1$ spiral in the square domain  $0 \leq x,y \leq 200$ with grid
     resolution $\Delta x = 0.5$. For each value of $q$ the $x$ and $y$ velocities for two different trajectories (i.e. two different starting positions) are shown. The asymptotic results are shown for the two choices of $\eps$ described in \S\ref{sec:eps}. We see that there are still some initial transients in the velocities, but that on the whole the asymptotic approximation does quite well. The approximation gets better as $q$ increases, which seems slightly counter-intuitive since the asymptotic approximation is in the limit $q \ra 0$. This can be explained by the fact that in the near-field scaling the $1/\log \eps$ correction terms play a more significant role than they do in the canonical scaling (in the canonical region $\eps$ disappears from the law of motion when written in terms of the original variables, while in the near field region it does not). From Figure \ref{fig:velocity} we also see that the green curves, corresponding to choosing $\eps=0.01$, fit less well at higher values of $q$, particularly near the end of the trajectory in which the spiral is approaching the boundary. On the other hand the blue curves, which take the distance to  the boundary into account in $\eps$ through (\ref{epseqn}), fit very well.

\begin{figure}[p]
\begin{center}
  \begin{subfigure}[b]{0.3\textwidth}
    \includegraphics[width=\textwidth]{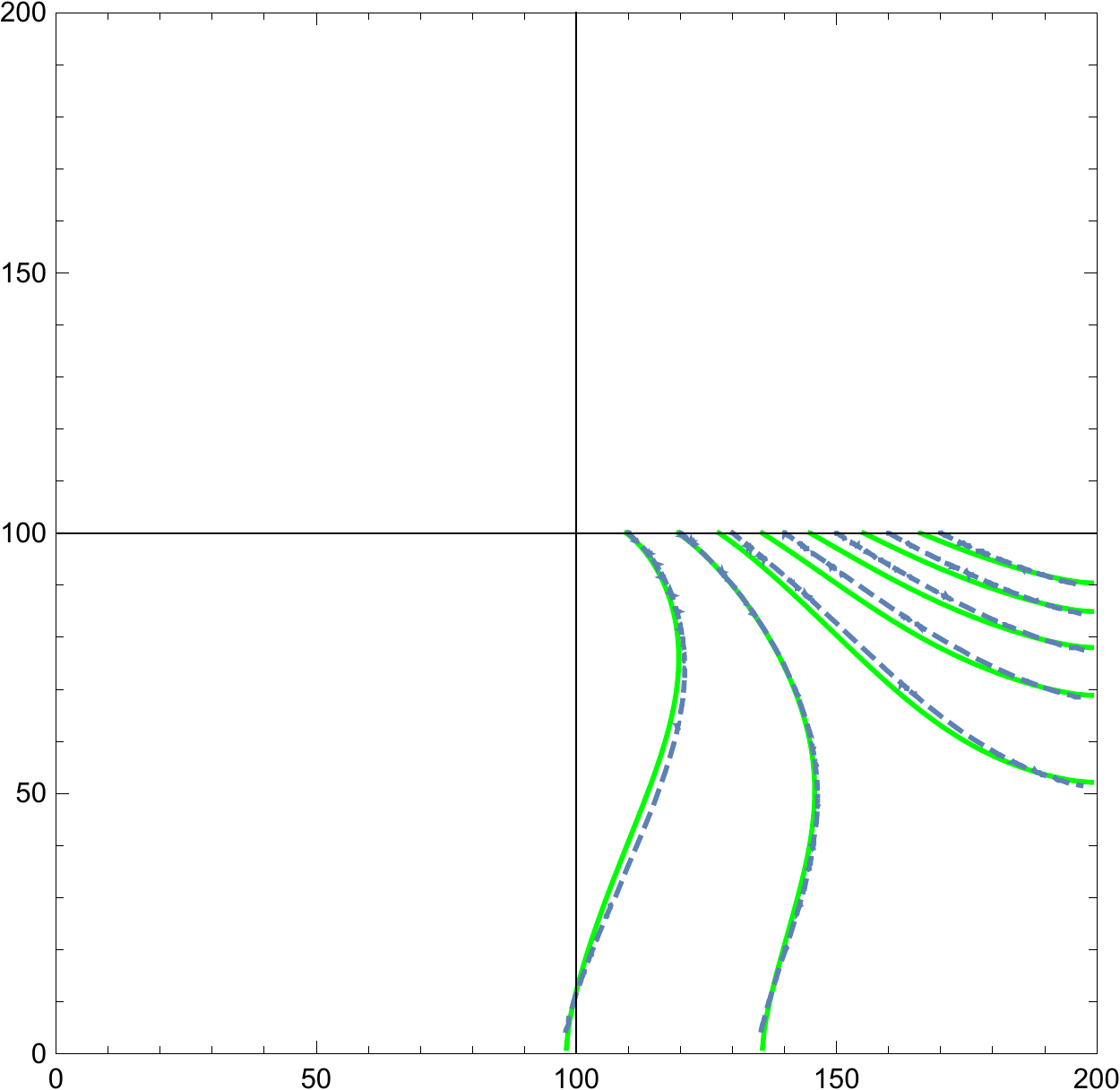}
    \caption{$q=0.1$ Near field IC}
        \label{fig:q100p2la1000}
    \end{subfigure}
   \begin{subfigure}[b]{0.3\textwidth}
    \includegraphics[width=\textwidth]{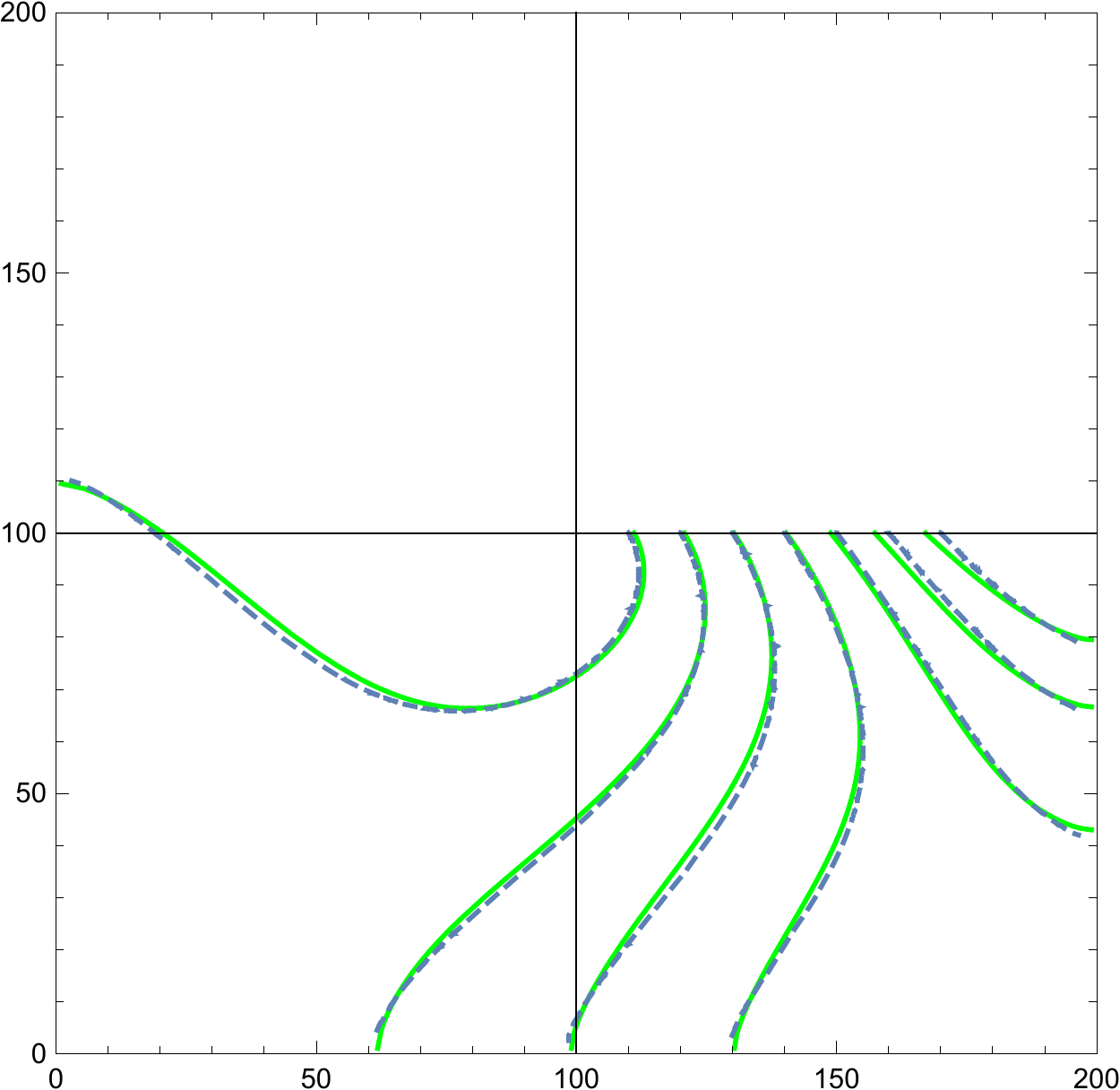}
    \caption{$q=0.2$ Near Field IC}
        \label{fig:q200p2la1000}
    \end{subfigure}
 \begin{subfigure}[b]{0.3\textwidth}
    \includegraphics[width=\textwidth]{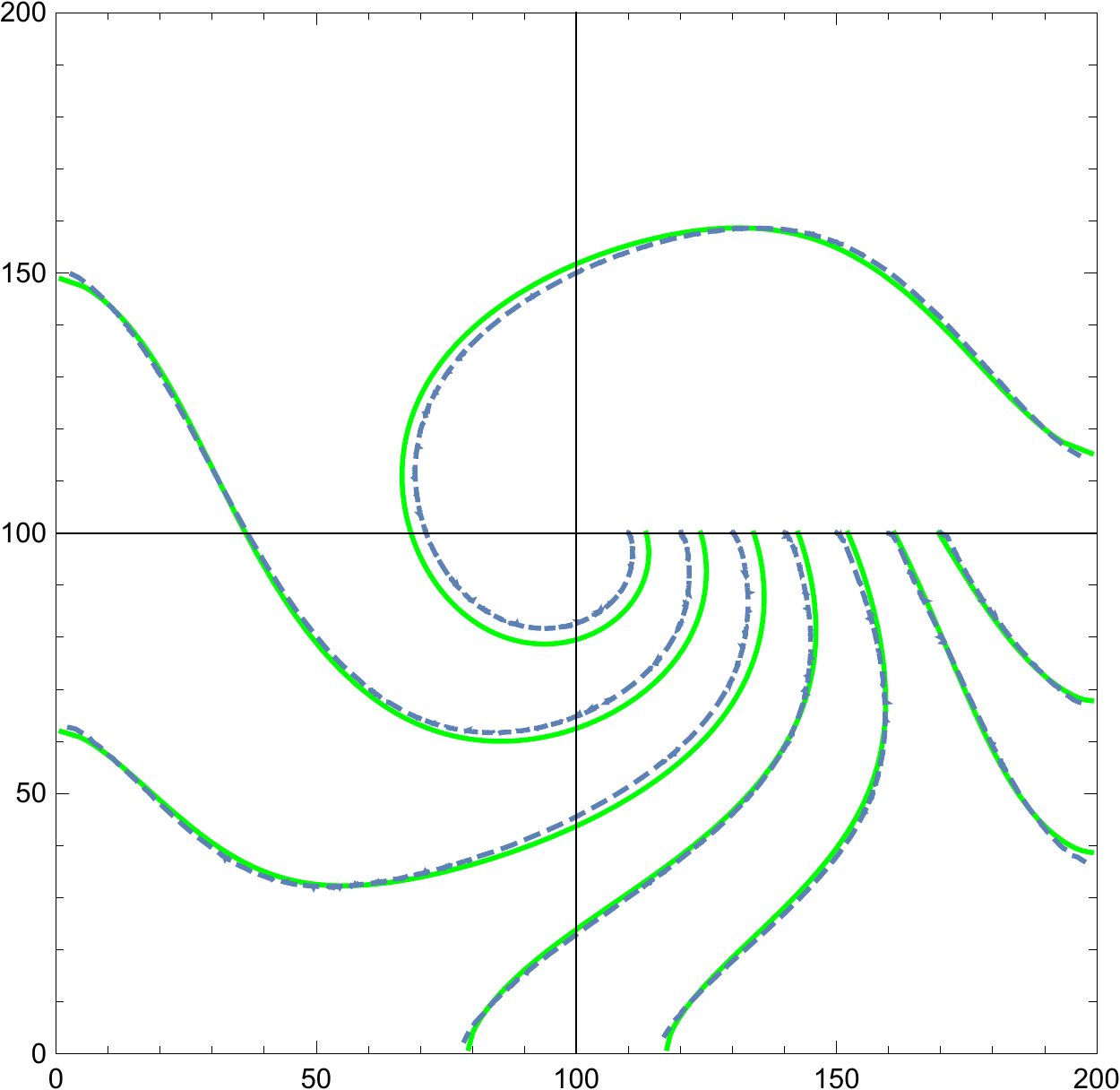}
    \caption{$q=0.3$ Near Field IC}
        \label{fig:q300p2la1000}
    \end{subfigure}
  \begin{subfigure}[b]{0.3\textwidth}
    \includegraphics[width=\textwidth]{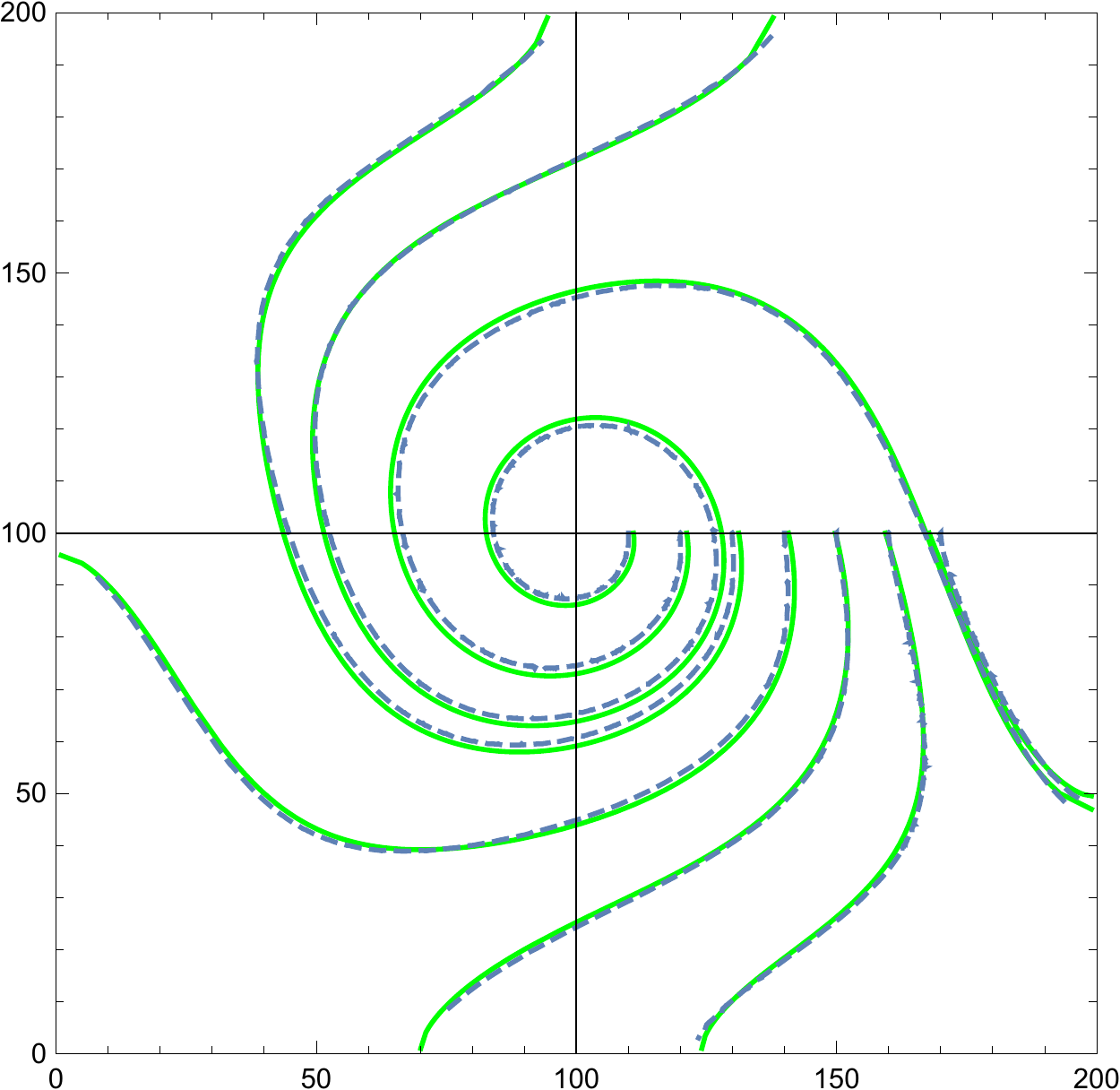}
    \caption{$q=0.35$ Canonical IC}
        \label{fig:q350p2la1000}
    \end{subfigure}
  \begin{subfigure}[b]{0.3\textwidth}
    \includegraphics[width=\textwidth]{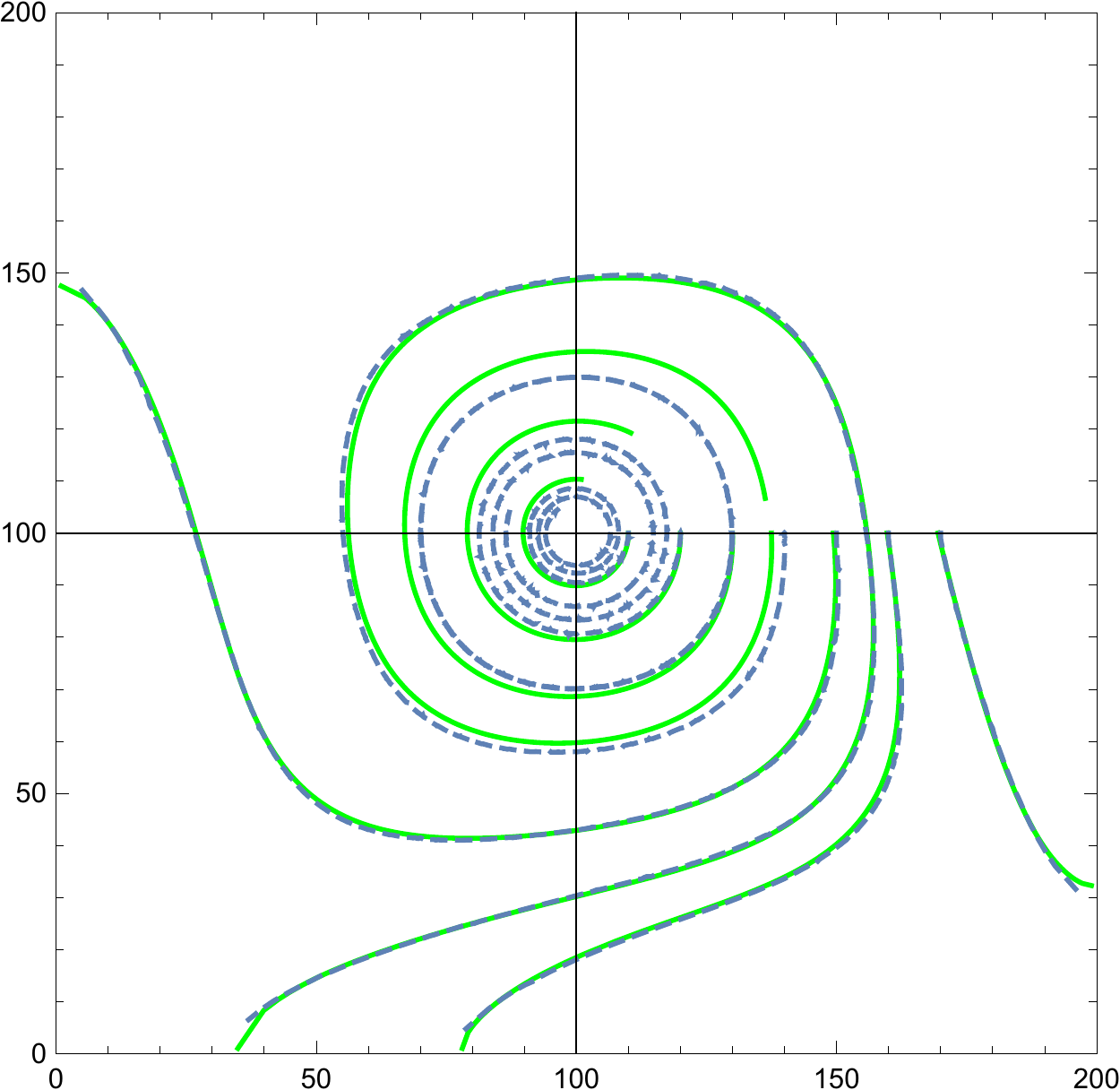}
    \caption{$q=0.4$ Canonical IC}
        \label{fig:q400p2la1000}
    \end{subfigure}
  \begin{subfigure}[b]{0.3\textwidth}
    \includegraphics[width=\textwidth]{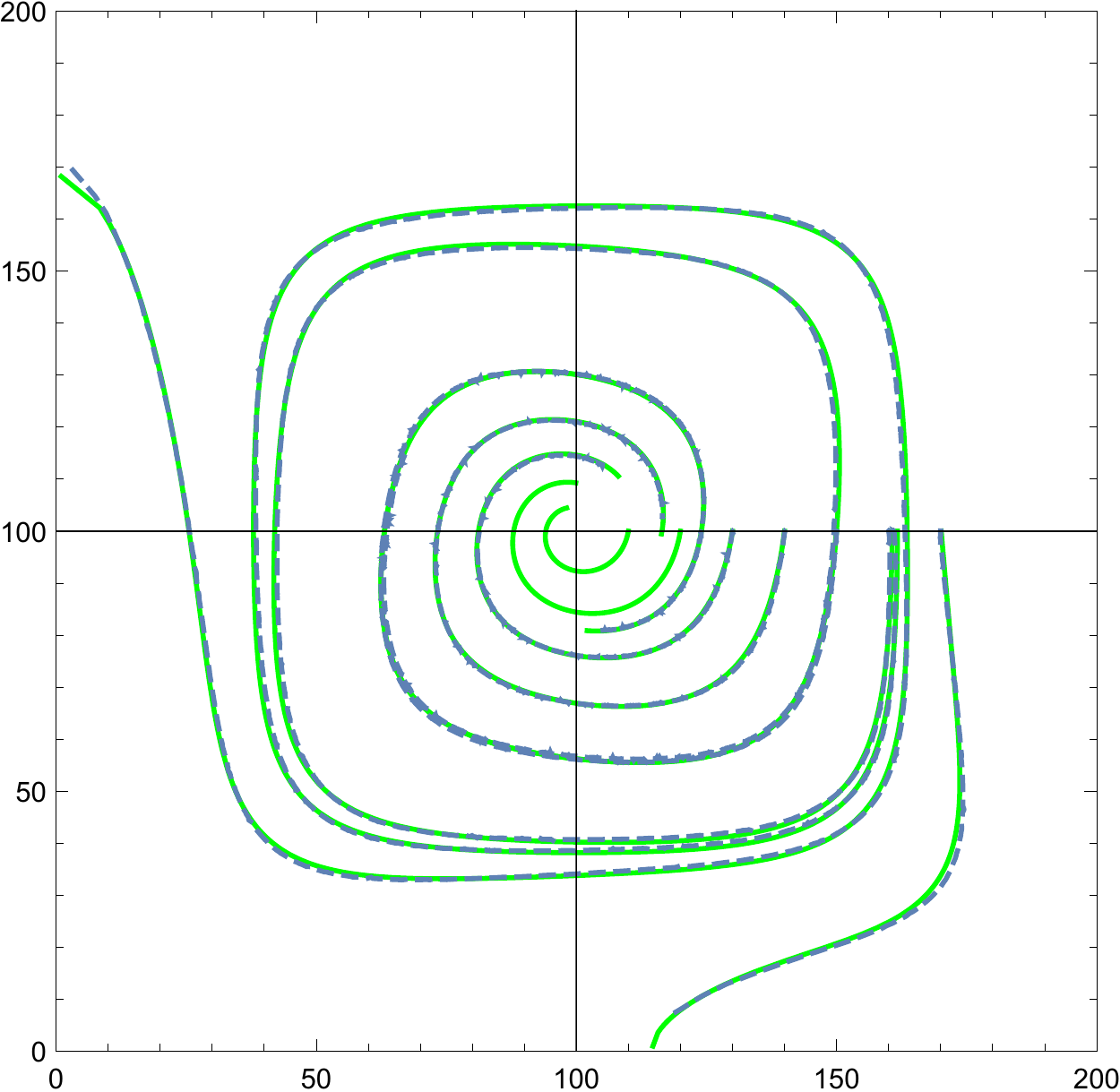}
    \caption{$q=0.45$ Canonical IC}
        \label{fig:q450p2la1000}
    \end{subfigure}
   \caption{Comparison between the trajectories provided by a direct
     numerical simulation of (\ref{eq70}) (dashed lines) and the
     uniform asymptotic approximation of \S\ref{sec:composite}  (solid
     lines) for a 
     single spiral    in a square domain of  side
     200. Numerical trajectories starting from   positions
     $(110,100),(120,100),\ldots,(170,100)$ are shown; $\eps$ is given by
     (\ref{epseqn}). 
       Note the appearance of an unstable periodic orbit in
   (\subref{fig:q400p2la1000}) and (\subref{fig:q450p2la1000}) which is
   captured by the asymptotic law of motion.  An extra orbit starting
   from position (161,100) is shown in (f)---the periodic orbit crosses
   the line $x=100$ somewhere between 160 and 161.}   
   \label{singnew}
\end{center}
\end{figure}

In Figure \ref{singnew} we compare directly  the numerical trajectories 
 (dashed lines) and those
     given by the
     uniform asymptotic approximation  with $\eps$ given by (\ref{epseqn}) (solid      lines), for a 
     single $+1$ spiral  in a square domain $0 \leq x,y \leq 200$.
     Numerical trajectories are shown starting from positions
     $(110,100),(120,100),\ldots,(170,100)$. Because of the initial transients in the numerical results, and to mitigate the effects of diverging trajectories mentioned earlier, we solve the asymptotic trajectories backwards from a point on the numerical trajectory near the boundary of the domain\footnote{For trajectories which do not leave the domain we solve forwards from the initial position.}. Specifically we find the asymptotic trajectory which coincides with the numerical trajectory on the smooth closed curve $(x-100)^4+(y-100)^4 = 90^4$.

    For small $q$ we see that the spiral is attracted to the boundary
    whatever its inital position. However, as $q$ is increased there is
    a Hopf bifurcation with the appearance of an unstable periodic
    orbit. Trajectories starting outside this periodic orbit are
    attracted to the boundary of the domain, but those starting inside it
    spiral in to the origin. As $q$ is increased further the periodic
    orbit grows in size and develops a more squareish shape. This can be
    understood as the spiral interacting with its images predominantly
    in the near-field limit, in which the motion is perpendicular to
    the line of centres. 
With the motion dominated by the nearest image the spiral will move
parallel to the nearest boundary until it nears the 
corner, when a second image takes over.
We see that the asymptotic law of motion captures the appearance of the perioidic orbit. In Fig.~\ref{fig:q400p2la1000} the amplitude of the asympotic periodic orbit is not quite right (it crosses the line $y=100$ close to $x=110$ rather than $x=130$), but in Fig.~\ref{fig:q450p2la1000} the periodic orbit is captured well quantitatively as well as qualitatively.

\begin{figure}[p]
\begin{center}
   \begin{subfigure}[b]{0.3\textwidth}
    \includegraphics[width=\textwidth]{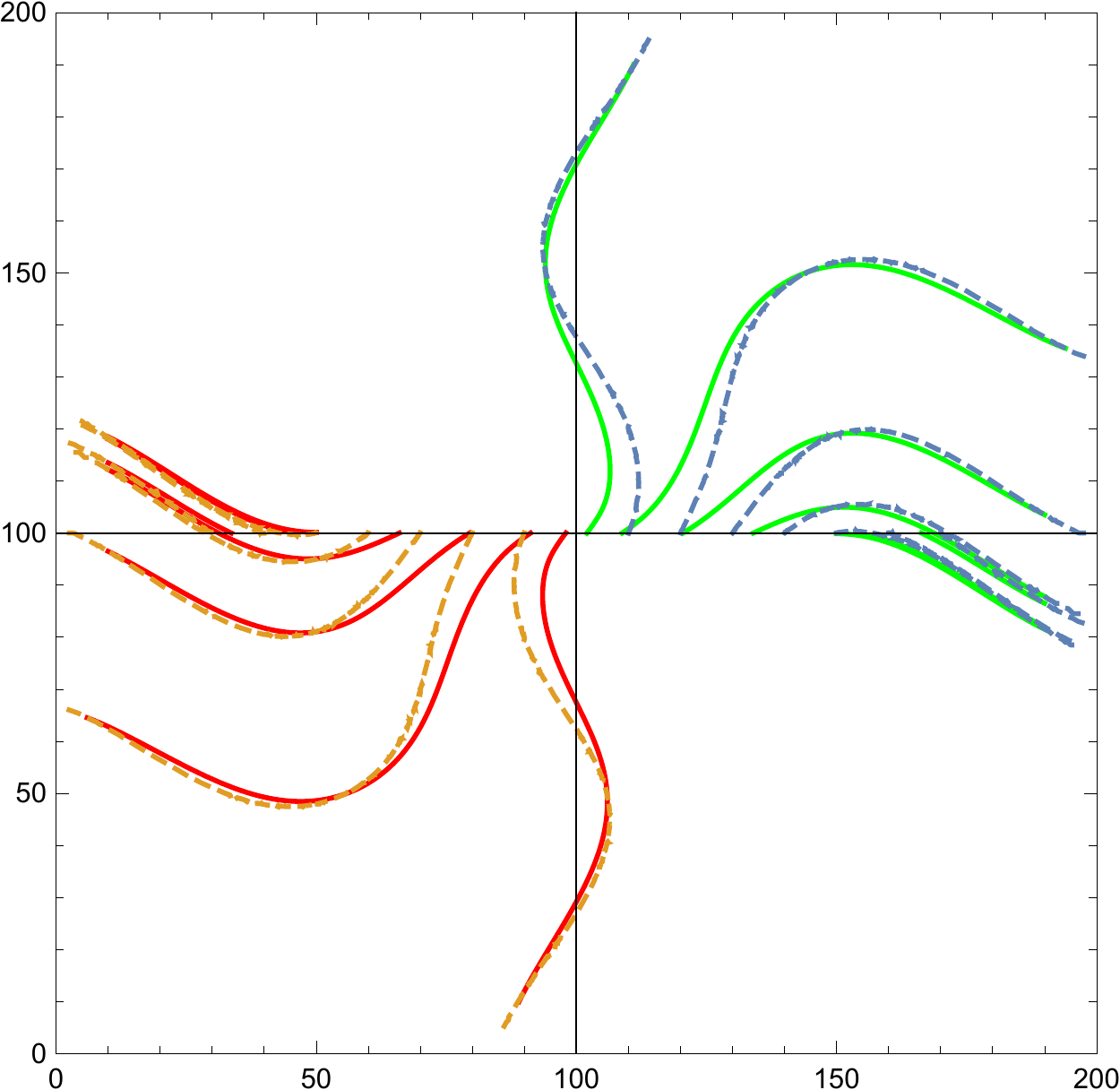}
    \caption{$q=0.2$ near field}
        \label{fig:doubleq200}
    \end{subfigure}
    \begin{subfigure}[b]{0.3\textwidth}
    \includegraphics[width=\textwidth]{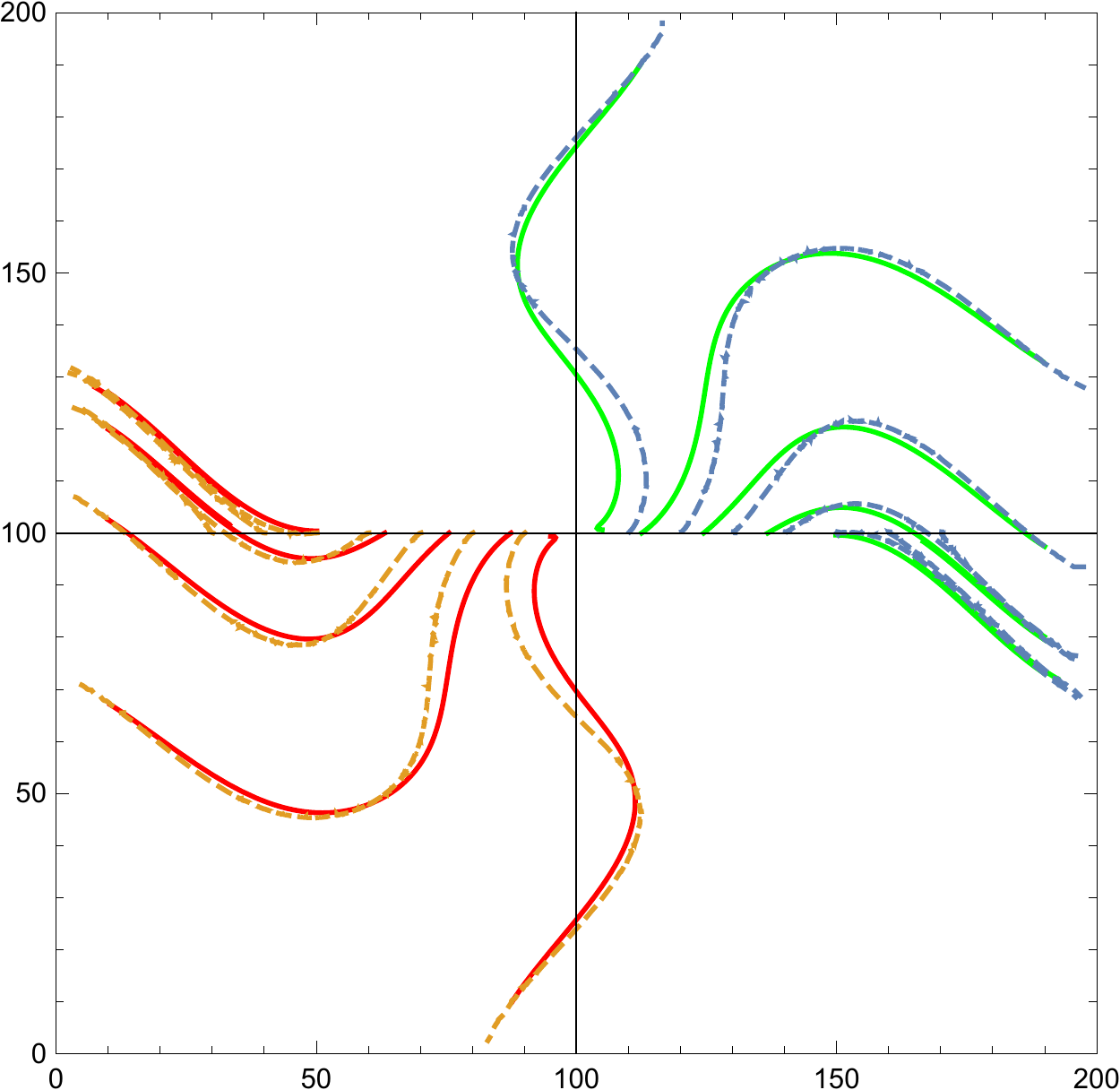}
    \caption{$q=0.25$}
        \label{fig:doubleq250}
    \end{subfigure}
 \begin{subfigure}[b]{0.3\textwidth}
    \includegraphics[width=\textwidth]{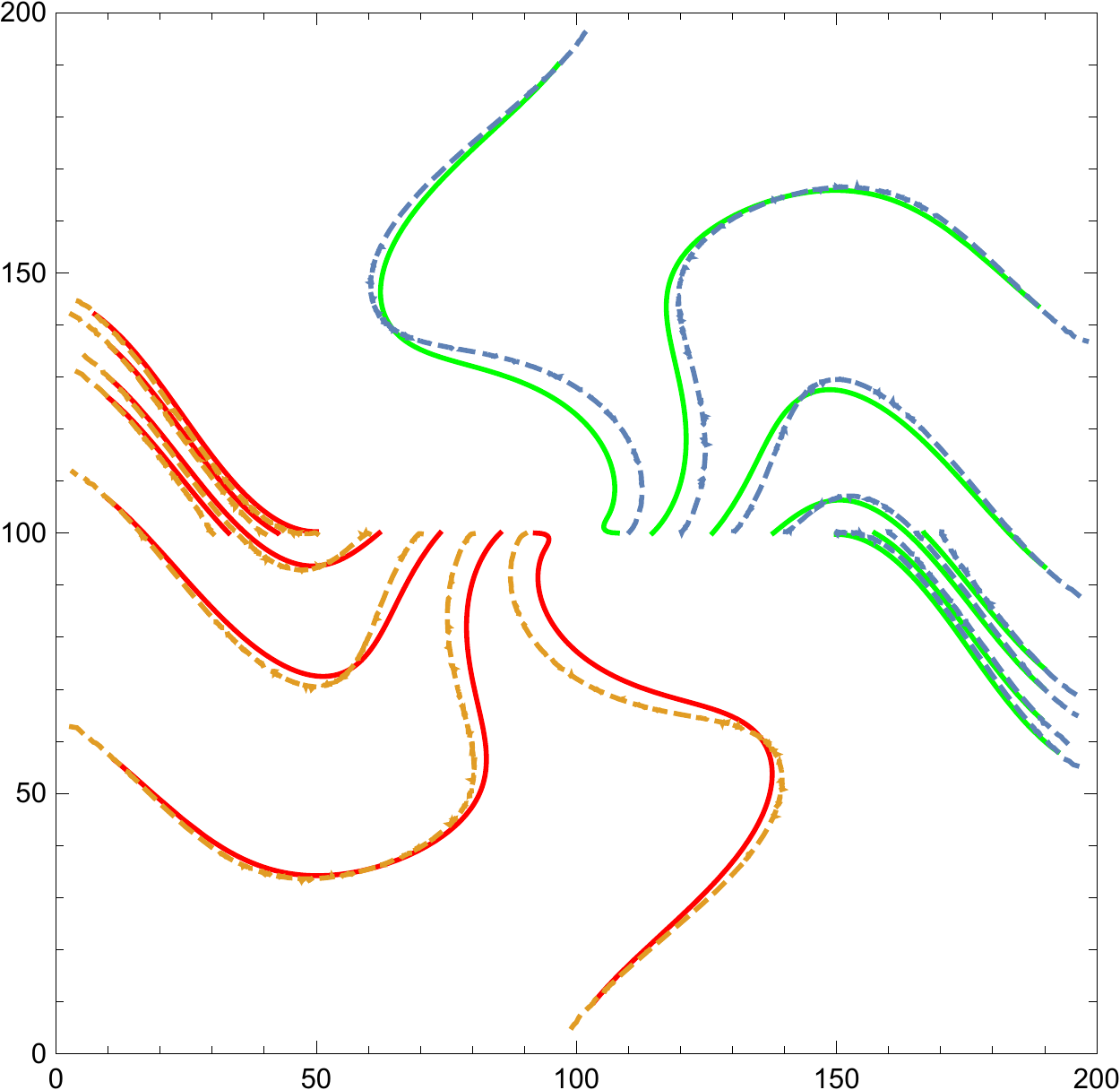}
    \caption{$q=0.3$}
        \label{fig:doubleq300}
    \end{subfigure}
  \begin{subfigure}[b]{0.3\textwidth}
    \includegraphics[width=\textwidth]{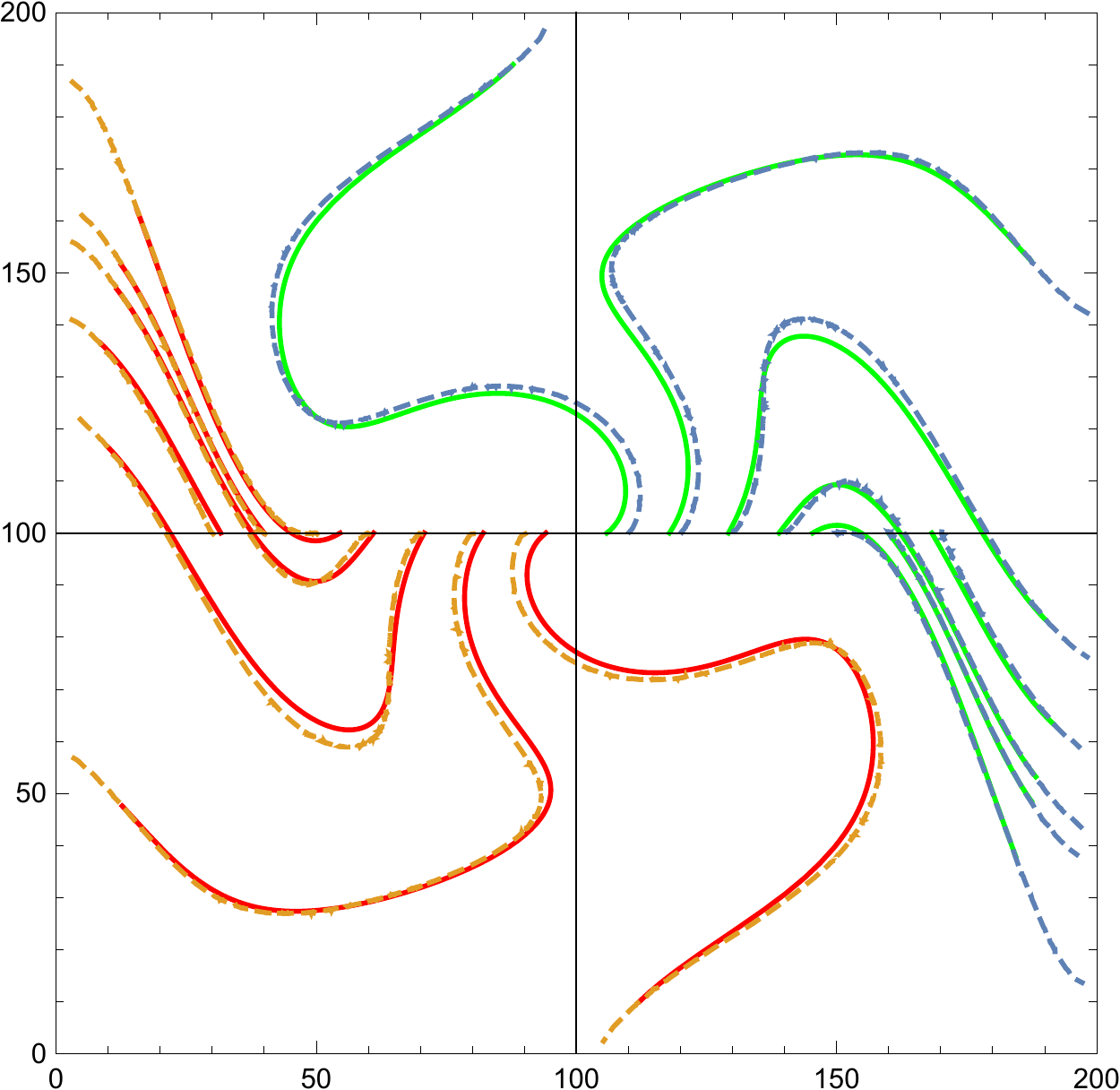}
    \caption{$q=0.35$}
        \label{fig:doubleq350}
    \end{subfigure}
  \begin{subfigure}[b]{0.3\textwidth}
    \includegraphics[width=\textwidth]{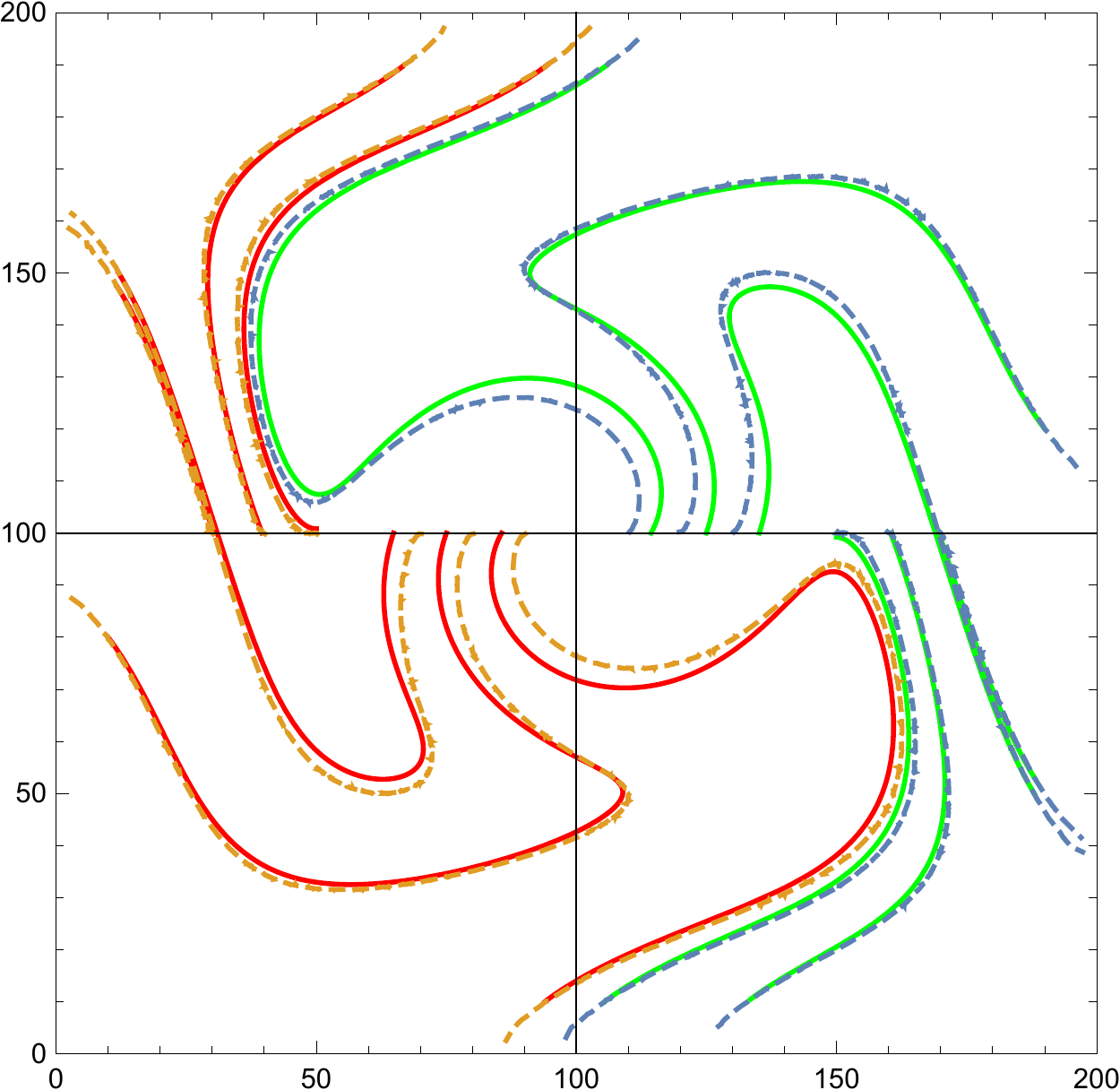}
    \caption{$q=0.4$}
        \label{fig:doubleq400}
    \end{subfigure}
    \caption{Comparison between the trajectories provided by a direct
     numerical simulation of (\ref{eq70}) (dashed lines) and the
     uniform asymptotic approximation of \S\ref{sec:composite}  (solid
     lines) for a 
pair of spirals    in a square domain of  side
     200. Spirals are placed symmetrically at positions $(100-x,100)$ and $(100+x,100)$ with $x
     = 10,20,\ldots,70$; $\eps$ is given by
     (\ref{epseqn2}). }
   \label{doublenew}
\end{center}
\end{figure}

 In Figure \ref{doublenew} we compare  the trajectories provided by a direct
     numerical simulation of (\ref{9}) (dashed lines) and those
     given by the
     uniform asymptotic approximation (solid
     lines) for a pair of +1 spirals  in the same square domain $0
     \leq x,y\leq 200$. We position the spirals symmetrically at
     positions $(100-x,100)$ 
     and $(100+x,100)$, where we choose $x = 10, 20, \ldots, 70$.
     We see that the agreement is qualitatively very good, again improving as $q$ increases. The spirals
attempt to circle around each other, as the near-field interaction
would indicate, but gradually drift apart until the image spirals take
over and force the pair to rotate in the opposite direction.

\section{Conclusions}
\label{conclusions}

We have developed a law of motion for interacting spiral waves in a
bounded domain in the limit that the twist parameter $q$ is small. We
find that the size of the domain is crucial in 
determining the form of this law of motion.
Our main results can be summarised as follows.
For $0<q\ll 1$, given a set of $\pm1$-armed spirals in a domain of diameter $O(1/\eps)$, the positions $\bx_\ell$ of the spirals evolve according to the following laws of motion:
\begin{description}
\item{(i) } For a so-called {\em canonical} domain size, which corresponds to $q |\log \eps| = \pi/2 + q \nu$ with $\nu = O(1)$ as $q, \eps \ra 0$,
\beqa
 \frac{\d\bx_{\ell}}{\d t}
&=&\frac{4 \pi q n_\ell  }{\beta_{\ell}}
\sum_{j=1,j \not = \ell}^N \beta_j \del^{\perp} \Gpn(\bx_\ell;\bx_j) 
+4 \pi q  n_\ell
\del^{\perp}
\Gpnreg(\bx_\ell;\bx_\ell) 
\eeqa
where $\del^{\perp} = (-\dd_y,\dd_x)$ and $\Gpn(\bx;\by)$ is the Neumann Green's function for the modified Helmholtz equation on $\Omega$, satisfying 
\beq
\lap \Gpn-q^2k^2 \Gpn=\delta(\bx-\by)\quad\textrm{in}\quad\Omega,\qquad
\frac{\dd \Gpn}{\dd n} = 0\quad\textrm{on}\quad \dd\Omega.
\eeq
with 
\[
\Gpnreg(\bx;\by) =\Gpn(\bx;\by)-
\frac{1}{2
  \pi}\log|\bx-\by|.\]

The coefficients $\beta_\ell$ are given (up to an arbitrary and irrelevant scaling factor) as solutions of the linear system of equations
\begin{equation}
 2 \pi \beta_\ell  \Gpnreg(\bx_\ell;\bx_\ell)+  2 \pi
 \sum_{j=1,j \not = \ell}^N \beta_j
 \Gpn(\bx_\ell;\bx_j)-\beta_{\ell}(c_1 - \pi/2 q)=0,\label{con:beta}
\end{equation}
where $c_1 \approx 0.098$,  whose solvability condition (the condition for a non-zero solution) determines the eigenvalue $k$.

\item{(ii) }For a so-called {\em near-field} domain size, which corresponds to $0<q |\log \eps| < \pi/2$,

\beqa
\frac{\d\bx_{\ell}}{\d t} & = & 
4 \pi q n_\ell \sum_{j=1,j \not = \ell}^N 
  \del^{\perp}\oGpn(\bx_{\ell};\bx_{j})  + 4\pi q n_\ell\del^{\perp} \oGpnreg(\bx_{\ell};\bx_{\ell})\nonumber \\ && \mbox{ }-4 \pi q \cot(q\log\e) \del\oGpdreg(\bx_{\ell};\bx_{\ell}) 
- \frac{4 \pi q\cot(q\log\e)}{n_\ell}\sum_{j=1,j \not = \ell}^N n_j\del
\oGpd(\bx_{\ell};\bx_{j}),\qquad
\eeqa
where $\oGpn(\bx;\by)$ and $\oGpn(\bx;\by)$ are the Neumann and Dirichlet Green's functions for Laplace's equation on $\Omega$, satisfying
\begin{align}
  \del^2 \oGpn  &=  \delta(\bx-\by) - \frac{1}{|\Omega|} \quad\textrm{
                  in }\Omega,&\qquad\quad
\frac{\dd \oGpn}{\dd{ n}}&=0\quad\textrm{ on
}\dd\Omega,\\
  \del^2 \oGpd & =  \delta(\bx-\by) \quad\textrm{
                  in }\Omega,&\qquad\quad
\oGpd&=0\quad\textrm{ on }\dd\Omega,
\end{align}
and
\[
\oGpnreg(\bx;\by) =\oGpn(\bx;\by)-
\frac{1}{2
  \pi}\log|\bx-\by|, \qquad \oGpdreg(\bx;\by) =\oGpd(\bx;\by)-
\frac{1}{2
  \pi}\log|\bx-\by|.\]

\item{(iii) }A uniform approximation, valid in both regions, is given by
\beqa
\frac{\d\bx_{\ell}}{\d t} & = & 
\frac{4 \pi q n_\ell}{\beta_\ell} \sum_{j=1,j \not = \ell}^N \beta_j
  \del^{\perp}\Gpn(\bx_{\ell};\bx_{j})  + 4\pi q n_\ell\del^{\perp} \Gpnreg(\bx_{\ell};\bx_{\ell})\nonumber \\ && \mbox{ }-4 \pi q \cot(q\log\e) \del\Gpdreg(\bx_{\ell};\bx_{\ell}) 
- \frac{4 \pi q\cot(q\log\e)}{n_\ell}\sum_{j=1,j \not = \ell}^N n_j\del
\Gpd(\bx_{\ell};\bx_{j}),\qquad
\eeqa
where $\Gpd$ is the Dirichlet Green's function for the modified Helmholtz equation given by 
\begin{align}
  \lap \Gpd-q^2 k^2 \Gpd&=\delta(\bx-\by)\quad\textrm{in}\quad\Omega,&\qquad
\Gpd &= 0\quad\textrm{on}\quad \dd\Omega,
  \end{align}
with
\[
\Gpdreg(\bx;\by) =\Gpd(\bx;\by)-
\frac{1}{2
  \pi}\log|\bx-\by|,\]
and $\beta_\ell$ and $k$  given by (\ref{con:beta}).

  \end{description}

Although we have focussed on Neumann boundary conditions  for
the complex Ginzburg-Landau equation (\ref{CGL}), the extension to
periodic boundary conditions is straightforward. 

\paragraph{Acknowledgements} M. Aguareles is part of the Catalan
Research group 2017 SGR 1392 and has been supported by the MINECO
grant MTM2017-84214-C2-2-P (Spain). She would also like to
thank the Oxford Centre for Industrial and Applied Mathematics,  where
part of this research was carried out.   

\newpage
\appendix
\section{Comment on \cite{pismen92}}
\label{appA}
As the only published work (other than our previous papers \cite{AgChWi08,AgCh09}) to consider the motion of spirals whose separation is not large by comparison to $\ee^{\pi/2q}$, \cite{pismen92} is an important reference in the field, even though their approach does not generalise easily to more than two spirals or to bounded domains.
Unfortunately it seems \cite{pismen92} make a number of mistakes right at the beginning of their paper.
Equations (8) and (9) in \cite{pismen92} are incorrect, and 
  should read
  \beqas
 \eta
  {\rho} {\bv}\cdot \del {\theta}
&=&\del^2{\rho} +\left( 
1- |\del
{\theta}|^2  -{\rho}^2 -\frac{v^2 \eta^2}{4} \right) {\rho}
,\\
-\eta
  {\bv}\cdot \del {\rho} 
&=& 2  \del {\theta}
\cdot \del {\rho} +  {\rho} \del^2
{\theta} -  q \left(1-k_0^2 -{\rho}^2 \right){\rho},
\eeqas
It seems that in \cite{pismen92} the authors mistakenly applied the transformation $u \ra u \ee^{-\ii \omega t + \eta \hat{\bv}\cdot \hat{\bx}/2}$ rather than $u \ra u \ee^{-\ii \omega t + \ii \eta   \hat{\bv}\cdot \hat{\bx}/2}$ which they had intended. 
Equation (10) in \cite{pismen92}  also seems incorrect, and should instead read
  \[ 
    \omega = \frac{\eta - q(1-k_0^2)}{1+\eta
    q (1-k_0^2)}.\]
    There are further errors in deriving (14) and (15) from (8) and (9) (two sign errors in (14) and a missing term in (15)), but since (8) and (9) are themselves incorrect that is rather academic. We do not follow through the implications of these mistakes, since the configuration of spirals they consider is a special case of the much more general setting considered here, so that their results are in any case superceded by ours.

\section{The law of motion when $b\neq 0$}
\label{secApA}
If $b\neq 0$, equation \eqref{pdeCGL2} reads
\begin{equation}
(1-\ii b)\frac{\dd \p}{\dd t} = \lap \p+(1-|\p|^2)\p+\ii q\p(1-k^2-|\p|^2),
\label{CGLb}
\end{equation}
which after writing  $\psi=fe^{\ii \ki}$ and $\alpha=kq/\e$ yields the system
\begin{eqnarray}
f_t+b f \ki_t &=& \delX^2 f- f|\delX\ki|^2+f(1-f^2)\label{fb0},\\
f^2\ki_t-b f f_t&=& \delX\cdot(f^2 \delX
   \ki)+qf^2(1-f^2)-\frac{\e^2\al^2}{q} f^2. \label{kib0} 
\end{eqnarray}
Writing $T=\m\e^2 t$ and $\bX=\e\bx$ the outer equations (\ref{11})-(\ref{12}) now read 
\begin{eqnarray}
\m\e^2(f_T+b f \ki_T) &=& \e^2\delX^2 f-\e^2 f|\delX\ki|^2+f(1-f^2)\label{fb},\\
\m\e^2(f^2\ki_T-b f f_T)&=& \e^2\delX\cdot(f^2 \delX
   \ki)+qf^2(1-f^2)-\frac{\e^2\al^2}{q} f^2, \label{kib} 
\end{eqnarray}
which, expanding in asymptotic power series in $\e$ as $f\sim f_0+\e^2 f_1+\cdots$ and $\ki\sim \ki_0+\e^2 \ki_1+\cdots$, gives,  in place of (\ref{P11}), 
\[ f_0=1, \quad f_1=-\frac{1}{2}\left(\m b\ki_{0T}+|\del\ki_0|^2\right),\]
so that equation (\ref{chi0eqn}) for the leading-order (in $\e$) phase becomes 
\[
\m\ki_{0T}(1-qb)=\delX^2\ki_0+q|\del\ki_0|^2-\alpha^2/q.
\]
We see that the correction due to nonzero $b$ is of $O(\mu q b)$, so that the equations for $\chi_{00}$ and $\chi_{01}$ in both the canonical scaling and the near field scaling are unchanged if $b=O(1)$.

In the inner region, if $b \not =0$,  equation (\ref{innerpsi}) becomes
\begin{equation*}
\e\m(1-\ii b)\left(\e\p_T-\frac{\d\bX_{\ell}}{\d T}\cdot\del\p\right) =
\del^2\p+(1+\ii q)(1-|\p|^2)\p-
\ii \frac{\e^2\al^2}{q}\p.
\end{equation*}
The leading order equation (\ref{psi0eqn}) is unchanged, while
the first order equation (\ref{1orderinner}) becomes
\begin{equation*}
-\m (1-\ii b)\frac{\d\bX_{\ell}}{\d T} \cdot \del \p_0 = \del^2
 \p_1+(1+\ii q)(\p_1(1-2|\p_0|^2)-\p_0^2\p_1^*).
\end{equation*}
or equivalently, in terms of $f_1$ and $\ki_1$,
\begin{eqnarray*}
- \m \frac{\d\bX_{\ell}}{\d T}\cdot\left(\del f_0+bf_0\del\ki_0\right) &=& \del^2 f_1
-f_1|\del  \ki_0|^2
-2 f_0 \del  \ki_0 \cdot \del \ki_1+
f_1 - 3 f_0^2 f_1,\\
- \m f_0 \frac{\d\bX_{\ell}}{\d T}\cdot\left(f_0\del \ki_0-b \del f_0\right) &=& \del \cdot(f_0^2\del
  \ki_1)+\del \cdot(2 f_0 f_1\del
  \ki_0)
+2 q f_0 f_1 -4 q f_0^3 f_1.
\end{eqnarray*}
When calculating the outer limit of the first order inner equation (\ref{chi1a}) is now modified to
\[
- \m (1-qb) \frac{\d\bX_{\ell}}{\d T}\cdot\del \widehat{\ki}_{00} =\delX^2\widehat{\ki}_{01}+2q\del\widehat{\ki}_{00}\cdot\del\widehat{\ki}_{10}.
\]
It is now clear that when $b$ is $O(1)$ as $q \ra 0$, a non-zero $b$ modifies the
law of motion only at $O(q)$, not at leading order. To have an effect
on the leading-order law of motion $b$ needs to be $O(1/q)$.
We outline here the modification to our calculations in this latter
case, and the resulting modified law of motion.

Writing $b=\tb/q$, $\mu = \tm q$, the outer equation (\ref{10}) reads
\begin{equation}
\e^2\tm\left(q-\ii \tb\right)\p_T = \e^2
\del^2\p+(1+\ii q)(1-|\p|^2)\p-\frac{\ii\e^2\alpha^2}{q}\p\label{A.outeral1q}, 
\end{equation}
and (\ref{11})-(\ref{12}) in terms of the modulus $f$ and phase $\chi$ 
 become
\begin{align}
\tm \e^2 (q f_T+\tb f\ki_T) &= \e^2 \del^{2} f -\e^2 f |\del \ki|^2+ (1-f^2)f,\\
\tm \e^2(q f^2 \ki_T-\tb f f_T)&= \e^2\del\cdot (f^2\del\ki)+qf^2(1-f^2)-\e^2\frac{\alpha^2}{q}f^2.
\end{align}
Expanding $\ki$ and $f$ in powers of $\e$  we find that equation (\ref{chi0eqn}) for 
leading-order phase becomes
\[
q\tm  \ki_{0T}(1-\tb) = \del^2 \ki_0 + q|\del\ki_0|^ 2-\al^2/q.
\]
Expanding  in powers of $q$ as usual we find that the terms
involving $\tb$ still  do not contribute at the relevant order in either the
near-field or canonical separation.

In the inner region the leading-order equation (\ref{psi0eqn}) is unchanged, while
the first-order equation (\ref{1orderinner}) becomes
\begin{equation*}
-\tm (q-\ii \tb)\frac{\d\bX_{\ell}}{\d T} \cdot \del \p_0 = \del^2
 \p_1+(1+\ii q)(\p_1(1-2|\p_0|^2)-\p_0^2\p_1^*).
\end{equation*}
or equivalently, in terms of $f_1$ and $\ki_1$,
\begin{eqnarray*}
- \tm \frac{\d\bX_{\ell}}{\d T}\cdot\left(q\del f_0+\tb f_0\del\ki_0\right) &=& \del^2 f_1
-f_1|\del  \ki_0|^2
-2 f_0 \del  \ki_0 \cdot \del \ki_1+
f_1 - 3 f_0^2 f_1,\\
- \tm f_0 \frac{\d\bX_{\ell}}{\d T}\cdot\left(q f_0\del \ki_0-\tb \del f_0\right) &=& \del \cdot(f_0^2\del
  \ki_1)+\del \cdot(2 f_0 f_1\del
  \ki_0)
+2 q f_0 f_1 -4 q f_0^3 f_1.
\end{eqnarray*}
When calculating the outer limit of the first-order inner equation (\ref{chi1a}) is now modified to
\[
- q \tm (1-\tb) \frac{\d\bX_{\ell}}{\d T}\cdot\del \widehat{\ki}_{00} =\delX^2\widehat{\ki}_{01}+2q\del\widehat{\ki}_{00}\cdot\del\widehat{\ki}_{10}.
\]
so that the solution
(\ref{h1solution}) is modified to
\beqa
\widehat{h}_1 &=&   -\frac{q \tm(1-\tb)  A_{\ell} \e^{-\ii  qn_{\ell}}(V_1-\ii
  V_2)}{4  }(R^{ \ii qn_{\ell} +1}+\gamma_1R^{1-\ii qn_{\ell}})\,
\ee^{(qn_{\ell}+\ii )\phi} \nonumber \\
&&\mbox{ }- \frac{q \tm(1-\tb) B_{\ell} \e^{\ii  qn_{\ell}}( V_1+ \ii  V_2)}{4  } (R^{
  -\ii qn_{\ell} +1}+\gamma_2R^{1+\ii qn_{\ell}})\, 
\ee^{(qn_{\ell}-\ii )\phi}.\label{A.outinh1q}
\eeqa
where
\[ \widehat{\ki}_{10}  = \frac{\widehat{h}_1 e^{- q
    \widehat{\ki}_{00}}}{q }.
\]
At this point the analysis for spirals at canonical separation and
those at near field separation differs, and we treat the two cases
separately.

\paragraph{Canonical separation} Since the leading-order inner equation is
independent of $\tb$, the leading-order matching 
is the same, giving $q\log(1/\eps)=\pi/2+\nu q$ as
before.
Matching the new solution \eqref{A.outinh1q} to the inner limit of the outer as  in \S\ref{sec:2.6} we find that (\ref{ki10}) becomes
\[
\ki_{10}\sim -\frac{\tm(1-\tilde{b}) r}{2}(V_1 \cos\phi+V_2\sin\phi)+\frac{n_\ell
  r}{\beta_{\ell}}\del \GG^\ell_{\mathrm{reg}}(\bX_{\ell})\cdot {\bf
  e}_{\phi} \qquad \mbox{ as } r \ra \infty.
\]
The solvability condition (\ref{compatibility}) is modified to
\begin{equation}
-\tm\tb  \pi \frac{\d{\bf X}_{\ell} }{\d T}\cdot {\bf d^{\perp}}
=\lim_{r \ra \infty} \int_0^{2\pi}({\bf e}_{\phi }\cdot {\bf d})\left(\frac{\partial\chi_{10}}{\partial 
 r}+\frac{\chi_{10}}{r}\right) \,
\d \phi,\label{newsolv} 
\end{equation}
where ${\bf d}^\perp = (-d_2,d_1)$. The terms in $\tb$ cancel, leaving the law of motion unchanged as 
\begin{equation*}
\frac{\d {\bf X}_{\ell}}{\d T}=-\frac{2 n_{\ell}}{\tm \beta_{\ell}}\, \del^\perp \GG^\ell_{\mathrm{reg}}(\bX_{\ell}).
\end{equation*}
We note that in \cite{AgCh09} there is a sign error which resulted in the terms involving $\tb$ adding up rather than cancelling, leading to an incorrect factor $1-2 \tb$ in the law of motion. 

\paragraph{Near field separation} 
In this case the computations follow similarly to the ones shown for the canonical separation. In particular, with $\tb$ non-zero the first order matching between the inner limit of the outer and outer limit of the inner
(\ref{ji1r}) becomes
\beqas
\inx\cdot\del \oGG^\ell_{\mathrm{reg}}(\bX_{\ell})&\sim&-\frac{\tm(1-\tb)}{4}\left(\frac{\ee^{-\ii 
    qn_\ell\log\e}(V_{1}-\ii V_{2})(1+\gamma_{1})r\ee^{\ii\phi}}{\ee^{-\ii
    qn_\ell\log\e}+\ee^{\ii qn_\ell\log\e}} +\frac{\ee^{\ii qn_\ell\log\e}(V_{1}+\ii
  V_{2})(1+\gamma_{2})r\ee^{-\ii\phi}}{\ee^{-\ii
    qn_\ell\log\e}+\ee^{\ii qn_\ell\log\e}}\right)
.
\eeqas
Solving for $\gamma_1$ and $\gamma_2$ and matching in the same way as done in \S\ref{sec:3.6} now gives
\beqas
\chi_{10} & \sim&-\frac{\tilde{\m}(1-\tilde{b})r}{4}\left(V_{1}\cos \phi +
                  V_{2}\sin\phi\right) 
\nonumber\\  
&&+ \frac{\tilde{\m}(1-\tilde{b}) 
  r}{4}\left(V_{1}\cos(\phi-2qn_{\ell}\log\e)+V_{2}
\sin(\phi-2qn_{\ell}\log\e)\right)\nonumber\\
&&+r\cos(qn_{\ell}\log\e)\left(\frac{\dd \oGG^\ell_{\mathrm{reg}}}{\dd
   X}({\bf
  X}_{\ell})\cos(\phi-qn_{\ell}\log 
    \e)+\frac{\dd \oGG^\ell_{\mathrm{reg}}}{\dd Y}({\bf 
    X}_{\ell})\sin(\phi-qn_{\ell}\log \e)\right).
    \eeqas
Then, using the  solvability condition (\ref{newsolv}), the law of
motion reads 
\beqas
\frac{\d {\bf X}_{\ell}}{\d T}
& = &  \frac{2 \cos (q n_\ell \log \eps)}{\tm(\tb^2 \cos^2 (q n_\ell \log \eps) + \sin^2(q n_\ell \log \eps))} \left(
\tb \cos (q n_\ell \log \eps) \del\GG^\ell_{\mathrm{reg}}(\bX_{\ell})\right.
\\ && \left. \hspace{7cm}+ \sin (q n_\ell \log \eps)\del^\perp\GG^\ell_{\mathrm{reg}}(\bX_{\ell})\right).
\eeqas
Again we note that the corresponding expression for an infinite domain in \cite{AgCh09} is incorrect because of the aforementioned sign error.



\end{document}